\documentclass[preprint]{elsarticle}

\usepackage[T1]{fontenc}

\usepackage{fullpage}
\usepackage{graphicx}
\usepackage{amsmath,amssymb}
\usepackage{mathrsfs}
\usepackage{float}
\usepackage[utf8]{inputenc}
\usepackage{listings}
\usepackage{color}
\usepackage{amsfonts}
\usepackage{array}
\usepackage{algpseudocode}
\usepackage{algorithm}
\usepackage{subcaption}
\usepackage{enumitem}
\biboptions{numbers,sort}

\numberwithin{equation}{section}

\DeclareMathOperator{\divr}{div}

\DeclareMathOperator{\Reyn}{Re}

\begin{document}

\title{A coupled finite volume and material point method for two-phase simulation of liquid--sediment and gas--sediment flows}

\author[1]{Aaron S. Baumgarten}
\author[1]{Benjamin L. S. Couchman}
\author[2]{Ken Kamrin\corref{cor1}}
\ead{kkamrin@mit.edu}

\cortext[cor1]{Corresponding author}

\address[1]{Department of Aeronautics and Astronautics,}
\address[2]{Department of Mechanical Engineering,\\ Massachusetts Institute of Technology, Cambridge, MA 02139, USA}

\begin{abstract}
	Mixtures of fluids and granular sediments play an important role in many industrial, geotechnical, and aerospace engineering problems, from waste management and transportation (liquid--sediment mixtures) to dust kick-up below helicopter rotors (gas--sediment mixtures). These mixed flows often  involve bulk motion of hundreds of billions of individual sediment particles and can contain both highly turbulent regions and static, non-flowing regions. This breadth of phenomena necessitates the use of continuum simulation methods, such as the material point method (MPM), which can accurately capture these large deformations while also tracking the Lagrangian features of the flow (e.g.\ the granular surface, elastic stress, etc.).
	
	Recent works using two-phase MPM frameworks to simulate these mixtures have shown substantial promise; however, these approaches are hindered by the numerical limitations of MPM when simulating pure fluids. In addition to the well-known particle ringing instability and difficulty defining inflow/outflow boundary conditions, MPM has a tendency to accumulate quadrature errors as materials deform, increasing the rate of overall error growth as simulations progress. In this work, we present an improved, two-phase continuum simulation framework that uses the finite volume method (FVM) to solve the fluid phase equations of motion and MPM to solve the solid phase equations of motion, substantially reducing the effect of these errors and providing better accuracy and stability for long-duration simulations of these mixtures.
	
\end{abstract}

\maketitle

\section{Introduction}
In this work, we are concerned with the numerical simulation of mixtures of fluids and grains in a wide range of applications and geometries. These mixtures are primarily composed of two materials: a porous granular solid (e.g.\ sand, powder, soil, grains) and a fluid (e.g.\ water, gas, air). Mixtures of relatively dense fluids and sediments play an important role in many industrial and geotechnical engineering problems, from transporting large volumes of industrial wastes \cite{turian1977} to building earthen levees and dams \cite{pailha2009}. On the other hand, mixtures of relatively light fluids (e.g.\ gases) and sand-like materials appear in vastly different engineering problems, from understanding dust kick-up below helicopter rotors \cite{keller2006} to modeling locomotion of wheels on rough terrain \cite{agarwal2019}.

To solve these problems, engineers have traditionally relied on a myriad of empirical models developed over the last century (e.g.\ models for the effective viscosity of mixtures \cite{einstein1906, batchelor1972, chong1971}, the permeability of granular beds \cite{carman1937}, and the various regimes of slurry transport \cite{turian1977}). These empirical models are derived by coupling relevant experimental observations to an understanding of the underlying physics governing the behavior of these mixtures; however, such models can only describe specific regimes of mixed flow. To address an engineering problem that involves complex interactions of fluids and grains spanning many flow regimes or in a complex geometry, engineers require a more general modeling approach.

A popular solution to this problem is modeling the microscopic behavior of these mixtures directly by simulating each grain--grain collision and calculating the exact motion of the pore fluid numerically (e.g.\ LBM--DEM \cite{cook2004}, CFD--DEM \cite{zhou2010}). This type of approach is highly accurate and has been used to develop complex constitutive relations for mixtures \cite{seto2013, mari2015discontinuous, amarsid2017}; to predict wave generation from landslides \cite{mao2020}; and to model consolidation, shearing, and collapse of saturated soil columns \cite{guo2016,kumar2017}. The primary limitation of this approach is difficulty simulating engineering problems involving large volumes of material. Since each grain and its surrounding fluid must be tracked over time, simulations involving hundreds of billions of sediment particles are computationally intractable. We therefore turn to a continuum modeling approach, where small scale phenomena are homogenized into bulk properties and behaviors.

Recent work simulating fluid-grain mixtures as continua (see \cite{soga2015}) presents a versatile foundation, but substantial improvement is necessary before existing simulation frameworks can be used for the range of engineering applications we seek to address. There are two primary challenges associated with such frameworks: i) accurate prediction of the stresses in the grains and fluid (i.e.\ the constitutive response of the mixture; see \cite{ceccato2016granular,fern2016}) and ii) accurate numerical approximation of the equations of motion for mixed flows (see \cite{bandara2015, baumgarten2019a}). It is this latter challenge that we seek to address in the present work.

When represented as continua, the individual grains and fluid-filled pores of these mixtures are homogenized into two independent material phases: the granular (or solid) phase and the fluid phase; each phase has its own velocity ($\boldsymbol{v}_s$ and $\boldsymbol{v}_f$), density ($\bar{\rho}_s$ and $\bar{\rho}_f$), and internal state ($\boldsymbol{\sigma}_s$, $\boldsymbol{\sigma}_f$, $\phi$, etc.). This requires that two sets of mass, momentum, and energy conservation laws be satisfied simultaneously and that the motion of the phases be tracked independently (see \cite{coussy2004, drumheller2000, jackson2000, klika2014}). Further, these mixtures are known to undergo extremely large deformations (as in riverbed erosion; see \cite{aussillous2013}) with internal flows that are highly coupled to surface topography changes (as in near-shore wave breaking; see \cite{mieras2019}), requiring both Lagrangian-like feature tracking for the deforming surfaces and Eulerian-like evaluation of the governing equations to avoid the issues associated with entangling meshes on the interior.


One continuum approach that has shown success in simulating these challenging problems is the two-phase material point method (MPM; see \cite{sulsky1994,abe2013,bandara2015, baumgarten2019a}) which has been used to model water-jet--soil interactions in \cite{liang2017}, fluidization of granular beds in \cite{redaelli2017}, submerged cone penetration in \cite{ceccato2016twophase}, slope failures in \cite{fern2016}, and shear thickening of dense suspensions in \cite{baumgarten2019b}. A key advantage of MPM is that it naturally captures Lagrangian flow features by using persistent material point tracers while solving the equations of motion on a temporary Eulerian grid. Despite its proven robustness in modeling many solid-like materials through large deformations, MPM is known to develop substantial integration errors when these deformations continue to grow, as occurs with pure fluids and mixtures with negligible solid volume fractions; these quadrature errors increase the rate of overall error growth during a simulation, limiting the use of MPM for long-duration fluid problems (see \cite{yang2018,bardenhagen2002,steffen2008examination,steffen2010,sulsky2016}). Although substantial work has been done to reduce these errors in MPM (see \cite{sadeghirad2011,zheng2013,sadeghirad2013,yue2015,nguyen2017,kularathna2017,moutsanidis2020}), it is desirable to avoid them completely where possible.

{The finite volume method (FVM) and finite element method (FEM) have also been applied to these types of mixture problems, though usually in situations where both phases of the mixture are fully fluidized (e.g.\ sediment transport; see \cite{david2011, wu2008, zhang2011two,phillips2009,ejtehadi2020}) or the solid phase remains static (e.g.\ fluid flow through subterranean faults and groundwater motion; see \cite{huyakorn1978,geiger2004,rees2004}). These approaches are frequently formulated in the Eulerian frame and are well suited to fluid dynamics problems.
In this work, we are particularly interested in FVM, which has the key advantage of naturally conserving mass, momentum, and energy by calculating fluxes between neighboring cells. Since these cells do not move with the material, FVM does not accumulate the kind of quadrature errors observed in MPM.} As a result, FVM can be applied to many fluid problems that are currently extremely difficult for MPM to model, such as turbulent flows in pipes and simulations involving gases. FVM can also be augmented to track Lagrangian features of fluid flows numerically (see \cite{hirt1981, sethian2003, udaykumar2001,sun2010}). Its long history of development and use for solving the equations of motion for fluids on both Eulerian and arbitrary Lagrangian-Eulerian grids (see \cite{roe1981,jameson1986,barth1989, trepanier1991}) make it an ideal choice for replacing MPM when substantial material deformation is expected. 
{(FEM could also be used for simulating the fluid phase, as in IGA-MPM \cite{moutsanidis2018}, but such an approach is not considered in this work.)}


In this work, we introduce a new approach for simulating fluid--sediment mixtures that combines the advantages of MPM for simulating the granular phase of the mixture with the proven robustness of FVM for simulating pure fluids. This approach discretizes the governing equations described in \cite{baumgarten2018,baumgarten2019a} on two overlapping simulation domains. Similar coupled approaches have shown success for fluid--structure interaction and solid--solid collision simulations (see \cite{moutsanidis2018,sun2010,gilmanov2008,chen2015}). Here the granular phase is represented on a set of material point tracers with kinematic fields (e.g.\ velocity) defined on an associated finite element grid, and the fluid phase is represented by average field quantities within a set of Eulerian finite volumes. In this way, we can use MPM to track the important Lagrangian features of the granular flow (e.g.\ solid volume fraction, elastic stress, granular free surface) and use FVM to solve the fluid equations of motion without sacrificing accuracy in long-running turbulent flows.

\section{Governing Equations}
We begin describing our approach for simulating fluid--sediment mixtures by reviewing the system of governing equations that we seek to solve. These equations derive from the continuum mixture theories of \cite{jackson2000, drumheller2000, coussy2004} (see Appendix \ref{sec:mixture_theory}) and can also be found in \cite{baumgarten2019a} where they are simulated using the two-phase MPM approach. In this formulation, we assume that the grains are rough (i.e.\ true contact can occur between grains; see \cite{zhao2002}), incompressible with density $\rho_s$, and essentially spherical with mean diameter $d$. Additionally, we assume that the grains are quasi-mono-disperse (no size-segregation during flow) and fully immersed in a compressible fluid having density $\rho_f$ and viscosity $\eta_0$.

To capture the dynamic behavior of the coupled flows we are interested in, we consider the materials that constitute our mixture independently. The individual solid grains are homogenized into a single continuum body called the \textit{solid} (or \textit{granular}) \textit{phase}, which has an effective density $\bar{\rho}_s$, velocity $\boldsymbol{v}_s$, and specific internal energy $\varepsilon_s$. Similarly, the fluid that fills the pore space between grains is homogenized into a single continuum body called the \textit{fluid phase}, which has an effective density $\bar{\rho}_f$, velocity $\boldsymbol{v}_f$, and specific internal energy $\varepsilon_f$. 
%

If we let $n$ denote the volume fraction of fluid in the mixture (also called the \textit{porosity} of the mixture) and $\phi$ denote the volume fraction of solid (also called the granular \textit{packing fraction}) then the phase-wise effective densities ($\bar{\rho}_s$ and $\bar{\rho}_f$) can be defined by the following relations for all spatial points $\boldsymbol{x}$ in the mixed domain $\Omega$,
\begin{equation}
\phi = 1 - n, \qquad \bar{\rho}_s = \phi \rho_s, \qquad \bar{\rho}_f = n \rho_f.
\label{eqn:porosity}
\end{equation}

Since we are interested in tracking Lagrangian flow features during our simulations, it is useful to express the equations governing the behavior of the solid phase in their Lagrangian form {(i.e.\ in the \textit{material} reference frame)}. To do this, we define the solid phase material derivative, $D^s/Dt$, as follows,
\begin{equation}
\label{eqn:material_derivative_solid}
\frac{D^s \psi}{Dt} \equiv \frac{\partial \psi}{\partial t} + \boldsymbol{v}_s \cdot \nabla \psi,
\end{equation}
for an arbitrary scalar field $\psi$ and with $\nabla$ the \textit{spatial} {(or Eulerian)} gradient operator (i.e.\ $\nabla \equiv \partial/\partial \boldsymbol{x}$).
Mass conservation in the solid phase takes the following form,
\begin{equation}
\label{eqn:mass_conservation_solid}
\frac{D^s \bar{\rho}_s}{Dt} = - \bar{\rho}_s \divr(\boldsymbol{v}_s),
\end{equation}
with $\divr()$ the \textit{spatial} divergence operator (i.e.\ $\divr(\boldsymbol{v}_s) \equiv \nabla \cdot \boldsymbol{v}_s$). Since we assume that $\rho_s$ is constant and uniform everywhere, this equation also allows us to track the evolution of the granular packing fraction and mixture porosity over time.
Momentum conservation in the solid phase takes the following form,
\begin{equation}
\label{eqn:momentum_conservation_solid}
\bar{\rho}_s \frac{D^s \boldsymbol{v}_s}{Dt} = \divr(\boldsymbol{\tilde{\sigma}}) + \bar{\rho}_s \boldsymbol{g} - \boldsymbol{f_d} - \phi \nabla p_f,
\end{equation}
with $\boldsymbol{\tilde{\sigma}}$ the \textit{effective granular stress} tensor, $\boldsymbol{g}$ the gravitational acceleration vector, $\boldsymbol{f}_d$ the inter-phase drag vector, and $p_f$ the fluid phase \textit{pore pressure}.

The effective granular stress tensor, $\boldsymbol{\tilde{\sigma}}$, has many different admissible forms depending both on the type of material used and the flow regimes under consideration. Commonly referenced examples include models for wet clays and dense sediments \cite{roscoe1968,terzaghi1943}, models for dry granular materials \cite{jop2006,dunatunga2015}, models for dense slurries \cite{amarsid2017,boyer2011}, and kinetic theories for fast moving granular gases \cite{jenkins1983,brilliantov2010}. In this work, we assume that each of these models can be expressed generally as,
\begin{equation}
\label{eqn:stress_equation_solid}
\boldsymbol{\tilde{\sigma}} = \boldsymbol{\tilde{T}}\big(\nabla \boldsymbol{v}_s, \boldsymbol{\bar{\xi}}\big),
\end{equation}
with $\boldsymbol{\tilde{T}}$ a model-specific function of the solid phase velocity gradient and the current solid phase state vector $\boldsymbol{\bar{\xi}}$. This state vector fully describes the relevant properties of the granular material (e.g.\ deformation gradient, packing fraction, internal energy, fabric tensor, etc.) and can evolve over time. This equation of state often takes the place of an explicit energy conservation equation when isothermal or adiabatic assumptions are made. In this work, we will primarily use the granular material models described in \cite{dunatunga2015,baumgarten2019a}.

The inter-phase drag vector, $\boldsymbol{f_d}$, captures the tractions on particle surfaces that arise due to the relative motion of the particles and pore fluid (e.g.\ Stokes drag, Darcy drag, etc.). This force (along with buoyancy) couples the behavior of the mixed phases. Following the works of \cite{vanderhoef2005,beetstra2007}, we let the drag vector take the form,
\begin{equation}
\boldsymbol{f_d} = \frac{18 \phi (1-\phi) \eta_0}{d^2}\ \hat{F}(\phi,\Reyn)\  (\boldsymbol{v}_s - \boldsymbol{v}_f),
\label{eqn:drag_force}
\end{equation}
with $\Reyn \equiv (n \rho_f d\ \|\boldsymbol{v}_s - \boldsymbol{v}_f\|) / \eta_0$ (see \cite{dupuit1863}) and $\hat{F}(\phi,\Reyn)$ a function given in those works.

The fluid pore pressure, $p_f$, is commonly given through an equation of state that has the following form,
\begin{equation}
\label{eqn:pressure_equation_fluid}
p_f = \hat{p}_f \big( \rho_f, \varepsilon_f \big),
\end{equation}
with $\hat{p}_f$ a model-specific function 
{of the fluid density and specific internal energy.}
In \cite{bandara2015, baumgarten2019a} an isothermal, weakly-compressible, barotropic model is used; however, an ideal gas law or thermo-fluid model is also valid.

Mass conservation in the fluid phase, expressed in an Eulerian frame, takes the following form,
\begin{equation}
\label{eqn:mass_conservation_fluid}
\frac{\partial \bar{\rho}_f}{\partial t} = - \divr(\bar{\rho}_f \boldsymbol{v}_f).
\end{equation}
Since we assume a compressible fluid fills the pores of the granular skeleton, we can use equations \eqref{eqn:porosity}, \eqref{eqn:mass_conservation_solid}, and \eqref{eqn:mass_conservation_fluid} to uniquely determine the \textit{true} fluid density, $\rho_f$.

Momentum conservation in the fluid phase, expressed in an Eulerian frame, takes the following form,
\begin{equation}
\label{eqn:momentum_conservation_fluid}
\frac{\partial \bar{\rho}_f \boldsymbol{v}_f}{\partial t} = - \divr \big( n p_f \boldsymbol{I} + \bar{\rho}_f\boldsymbol{v}_f \otimes \boldsymbol{v}_f \big)  + \divr(\boldsymbol{\tau_f}) + \bar{\rho}_f \boldsymbol{g} + \boldsymbol{f_d} + p_f \nabla n,
\end{equation}
with $\boldsymbol{\tau_f}$ the \textit{effective fluid shear stress} tensor and $\otimes$ the tensor product.

The fluid shear stress tensor, $\boldsymbol{\tau_f}$, is often expressed as,
\begin{equation}
\label{eqn:shear_stress_equation_fluid}
\boldsymbol{\tau_f} = 2 \eta_r \boldsymbol{D}_{f0},
\end{equation}
with $\boldsymbol{D}_{f0}$ the deviatoric component of the fluid strain-rate tensor, $\boldsymbol{D}_f \equiv \tfrac{1}{2}(\nabla \boldsymbol{v}_f  + \nabla \boldsymbol{v}_f ^\top)$, and $\eta_r$ an effective mixture viscosity. Although there is no universally accepted model for this component of stress across all fluid flow regimes (see \cite{stickel2005}), it is generally agreed that the effective viscosity varies with the solid volume fraction $\phi$; first-order (see \cite{einstein1906}), second-order (see \cite{batchelor1972}), and higher-order (see \cite{jackson2000}) forms appear throughout the literature. As in \cite{baumgarten2019a}, we will consider only the first-order form of this stress given in \cite{einstein1906}.

Energy conservation in the fluid phase, expressed in an Eulerian frame, takes the following form,
\begin{equation}
\label{eqn:energy_conservation_fluid}
\frac{\partial \bar{\rho}_f E_f}{\partial t} = -\divr \big((\bar{\rho}_f E_f + np_f) \boldsymbol{v}_f \big) + \divr(\boldsymbol{\tau_f} \boldsymbol{v}_f - \boldsymbol{q}_f) +  (\bar{\rho}_f \boldsymbol{g}) \cdot \boldsymbol{v}_f + \big(p_f \nabla n + \boldsymbol{f_d}\big) \cdot \boldsymbol{v}_s - {\phi p_f \divr(\boldsymbol{v}_s)},
\end{equation}
with $\boldsymbol{q}_f$ the \textit{effective fluid heat flow} vector and $E_f$ the \textit{specific total energy} of the fluid with $E_f \equiv \varepsilon_f + \tfrac{1}{2} \boldsymbol{v}_f \cdot \boldsymbol{v}_f$. 
(Note that this form of the specific total energy does not include the contributions of microscopic fluid velocity fluctuations that may arise in mixed flows, such as from tortuosity; see \cite{batchelor1970, coussy2004, wilmanski2005, kosinski2002} and Appendix \ref{sec:mixture_theory} for further discussion.)

The effective fluid heat flow vector, $\boldsymbol{q}_f$, takes the following form,
\begin{equation}
\label{eqn:heat_flux_equation_fluid}
\boldsymbol{q}_f = - n k_f \nabla \vartheta_f,
\end{equation}
with $k_f$ the \textit{effective coefficient of thermal conductivity} in the fluid and $\vartheta_f$ the fluid phase temperature.

\section{Mixed Weak- and Integral-Form of Governing Equations}
In this section, we introduce a numerical approximation of the governing equations described above that can be evaluated using the two-phase, finite volume and material point method (FV-MPM). The governing equations are summarized as follows:
\begin{equation}\label{eqn:mixture_equations}
	\begin{aligned}
		\frac{D^s \bar{\rho}_s}{D t} &= -\bar{\rho}_s \divr(\boldsymbol{v}_s),\\
		\bar{\rho}_s \frac{D^s \boldsymbol{v}_s}{Dt} &= \divr(\boldsymbol{\tilde{\sigma}}) + \bar{\rho}_s \boldsymbol{g} - \boldsymbol{f_d} - \phi \nabla p_f,\\
		\frac{\partial \bar{\rho}_f}{\partial t} &= -\divr (\bar{\rho}_f \boldsymbol{v}_f),\\
		\frac{\partial \bar{\rho}_f \boldsymbol{v}_f}{\partial t} &= - \divr \big( \bar{\rho}_f\boldsymbol{v}_f \otimes \boldsymbol{v}_f + n p_f \boldsymbol{I} \big)  + \divr(\boldsymbol{\tau_f}) + \bar{\rho}_f \boldsymbol{g} + \boldsymbol{f_d} + p_f \nabla n,\\
		\frac{\partial \bar{\rho}_f E_f}{\partial t} &= -\divr \big((\bar{\rho}_f E_f + np_f) \boldsymbol{v}_f \big) + \divr(\boldsymbol{\tau_f} \boldsymbol{v}_f - \boldsymbol{q}_f) + (\bar{\rho}_f \boldsymbol{g}) \cdot \boldsymbol{v}_f + \big(p_f \nabla n + \boldsymbol{f_d}\big) \cdot \boldsymbol{v}_s - {\phi p_f \divr(\boldsymbol{v}_s).}
	\end{aligned}
\end{equation}
{Note that the two material phases of the mixture interact through drag and buoyant forces. Interactions between a fluid and a non-porous solid body require additional compatibility conditions.}

Suppose we multiply the first two equations in \eqref{eqn:mixture_equations} by a scalar test function, $w$, and a vector test function, $\boldsymbol{w}$, respectively, and integrate both expressions over the same simulated domain, $\Omega$. Suppose, further, that we integrate the final three equations in \eqref{eqn:mixture_equations} over an arbitrary subdomain, $\Omega_\alpha$, with boundary $\partial \Omega_\alpha$. After applying the divergence theorem where relevant, the resulting set of equations can be separated into three distinct groups:
i) a weak expression of mass conservation in the solid phase to be evaluated on the material points of MPM,
\begin{equation}
	\label{eqn:weak_mass_conservation_solid}
	\int_\Omega w \frac{D^s \bar{\rho}_s}{D t}\ dv = -\int_\Omega w \bar{\rho}_s \divr(\boldsymbol{v}_s)\ dv;
\end{equation}
ii) a weak expression of momentum conservation in the solid phase to be evaluated on the background finite element grid of MPM,
\begin{equation}
	\label{eqn:weak_momentum_conservation_solid}
	\int_\Omega \bar{\rho}_s \boldsymbol{a}_s \cdot \boldsymbol{w}\ dv = \int_\Omega \big(-\boldsymbol{\tilde{\sigma}}:\nabla \boldsymbol{w} + \bar{\rho}_s \boldsymbol{g} \cdot \boldsymbol{w} - \boldsymbol{f_d} \cdot \boldsymbol{w} + { p_f \boldsymbol{I} : \nabla(\phi \boldsymbol{w})\big)} dv + \int_{\partial \Omega} \boldsymbol{\tau_s} \cdot \boldsymbol{w}\ da;
\end{equation}
and iii) integral expressions for mass, momentum, and energy conservation in the fluid phase to be evaluated on each subdomain of the finite volume grid of FVM,
\begin{equation}\label{eqn:weak_mixture_equations_fluid}
	\begin{aligned}
		\frac{\partial}{\partial t}\bigg(\int_{\Omega_\alpha} \bar{\rho}_f\ dv\bigg) =& -\int_{\partial \Omega_\alpha} \bar{\rho}_f \boldsymbol{v}_f \cdot \hat{\boldsymbol{n}}\ da,\\
		\frac{\partial}{\partial t}\bigg(\int_{\Omega_\alpha} \bar{\rho}_f \boldsymbol{v}_f\ dv\bigg) =&- \int_{\partial \Omega_\alpha}{ \big( \bar{\rho}_f\boldsymbol{v}_f (\boldsymbol{v}_f \cdot \hat{\boldsymbol{n}}) + n p_f \hat{\boldsymbol{n}} \big)\ da}\\
		& +\int_{\partial \Omega_\alpha}{ \boldsymbol{\tau_f} \hat{\boldsymbol{n}}\ da}\\
		& + \int_{\Omega_\alpha}{ \big(\bar{\rho}_f \boldsymbol{g} + \boldsymbol{f_d} + p_f \nabla n\big)\ dv},\\
		\frac{\partial}{\partial t} \bigg(\int_{\Omega_\alpha} \bar{\rho}_f E_f\ dv \bigg) =&-\int_{\partial \Omega_\alpha} {(\bar{\rho}_f E_f + np_f) (\boldsymbol{v}_f \cdot \hat{\boldsymbol{n}})\ da}\\
		& + \int_{\partial \Omega_\alpha}{(\boldsymbol{\tau_f} \boldsymbol{v}_f - \boldsymbol{q}_f) \cdot \hat{\boldsymbol{n}}\ da}\\
		& + \int_{\Omega_\alpha}{ (\bar{\rho}_f \boldsymbol{v}_f \cdot \boldsymbol{g})\ dv}\\
		& + \int_{\Omega_\alpha}{\big(p_f \nabla n + \boldsymbol{f_d}\big) \cdot \boldsymbol{v}_s - {\phi p_f \divr(\boldsymbol{v}_s)}\ dv}.
	\end{aligned}
\end{equation}
(Note that $\boldsymbol{\tau_s}$ in equation \eqref{eqn:weak_momentum_conservation_solid} denotes the prescribed traction on the solid phase at the simulated domain boundary $\partial \Omega$.)

\begin{figure}[!ht]
	\centering
	\includegraphics[scale=0.45]{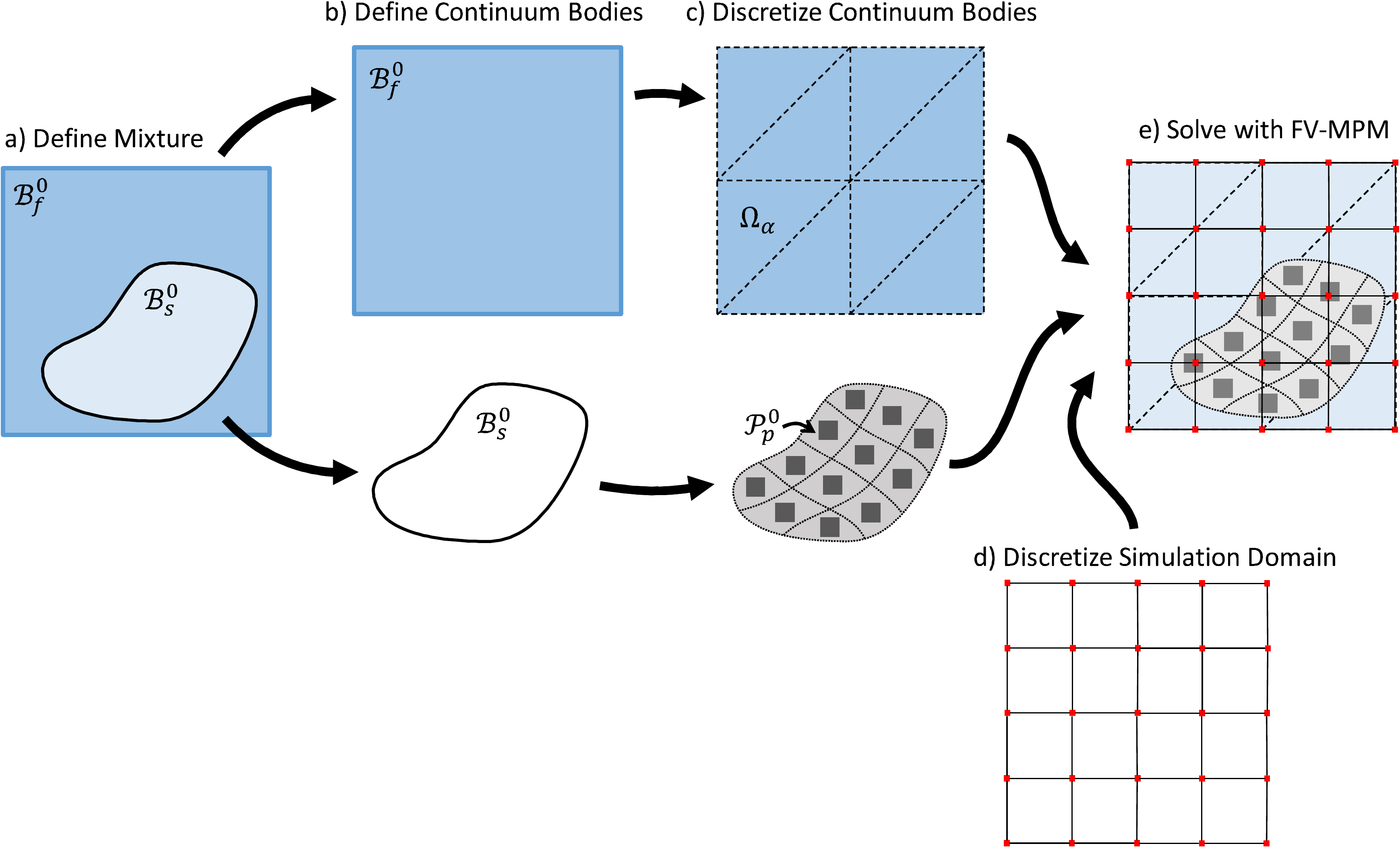}
	\caption{The basic method of solving mixture problems using the finite volume and material point method (FV-MPM). a) Define the mixture and initial configuration including densities, porosities, stresses. b) Define the initial solid and fluid phase continuum bodies ($\mathcal{B}_s^0$ and $\mathcal{B}_f^0$, respectively). c) Break the continuum bodies into piecewise-defined chunks of material: use material points to track the solid phase and finite volumes to track the fluid ($\mathcal{P}_p^0$ is the body part associated with the $p$th material point). d) Define finite element nodal basis functions and simulation domain. e) Solve the discrete equations of motion for the mixture using FV-MPM.}
	\label{fig:discretization}
\end{figure}

\begin{figure}[!ht]
	\centering
	\includegraphics[scale=0.4]{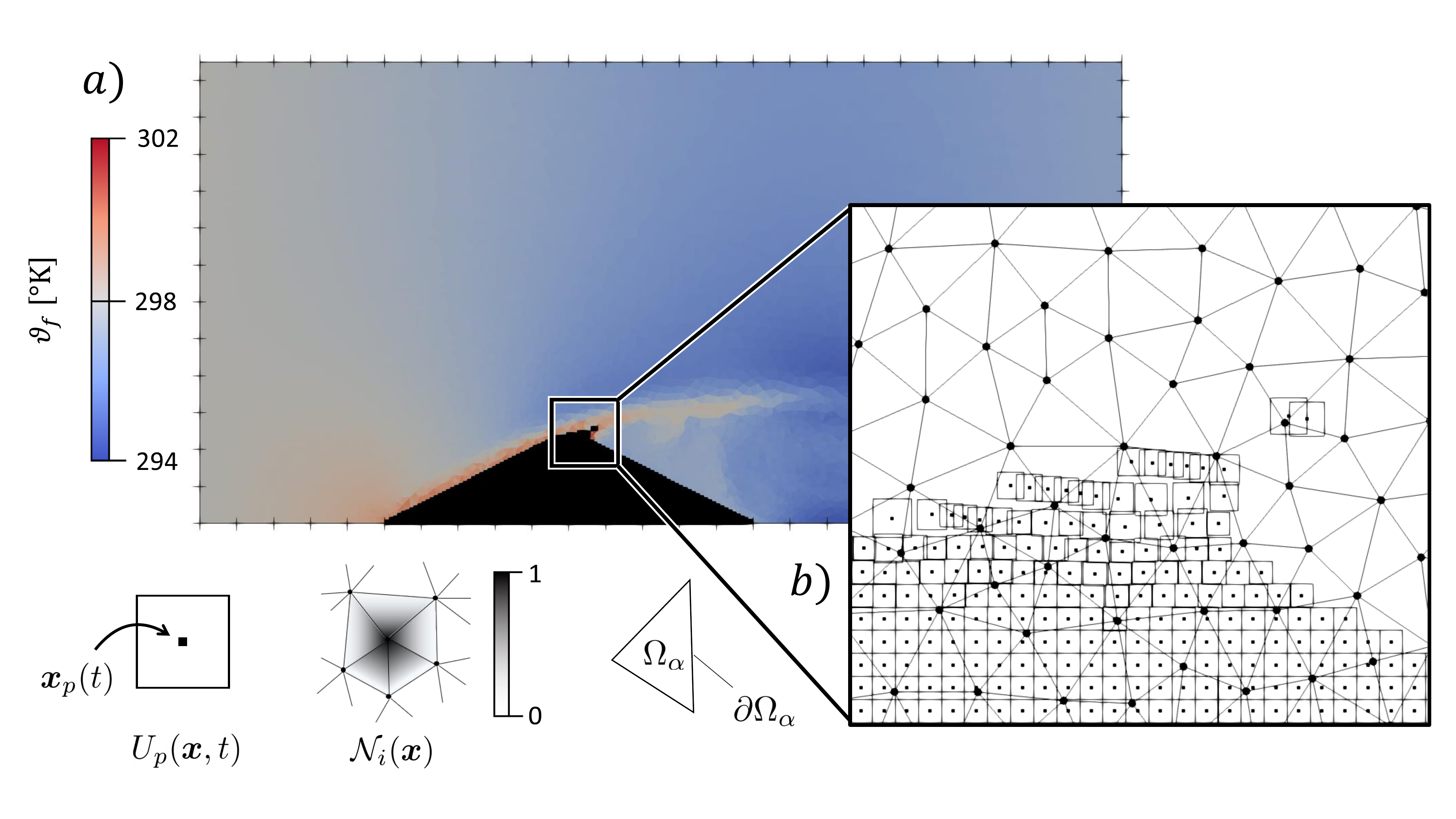}
	\caption{An example of a two-phase finite volume and material point method (FV-MPM) simulation highlighting the different solution spaces associated with each phase of the mixture. a) A snapshot from the simulation of a sand--air mixture with wind blowing from left to right; the granular phase is represented by black material tracers and the fluid phase is colored according to the local effective temperature, $\vartheta_f$. b) A region of this simulation showing the material point characteristic functions, $U_p(\boldsymbol{x},t)$, visualized using their GIMP representations (squares; see \cite{bardenhagen2004}); the nodal basis functions, $\mathcal{N}_i(\boldsymbol{x})$, and associated nodes (black circles); and the finite volume subdomains, $\Omega_\alpha$ (triangles). Here, we chose nodal basis functions and finite volume subdomains that are defined on the same triangular grid.
	}
	\label{fig:discretization_example}
\end{figure}

In order to evaluate these expressions numerically, we choose to represent the mixture fields on three separate solution spaces associated with the three groups above:
\begin{enumerate}[label=\roman*)]
	\item a set of $N_m$ material point characteristic functions, $\{U_p(\boldsymbol{x},t)\ |\ p \in [1,N_m]\}$,
	\item a set of $N_n$ nodal basis functions, $\{\mathcal{N}_i(\boldsymbol{x},t)\ |\ i \in[1,N_n]\}$,
	\item and a set of $N_v$ finite volume domains, $\{\Omega_\alpha\ |\ \alpha\in[1,N_v]\}$,
\end{enumerate}
with $\boldsymbol{x}$ a spatial position within the simulated domain $\Omega$ and $t$ a point in time within the simulated period $[0,t_{\text{end}}]$. A visual representation of these different spaces is shown in Figure \ref{fig:discretization} for general problems and in Figure \ref{fig:discretization_example} for the sand--air mixture simulation shown in Figure \ref{fig:erosion_2D}a.

\subsection{Material Point Characteristic Functions}
Consider first the material points. Suppose the initial domain of the solid phase continuum, $\mathcal{B}_s^0$, is composed of $N_m$ distinct subdomains, $\{\mathcal{P}_p^0\ |\ p\in[1,N_m]\}$, as in Figure \ref{fig:discretization}c. We can then use these subdomains to define the material points characteristic functions as follows,
\begin{equation}
	\sum_{p = 1}^{N_m}U_p(\boldsymbol{x},0) = \bigg\{ \begin{matrix}
		1&\boldsymbol{x} \in \mathcal{B}_s^0,\\
		0&\text{else}
	\end{matrix} \quad \text{and} \quad
	U_p(\boldsymbol{x},0) = \bigg\{ \begin{matrix}
		1&\boldsymbol{x} \in \mathcal{P}_p^0,\\
		0&\text{else}
	\end{matrix} \quad \forall p \in [1,N_m].
	\label{eqn:point_characteristic_function}
\end{equation}
{(Note that these `top-hat' characteristic functions are defined as in the GIMP method proposed in \cite{bardenhagen2004} and replace the Dirac delta functions originally used in \cite{sulsky1994}.)} 
During a simulation, the solid phase domain and its $p$th subdomain will evolve to $\mathcal{B}_s^t$ and $\mathcal{P}_p^t$, respectively, and this evolution can be tracked using the characteristic functions if we require that they be \textit{co-moving with the material} (i.e\ $D^s(U_p(\boldsymbol{x},t))/Dt = 0$). In practice, it is also useful to define a set of material point weights, $v_{p}(t)$, and centroids, $\boldsymbol{x}_p(t)$, as follows,
\begin{equation}
	\label{eqn:point_volume_and_centroid}
	v_{p}(t) = \int_{\Omega}{ U_p(\boldsymbol{x},t) dv}, \quad
	\boldsymbol{x}_p(t) = \frac{1}{v_p(t)} \int_{\Omega}{\boldsymbol{x}\  U_p(\boldsymbol{x},t) dv}.
\end{equation}
As is common in MPM, we let these characteristic functions define the solid phase material fields ($\bar{\rho}_s$, $\boldsymbol{\tilde{\sigma}}$, and $\boldsymbol{\bar{\xi}}$) as,
\begin{equation}
	\begin{aligned}
		\bar{\rho}_s = \sum_{p=1}^{N_m}\bar{\rho}_{sp}(t) \ U_p(\boldsymbol{x},t), \quad
		\boldsymbol{\tilde{\sigma}} = \sum_{p=1}^{N_m}\boldsymbol{\tilde{\sigma}}_p(t)\ U_p(\boldsymbol{x},t), \quad
		\boldsymbol{\bar{\xi}} = \sum_{p=1}^{N_m}\boldsymbol{\bar{\xi}}_p(t)\ U_p(\boldsymbol{x},t),
	\end{aligned}
	\label{eqn:solid_continuum_discretization}
\end{equation}
with the test function $w$ defined similarly,
\begin{equation}
	w = \sum_{i=1}^{N_m} w_p U_p(\boldsymbol{x},t).
	\label{eqn:solid_test_function}
\end{equation}
Here $\bar{\rho}_{sp}(t)$, $\boldsymbol{\tilde{\sigma}}_p(t)$, $\boldsymbol{\bar{\xi}}_p(t)$, and $w_p$ are the coefficients associated with the $p$th characteristic function and, by extension, the value of these fields at the $p$th material point centroid, $\boldsymbol{x}_p(t)$.

\subsection{Finite Element Basis Functions}
Consider now the finite element nodal basis functions. This set of $N_n$ continuous ($\mathcal{C}^0$) functions is defined on an Eulerian computational grid, see Figure \ref{fig:discretization}d, with the property that 
$
\sum_{i=1}^{N_n} \mathcal{N}_i(\boldsymbol{x}) = 1.
$
Since these functions are \textit{not} co-moving with the material, we have,
$
\partial \big(\mathcal{N}_i(\boldsymbol{x})\big)/\partial t = 0.
$
As in MPM, we use these functions to define the solid phase kinematic fields, $\boldsymbol{a}_s$ and $\boldsymbol{v}_s$, as follows,
\begin{equation}\label{eqn:solid_grid_discretization}
	\begin{aligned}
		\frac{D^s\boldsymbol{v}_s}{Dt} = \boldsymbol{a}_s = \sum_{i=1}^{N_n} \boldsymbol{a}_{si}(t)\ \mathcal{N}_i(\boldsymbol{x}), \quad
		\boldsymbol{v}_s = \sum_{i=1}^{N_n} \boldsymbol{v}_{si}(t)\ \mathcal{N}_i(\boldsymbol{x}),
	\end{aligned}
\end{equation}
along with the test function $\boldsymbol{w}$ and the mixture porosity field $n$,
\begin{equation}\label{eqn:test_function_and_porosity}
	\boldsymbol{w} = \sum_{i=1}^{N_n} \boldsymbol{w}_i\ \mathcal{N}_i(\boldsymbol{x}), \quad
	n = \sum_{i=1}^{N_n} n_i(t)\ \mathcal{N}_i(\boldsymbol{x}).
\end{equation}
Here $\boldsymbol{a}_{si}(t)$, $\boldsymbol{v}_{si}(t)$, $\boldsymbol{w}_i$, and $n_i(t)$ are the coefficients associated with the $i$th nodal basis function but not necessarily the value of these fields at the position $i$th node (e.g.\ B-spline coefficients; see \cite{steffen2008analysis}).

\subsection{Finite Volume Subdomains}
The final space on which our mixture solutions are defined is the set of $N_v$ non-overlapping finite volumes. This set of volumes is defined on an Eulerian computational grid, which \textit{may be unique from the grid associated with the $N_n$ nodal bases}, with 
$
\cup_{\alpha=1}^{N_v} \Omega_\alpha = \Omega
$
and $\Omega_\alpha$ the domain encompassed by the $\alpha$th finite volume, see Figure \ref{fig:discretization}c. Each of these volumes has an associated boundary $\partial \Omega_\alpha$, an associated weight, $V_\alpha$, and an associated centroid, $\boldsymbol{X}_\alpha$, defined as,
\begin{equation}
	V_\alpha \equiv \int_{\Omega_\alpha}{1\ dv}, \quad 
	{
		\boldsymbol{X}_\alpha \equiv \frac{1}{V_\alpha} \int_{\Omega_\alpha} {\boldsymbol{x}\ dv},
	}
\end{equation}
which are fixed in the Eulerian frame, such that
$
\partial\big(V_\alpha\big)/\partial t = 0
$
and
$
\partial \big( \boldsymbol{X}_\alpha \big)/\partial t = \boldsymbol{0}.
$
As is common in FVM, we track the fluid phase effective density ($\bar{\rho}_f$), momentum ($\bar{\rho}_f \boldsymbol{v}_f$), and energy ($\bar{\rho}_f E_f$) by defining cell-wise averages for each of these fields over each of the $N_v$ finite volumes:
\begin{equation}
	\label{eqn:finite_volume_conservation}
	\begin{aligned}
		\langle\bar{\rho}_{f}\rangle_\alpha(t) = \frac{1}{V_\alpha} \int_{\Omega_\alpha}{\bar{\rho}_f\ dv}, \quad
		\langle\bar{\rho}_{f} \boldsymbol{v}_f\rangle_\alpha(t) = \frac{1}{V_\alpha} \int_{\Omega_\alpha}{\bar{\rho}_f \boldsymbol{v}_f\ dv}, \quad
		\langle\bar{\rho}_{f} E_f\rangle_\alpha(t) = \frac{1}{V_\alpha} \int_{\Omega_\alpha}{\bar{\rho}_f E_f \ dv},
	\end{aligned}
\end{equation}
with $\langle\bar{\rho}_{f}\rangle_\alpha(t)$, $\langle\bar{\rho}_{f} \boldsymbol{v}_f\rangle_\alpha(t)$, and $\langle\bar{\rho}_{f} E_f\rangle_\alpha(t)$ are time-dependent cell averages associated with the volume $\Omega_\alpha$.

\subsection{Discrete Weak- and Integral-Form Equations}
We now return to the weak- and integral-form governing equations defined in \eqref{eqn:weak_mass_conservation_solid}, \eqref{eqn:weak_momentum_conservation_solid}, and \eqref{eqn:weak_mixture_equations_fluid}. Substitution of the solution fields described above into these equations can be simplified by using the mapping matrices $[\mathcal{S}]$, $[\mathcal{G}]$, $[\mathcal{A}]$, $[\mathcal{B}]$, and $[\mathcal{M}]$ with components defined as follows,
\begin{equation}
	\label{eqn:mapping_matrices}
	\begin{aligned}
		\mathcal{S}_{ip} &= \frac{1}{v_p} \int_{\Omega}{ \mathcal{N}_i(\boldsymbol{x}) U_p(\boldsymbol{x},t) \ dv},\\
		\mathcal{G}_{ip} &= \frac{1}{v_p} \int_{\Omega}{ \nabla \mathcal{N}_i(\boldsymbol{x}) U_p(\boldsymbol{x},t)\ dv},\\
		\mathcal{A}_{i\alpha} &= \frac{1}{V_\alpha} \int_{\Omega_\alpha}{\mathcal{N}_i(\boldsymbol{x})\ dv},\\
		\mathcal{B}_{ij} &= \int_{\Omega} \mathcal{N}_i(\boldsymbol{x}) \mathcal{N}_j(\boldsymbol{x})\ dv,\\
		\mathcal{M}_{ij} &= \sum_{p=1}^{N_m} \int_{\Omega}{ \bar{\rho}_{sp} U_p(\boldsymbol{x},t) \mathcal{N}_i(\boldsymbol{x}) \mathcal{N}_j(\boldsymbol{x})\ dv}.
	\end{aligned}
\end{equation}
$[\mathcal{S}]$ and $[\mathcal{G}]$ are the standard, time-varying MPM mapping matrices that allow for cross integration of fields between the material points and the finite element nodes, $[\mathcal{A}]$ is the FE-FVM mapping matrix that allows for integration of fields defined on the finite element bases over the set of finite volumes, $[\mathcal{B}]$ is the volume matrix of the finite element grid, and $[\mathcal{M}]$ is the time-dependent finite element mass matrix associated with the solid phase momentum balance equation.

Using these matrices and equations \eqref{eqn:point_characteristic_function}, \eqref{eqn:solid_continuum_discretization}, \eqref{eqn:solid_test_function}, and \eqref{eqn:solid_grid_discretization}, we can re-express the solid phase mass conservation expression in \eqref{eqn:weak_mass_conservation_solid} as follows,
\begin{equation}
	\label{eqn:density_evolution}
	\frac{d}{d t}\big(\bar{\rho}_{sp}\big) = -\sum_{i=1}^{N_n} \bar{\rho}_{sp} \boldsymbol{v}_{si} \cdot \mathcal{G}_{ip}, \qquad \forall p \in [1,N_m],
\end{equation}
though it is sometimes more useful to express this equation in terms of the fixed material point masses, $m_p = v_p \bar{\rho}_{sp}$, with,
\begin{equation}
	\label{eqn:volume_evolution}
	\frac{d}{d t}\big(m_p) = 0, \qquad
	\frac{1}{v_p}\frac{d}{d t}\big(v_p\big) = \sum_{i=1}^{N_n} \boldsymbol{v}_{si} \cdot \mathcal{G}_{ip}, \qquad \forall p \in [1,N_m].
\end{equation}
Similarly, recognizing that the components of $\boldsymbol{w}$ can be defined arbitrarily, we can substitute \eqref{eqn:solid_continuum_discretization}, \eqref{eqn:solid_grid_discretization}, and \eqref{eqn:test_function_and_porosity} into the solid phase momentum conservation expression in \eqref{eqn:weak_momentum_conservation_solid} to find,
\begin{equation}
	\label{eqn:momentum_evolution}
	\begin{aligned}
		\sum_{j=1}^{N_n} \mathcal{M}_{ij} \boldsymbol{a}_{sj}  &= \boldsymbol{f}_i^{\text{int}} + \boldsymbol{f}_i^{\text{ext}} + \boldsymbol{f}_i^{\text{drag}} + \boldsymbol{f}_i^{\text{buoy}} + \boldsymbol{f}_i^{\boldsymbol{\tau}}, \qquad \forall i \in [1,N_n]\\
		\boldsymbol{f}_i^{\text{int}} &= -\sum_{p=1}^{N_m} {\mathcal{G}_{ip}}^\top v_p \boldsymbol{\tilde{\sigma}}_p, \\
		\boldsymbol{f}_i^{\text{ext}} &= \sum_{p=1}^{N_m} \mathcal{S}_{ip} m_p \boldsymbol{g}, \\
		\boldsymbol{f}_i^{\text{drag}} &= - \int_{\Omega}{\boldsymbol{f_d}\ \mathcal{N}_i(\boldsymbol{x})\ dv},\\
		\boldsymbol{f}_i^{\text{buoy}} &= \sum_{j=1}^{N_n} (1 - n_j) \int_{\Omega}{\nabla \big( \mathcal{N}_i(\boldsymbol{x}) \mathcal{N}_j(\boldsymbol{x}) \big) p_f\ dv},\\
		\boldsymbol{f}_i^{\boldsymbol{\tau}} &= \int_{\partial \Omega}{\mathcal{N}_i(\boldsymbol{x}) \boldsymbol{\tau_s}\ da}.
	\end{aligned}
\end{equation}
$\boldsymbol{f}_i^{\text{int}}$, $\boldsymbol{f}_i^{\text{ext}}$, and $\boldsymbol{f}_i^{\boldsymbol{\tau}}$ are the common MPM nodal force vectors and can be calculated efficiently once the time-varying coefficients of the $[\mathcal{S}]$ and $[\mathcal{G}]$ matrices are determined.  $\boldsymbol{f}_i^{\text{drag}}$ and $\boldsymbol{f}_i^{\text{buoy}}$ are new coupling terms between the MPM and FVM domains and, in practice, require more careful calculation. In particular, the fluid pore pressure $p_f$ and fluid phase velocity $\boldsymbol{v}_f$, which are necessary to calculate these terms, must be reconstructed from the $N_v$ finite volume cell average fluid properties according to the procedure described in the next section. Some useful approximations of these terms can also be found in Appendix \ref{sec:numerical_approximation}.

Finally, consider the integral-form system of equations described in \eqref{eqn:weak_mixture_equations_fluid}. It is useful, particularly for evaluating the fluxes across volume boundaries, to re-express this system of equations using the fluid state vector, $\boldsymbol{u} = \{\rho_f;\ \rho_f \boldsymbol{v}_f;\ \rho_f E_f\}$, as follows,
\begin{equation}
	\label{eqn:fluid_state_vector_equation}
	\begin{aligned}
		\frac{d}{dt}
		{\small
			\begin{pmatrix}
				\langle \bar{\rho}_f \rangle_\alpha\\\
				\langle \bar{\rho}_f \boldsymbol{v}_f \rangle_\alpha\\
				\langle \bar{\rho}_f E_f \rangle_\alpha
			\end{pmatrix}
		}
		&= \boldsymbol{F}_\alpha^{\text{int}} + \boldsymbol{F}_\alpha^{\text{ext}} + \boldsymbol{F}_\alpha^{\text{drag}} + \boldsymbol{F}_\alpha^{\text{buoy}}, \qquad \forall \alpha \in [1,N_v],\\
		\boldsymbol{F}_\alpha^{\text{int}} &= - \frac{1}{V_\alpha} \int_{\partial \Omega_\alpha} \hat{\boldsymbol{f}}(\boldsymbol{u},n; \hat{\boldsymbol{n}})\ da,\\
		\boldsymbol{F}_\alpha^{\text{ext}} &= 
		{\small
			\begin{pmatrix}
				0\\
				\langle \bar{\rho}_f \rangle_\alpha\ \boldsymbol{g}\\
				\langle \bar{\rho}_f \boldsymbol{v}_f \rangle_\alpha \cdot \boldsymbol{g}
			\end{pmatrix}
		}
		,\\
		\boldsymbol{F}_\alpha^{\text{drag}} &= \frac{1}{V_\alpha} \int_{\Omega_\alpha}
		\sum_{i=1}^{N_n}
		{\small
			\begin{pmatrix}
				0\\
				\boldsymbol{f_d}\\
				\boldsymbol{f_d} \cdot \boldsymbol{v}_{si}
			\end{pmatrix}
		}
		\mathcal{N}_i(\boldsymbol{x})
		\ dv,\\
		\boldsymbol{F}_\alpha^{\text{buoy}} & = \frac{1}{V_\alpha} \int_{\Omega_\alpha}\sum_{i=1}^{N_n} \sum_{j=1}^{N_n}
		{ 
		{\small
			\begin{pmatrix}
				0\\
				n_i p_f \nabla \mathcal{N}_i(\boldsymbol{x})\\
				n_i p_f \nabla \mathcal{N}_i(\boldsymbol{x}) \cdot \boldsymbol{v}_{sj} &-& (1 - n_j) p_f \nabla \mathcal{N}_i(\boldsymbol{x}) \cdot \boldsymbol{v}_{si}
			\end{pmatrix}
		}}
		\mathcal{N}_j(\boldsymbol{x})\ dv.
	\end{aligned}
\end{equation}
with $\hat{\boldsymbol{f}}$ the oriented flux vector associated with the outward pointing face normal, $\hat{\boldsymbol{n}}$, defined as follows,
\begin{equation}
	\hat{\boldsymbol{f}}(\boldsymbol{u},n; \hat{\boldsymbol{n}}) = n
	{\small 
		\begin{pmatrix}
			\rho_f \boldsymbol{v}_f \cdot \hat{\boldsymbol{n}}\\
			\rho_f\boldsymbol{v}_f (\boldsymbol{v}_f \cdot \hat{\boldsymbol{n}}) + p_f \hat{\boldsymbol{n}}\\
			(\rho_f E_f + p_f) (\boldsymbol{v}_f \cdot \hat{\boldsymbol{n}})
	\end{pmatrix}} -
	{\small
		\begin{pmatrix}
			0\\
			\boldsymbol{\tau_f} \hat{\boldsymbol{n}}\\
			(\boldsymbol{\tau_f} \boldsymbol{v}_f - \boldsymbol{q}_f) \cdot \hat{\boldsymbol{n}}
	\end{pmatrix}}.
\end{equation}
Numerical approximation of the finite volume forces in \eqref{eqn:fluid_state_vector_equation} is possible using numerical quadrature, but it is also necessary to reconstruct the fluid phase fields within each cell.
Since this reconstruction may not be continuous across the cell boundaries, $\partial \Omega_\alpha$, we introduce a numerical flux function, $\hat{\boldsymbol{h}}(\boldsymbol{u}^+, \boldsymbol{u}^-, n; \hat{\boldsymbol{n}})$, as in \cite{barth1989}, such that,
\begin{equation}
	\label{eqn:discontinuous_flux_function}
	\boldsymbol{F}_\alpha^{\text{int}} \approx - \frac{1}{V_\alpha} \int_{\partial \Omega_\alpha} \hat{\boldsymbol{h}}(\boldsymbol{u}^+, \boldsymbol{u}^-, n; \hat{\boldsymbol{n}})\ da.
\end{equation}
Here $\boldsymbol{u}^+$ and $\boldsymbol{u}^-$ represent the two reconstructed fluid state vector values on each side of the boundary $\partial \Omega_\alpha$.

In this work, we implement the entropy adjusted Roe flux function (see \cite{roe1981,harten1997}), which has the following form,
\begin{equation}
	\label{eqn:roe_flux_function}
	\hat{\boldsymbol{h}}(\boldsymbol{u}^+, \boldsymbol{u}^-, n; \hat{\boldsymbol{n}}) = \tfrac{1}{2}\big[ \hat{\boldsymbol{f}}(\boldsymbol{u}^+, n, \hat{\boldsymbol{n}}) + \hat{\boldsymbol{f}}(\boldsymbol{u}^-, n, \hat{\boldsymbol{n}})- | \boldsymbol{A} | (\boldsymbol{u}^+ - \boldsymbol{u}^-) \big],
\end{equation}
where $|\boldsymbol{A}|$ is the positive definite matrix formed from the flux Jacobian, $\partial \hat{\boldsymbol{f}}/\partial \boldsymbol{u}$. To implement the entropy fix, we substitute $\boldsymbol{\bar{A}}$ for $\boldsymbol{A}$ as follows,
\begin{equation}
	(\boldsymbol{u}^+ - \boldsymbol{u}^-) = \sum_{k=1}^m b_k \boldsymbol{r}_k, \qquad \big|\boldsymbol{\bar{A}}\big|(\boldsymbol{u}^+ - \boldsymbol{u}^-) = \sum_{k=1}^m \hat{Q}(\lambda_k) b_k \boldsymbol{r_k},
\end{equation}
with $\{\lambda_k\}$ and $\{\boldsymbol{r}_k\}$ the sets of eigenvalues and right eigenvectors of $\boldsymbol{A}$, $\{b_k\}$ defined by the relationship above, and $\hat{Q}(\lambda_k)$ defined as:
\begin{equation}
	\hat{Q}(\lambda_k) = 
	\begin{cases}
		|\lambda_k| & \text{if}\quad |\lambda_k| \geq \delta_a,\\
		\tfrac{1}{2}\big((\lambda_k^2/\delta_a) + \delta_a) & \text{if}\quad |\lambda_k| < \delta_a,
	\end{cases}
\end{equation}
with $\delta_a = 0.1 \bar{a}$ for eigenvalues associated with acoustic waves and $\delta_a = 0$ for eigenvalues associated with advective waves ($\bar{a}$ is the Roe averaged speed of sound). (Note that $|\boldsymbol{\bar{A}}|$ and $|\boldsymbol{A}|$ are generally identical for the cases considered in this work with the exception of section \ref{sec:rocket}.)

In this way, we can fully describe the mixture fields and their evolution in terms of the discrete coefficients,
\begin{equation}
	\label{eqn:discrete_coefficients}
	\bar{\rho}_{sp}(t), \quad
	\boldsymbol{\tilde{\sigma}}_p(t), \quad
	\boldsymbol{\bar{\xi}}_p(t), \quad
	\boldsymbol{a}_{si}(t), \quad
	\boldsymbol{v}_{si}(t), \quad
	n_i(t), \quad
	\langle\bar{\rho}_{f}\rangle_\alpha(t), \quad
	\langle\bar{\rho}_{f}\boldsymbol{v}_f\rangle_\alpha(t), \quad
	\langle\bar{\rho}_{f}E_f\rangle_\alpha(t), \quad
\end{equation}
and the time-dependent deformations of the material point characteristic functions (further discussion below).

\section{Reconstructing Fluid Fields within Finite Volumes}
As noted above, in order to evaluate the integral expressions in \eqref{eqn:momentum_evolution} and \eqref{eqn:fluid_state_vector_equation}, we need to approximate the values of the fluid fields ($p_f$, $\bar{\rho}_f$, $\rho_f$, $\boldsymbol{v}_f$, $E_f$, $\boldsymbol{\tau_f}$, and $\boldsymbol{q}_f$) at all points in the simulated domain using the set of $N_v$ finite volume cell averages and the $N_n$ porosity coefficients. To do this, we reconstruct the \textit{true} fluid phase density ($\rho_f$), momentum ($\rho_f \boldsymbol{v}_f$), and energy ($\rho_f E_f$) fields using the second-order approach described in \cite{barth1989}. 

We begin by defining the average porosity of each finite volume, $\langle n \rangle_\alpha$, as follows,
\begin{equation}
\label{eqn:cell_porosity}
\langle n \rangle_\alpha = \sum_{i=1}^{N_n} \mathcal{A}_{i\alpha} n_i, \quad \forall \alpha \in [1,N_v]
\end{equation}
and approximate the cell-wise averages of these \textit{true} fluid fields as,
\begin{equation}\label{eqn:numerical_fvm_local_fields}
\begin{aligned}
\langle \rho_f \rangle_\alpha \approx \frac{\langle \bar{\rho}_f \rangle_\alpha}{\langle n \rangle_\alpha}, \quad
\langle \rho_f \boldsymbol{v}_f \rangle_\alpha \approx \frac{\langle \bar{\rho}_f \boldsymbol{v}_f \rangle_\alpha}{\langle n \rangle_\alpha}, \quad
\langle \rho_f E_f \rangle_\alpha \approx \frac{\langle \bar{\rho}_f E_f \rangle_\alpha}{\langle n \rangle_\alpha}, \quad \forall \alpha \in [1,N_v].
\end{aligned}
\end{equation}
If we take these cell-wise averages to be the value of these fields at the cell centroids, then we can approximate the fluid phase fields within each volume as follows,
\begin{equation}
\label{eqn:fluid_phase_reconstruction}
\begin{aligned}
\rho_f &\approx \langle \rho_f \rangle_\alpha + \langle \nabla \rho_f \rangle_\alpha \cdot (\boldsymbol{x} - \boldsymbol{X}_\alpha) & \forall \boldsymbol{x} \in \Omega_\alpha,\\
\rho_f \boldsymbol{v}_f &\approx \langle \rho_f \boldsymbol{v}_f \rangle_\alpha + \langle \nabla \rho_f \boldsymbol{v}_f \rangle_\alpha (\boldsymbol{x} - \boldsymbol{X}_\alpha) & \forall \boldsymbol{x} \in \Omega_\alpha,\\
\rho_f E_f &\approx \langle \rho_f E_f \rangle_\alpha + \langle \nabla \rho_f E_f \rangle_\alpha \cdot (\boldsymbol{x} - \boldsymbol{X}_\alpha) & \forall \boldsymbol{x} \in \Omega_\alpha,
\end{aligned}
\end{equation}
with $\langle \nabla \rho_f \rangle_\alpha$,  $\langle \nabla \rho_f \boldsymbol{v}_f \rangle_\alpha$, and $\langle \nabla \rho_f E_f \rangle_\alpha$ the reconstructed field gradients in $\Omega_\alpha$. To estimate these gradients, we use the flux limited reconstruction described by Barth \& Jespersen in \cite{barth1989},
\begin{equation}
\label{eqn:barth_and_jespersen_flux_limiter}
\begin{aligned}
\langle \nabla \rho_f \rangle_\alpha &= \Phi_{\rho_f} [\nabla \rho_f]_\alpha,\\
\langle \nabla \rho_f \boldsymbol{v}_f \rangle_\alpha &= \Phi_{\boldsymbol{v}_f} [\nabla \rho_f \boldsymbol{v}_f]_\alpha,\\
\langle \nabla \rho_f E_f \rangle_\alpha &= \Phi_{E_f} [\nabla \rho_f E_f]_\alpha,
\end{aligned}
\end{equation}
with $[\nabla \rho_f]_\alpha$, $[\nabla \rho_f \boldsymbol{v}_f]_\alpha$, and $[\nabla \rho_f E_f]_\alpha$ the gradients associated with a linear, least squares fit to the surrounding centroid data and $\Phi_{\rho_f}$, $\Phi_{\boldsymbol{v}_f}$, and $\Phi_{E_f}$ the maximum admissible values in the range $[0,1]$ that ensure the local field reconstructions are bounded by the surrounding cell maxima/minima. With the \textit{true} fluid fields and their gradients determined, we can approximate the effective fluid density field as,
\begin{equation}
	\label{eqn:fluid_effective_density_reconstruction}
	\bar{\rho}_f \approx \rho_f \sum_{i=1}^{N_n} n_i\ \mathcal{N}_i(\boldsymbol{x}),
\end{equation}
and use appropriate equations of state (e.g.\ the ideal gas law or a barotropic fluid model; see \cite{roe1981,baumgarten2019a}) to calculate the remaining fields ($p_f$, $\boldsymbol{\tau_f}$, and $\boldsymbol{q}_f$). Further discussion of this approach, along with potential improvements that explicitly account for local variations in the porosity field, can be found in Appendix \ref{sec:numerical_approximation}.

\section{Convection of Material Points}
\begin{figure}[!h]
	\centering
	\includegraphics[scale=0.5]{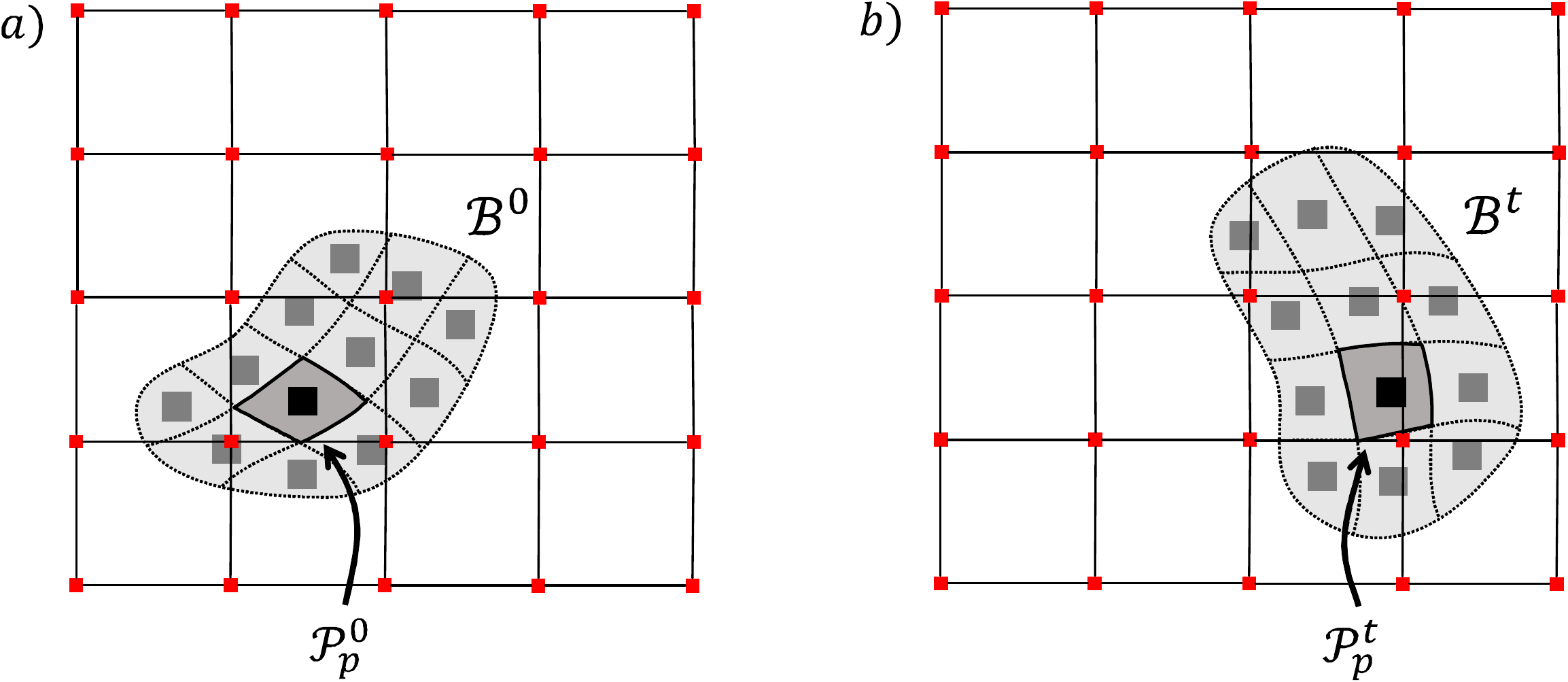}
	\caption{Schematic showing a representative finite element grid and material point discretization. The continuum body and the $p$th material subdomain are shown in their a) initial configurations ($\mathcal{B}^0$ and $\mathcal{P}_p^0$) and b) configurations at time $t$ ($\mathcal{B}^t$ and $\mathcal{P}_p^t$). The $p$th material point characteristic function, $U_p(\boldsymbol{x},t)$, is determined by the position of the $p$th material subdomain.}
	\label{fig:convection}
\end{figure}

One of the primary advantages of MPM is the ability to track Lagrangian flow features and history-dependent material properties on the material point characteristic functions. Since the material points themselves move with the material in a Lagrangian frame (see Figure \ref{fig:convection}), the convection of these additional properties occurs naturally and does not need to be explicitly calculated (i.e.\ there is no need to introduce additional advection equations). However, the time-dependent nature of the characteristic functions requires that the mapping matrices $[\mathcal{S}]$ and $[\mathcal{G}]$ be recalculated frequently and that this motion be numerically approximated, both of which come at significant computational cost. In this work, we approximate the motions and deformations of the material points using the original MPM approach from \cite{sulsky1994} in section \ref{sec:collapse} and the 
{uGIMP approach from \cite{bardenhagen2004} in sections \ref{sec:porous}, \ref{sec:terzaghi}, \ref{sec:2d_erosion}, \ref{sec:3d_erosion}, and \ref{sec:rocket}. Here, uGIMP denotes the use of unchanging approximations of the GIMP material point domains.} 
(uGIMP provides better stability when linear basis functions are chosen.) Both of these approaches track the changes to the material point weights as in equation \eqref{eqn:volume_evolution} and centroids as follows,
\begin{equation}
	\label{eqn:point_centroid_evolution}
	\frac{d}{d t}\big(\boldsymbol{x}_p\big) \approx \sum_{i=1}^{N_n} \mathcal{S}_{ip} \boldsymbol{v}_{si} \quad \forall p \in [1,N_m].
\end{equation}
Although higher-order approximations exist in the literature (e.g.\ cpGIMP \cite{bardenhagen2004}, CPDI \cite{sadeghirad2011}, and CPDI2 \cite{sadeghirad2013}), these methods are known to become {ill-conditioned} in problems involving {extreme shear deformations.} 

With the material point motion determined, we eliminate the need for an additional convection equation for the porosity field, $n$, by defining a weak, mass conserving mapping between the material point density field and finite element porosity field using the test function $\boldsymbol{w}$ as follows,
\begin{equation}
	\int_\Omega (1 - n) \rho_s \boldsymbol{w}\ dv = \int_\Omega \bar{\rho}_s \boldsymbol{w}\ dv,
\end{equation}
which is equivalently,
\begin{equation}\label{eqn:weak_porosity}
	\sum_{j=1}^{N_n} \mathcal{B}_{ij} (1 - n_j) \rho_s = \sum_{p=1}^{N_m} \mathcal{S}_{ip} m_p, \quad \forall i \in [1,N_n].
\end{equation}
Similarly, to convect solid phase momentum, we introduce an approximation of the velocity field, $\boldsymbol{v}_s^*(\boldsymbol{x},t)$, that is defined on the material point characteristic functions,
\begin{equation}
\label{eqn:approximate_velocity_field}
\boldsymbol{v}_s^*(\boldsymbol{x},t) = \sum_{p=1}^{N_m} \boldsymbol{v}_{sp}^*(t) U_p(\boldsymbol{x},t),
\end{equation}
and can be tracked independently from the true velocity field, $\boldsymbol{v}_s(\boldsymbol{x},t)$. We then define a weak, momentum conserving mapping between these fields using the test function $\boldsymbol{w}$ as follows,
\begin{equation}
\label{eqn:exact_velocity_projection}
\int_{\Omega} \bar{\rho}_s \boldsymbol{v}_s \cdot \boldsymbol{w}\ dv = \int_{\Omega} \bar{\rho}_s \boldsymbol{v}_s^* \cdot \boldsymbol{w}\ dv,
\end{equation}
which is equivalently,
\begin{equation}
\label{eqn:velocity_projection}
\sum_{j=1}^{N_n} \mathcal{M}_{ij} \boldsymbol{v}_{sj} = \sum_{p=1}^{N_m} \mathcal{S}_{ip} m_p \boldsymbol{v}_{sp}^*,  \qquad \forall i \in [1,N_n].
\end{equation}
In this way, we can define the coefficients associated with the node-based solid velocity field and mixture porosity field at any time using the persistent material point densities and velocities. The evolution of the material point densities is defined in \eqref{eqn:volume_evolution} while the evolution of the material point velocities is defined as,
\begin{equation}
\label{eqn:material_point_acceleration}
\frac{d}{dt}\big(\boldsymbol{v}_{sp}^*\big) = \sum_{i=1}^{N_n} \mathcal{S}_{ip} \boldsymbol{a}_{si}.
\end{equation}
This approach is sometimes referred to as the FLIP projection method (see \cite{brackbill1986,brackbill1988}), though there are several other valid projections that appear in the MPM literature (e.g\ PIC \cite{sulsky1994}, APIC \cite{jiang2015}, and XPIC \cite{hammerquist2017}).

\section{Numerical Algorithm}
In order to find time-accurate solutions to the system of equations in \eqref{eqn:mixture_equations}, we need to describe the time-history of the solution coefficients in \eqref{eqn:discrete_coefficients} in some consistent manner. Toward this end, we discretize the time domain, $[0,t_{\text{end}}]$, into a set of $N_s + 1$ discrete time markers, $\{t^k\ \forall k \in [0,N_s]\}$, with,
\begin{equation}
\label{eqn:time_discretization}
t^k = k \Delta t \quad \text{and} \quad \Delta t = t_{\text{end}}/N_s.
\end{equation}
{(Note that the use of uniformly distributed time markers is an implementation choice and not necessary for stability or accuracy of the FV-MPM scheme.)}
We can then denote the values of our solution coefficients at these time markers using the convention, $\psi_j^k = \psi_j(t^k)$, with $\psi_j(t)$ an arbitrary, time-dependent function, and determine their values using the numerical algorithm described below.
As in \cite{sulsky1994}, calculations in this numerical algorithm can be simplified by using diagonalized computational matrices. Here $[\mathcal{M}]^k$ and $[\mathcal{B}]$ are diagonalized by summing each row into $[\mathcal{M}_D]^k$ and $[\mathcal{B}_D]$. This substitution introduces some error but still produces a consistent method (see \cite{hughes1987}).

Suppose the following parameters are known at $t^k$: (i) the mapping matrices, $[\mathcal{S}]^k$, $[\mathcal{G}]^k$, $[\mathcal{A}]$, $[\mathcal{B}_D]$, and $[\mathcal{M}_D]^k$; (ii) the solution coefficients,
$\bar{\rho}_{sp}^k,\ 
\boldsymbol{\tilde{\sigma}}_p^k,\ 
\boldsymbol{\bar{\xi}}_p^k,\ 
\langle\bar{\rho}_{f}\rangle_\alpha^k,\ 
\langle\bar{\rho}_{f}\boldsymbol{v}_f\rangle_\alpha^k,\ \text{and}\ 
\langle\bar{\rho}_{f}E_f\rangle_\alpha^k;
$ (iii) the coefficients of the material point approximation of the velocity field, $\boldsymbol{v}_{sp}^{*k}$; and (iv) the numerical representation of the material point characteristic functions, $U_p(\boldsymbol{x},t^k)$, (e.g.\
$v_p^k$ and $\boldsymbol{x}_p^k$ for basic MPM \cite{sulsky1994} and uGIMP \cite{bardenhagen2004}). We can use a first-order, Forward Euler time integration method (with the update-stress-last MPM approach from \cite{bardenhagen2002,buzzi2008}) to determine the value of the solution coefficient at $t^{k+1}$ according to the following steps:
\begin{enumerate}[label=(\arabic*)]
	\item Determine the mixture porosity coefficients, as in \eqref{eqn:weak_porosity}, using the diagonal matrix $[\mathcal{B}_D]$:
	\begin{equation}
	{\mathcal{B}_{D}}_{ii} (1 - n^k_i) \rho_s = \sum_{p=1}^{N_m} \mathcal{S}^k_{ip} m_p \quad \forall i \in [1,N_n].
	\end{equation}
	
	\item Determine the solid phase velocity coefficients, as in \eqref{eqn:velocity_projection}, using the diagonal matrix $[\mathcal{M}_D]$:
	\begin{equation}
	{\mathcal{M}_D}_{ii}^k \boldsymbol{v}_{si}^k = \sum_{p=1}^{N_m} \mathcal{S}^k_{ip} m_p \boldsymbol{v}_{sp}^{*k},  \qquad \forall i \in [1,N_n].
	\end{equation}
	
	\item Calculate the nodal force vectors, $(\boldsymbol{f}_i^{\text{int}})^k$ and $(\boldsymbol{f}_i^{\text{ext}})^k$, as in \eqref{eqn:momentum_evolution}.
	
	\item Construct an approximation of the fluid phase fields, $\rho_f$, $\boldsymbol{v}_f$, $E_f$, and $\bar{\rho}_f$, within each finite volume cell, as in \eqref{eqn:barth_and_jespersen_flux_limiter}.
	
	\item Use the reconstructed fluid phase fields to calculate the fluid phase pressures and stresses, $p_f$ and $\boldsymbol{\tau_f}$; the fluid phase heat flux, $\boldsymbol{q}_f$; the nodal force vectors, $(\boldsymbol{f}_i^\text{drag})^k$ and $(\boldsymbol{f}_i^\text{buoy})^k$, as in \eqref{eqn:momentum_evolution}; and the finite volume force vectors, $(\boldsymbol{F}_\alpha^\text{int})^k$, $(\boldsymbol{F}_\alpha^\text{ext})^k$, $(\boldsymbol{F}_\alpha^\text{drag})^k$, and $(\boldsymbol{F}_\alpha^\text{buoy})^k$, as in \eqref{eqn:fluid_state_vector_equation} and \eqref{eqn:discontinuous_flux_function}. Numerical quadrature can be particularly helpful in these calculations. (See Appendix \ref{sec:numerical_approximation} for alternative approximations of the drag and buoyancy terms.)
	
	\item Use an explicit time integrator to obtain the updated fluid phase field coefficients from the equations of motion:
	\begin{equation}
	{\small
	\begin{pmatrix}
	\langle \bar{\rho}_f \rangle_\alpha^{k+1}\\
	\langle \bar{\rho}_f \boldsymbol{v}_f \rangle_\alpha^{k+1}\\
	\langle \bar{\rho}_f E_f \rangle_\alpha^{k+1}
	\end{pmatrix}}
	=
	{\small
	\begin{pmatrix}
	\langle \bar{\rho}_f \rangle_\alpha^{k}\\
	\langle \bar{\rho}_f \boldsymbol{v}_f \rangle_\alpha^{k}\\
	\langle \bar{\rho}_f E_f \rangle_\alpha^{k}
	\end{pmatrix}}
	+ \Delta t \big[ (\boldsymbol{F}_\alpha^{\text{int}})^k + (\boldsymbol{F}_\alpha^{\text{ext}})^k + (\boldsymbol{F}_\alpha^{\text{buoy}})^k + (\boldsymbol{F}_\alpha^{\text{drag}})^k \big], \quad \forall \alpha \in [1,N_v].
	\end{equation}
	
	\item Determine the solid phase acceleration coefficients associated with the finite element grid nodes from the equations of motion and appropriate boundary conditions:
	\begin{equation}
	{\mathcal{M}_D}_{ii}^k \boldsymbol{a}_{si}^k = (\boldsymbol{f}_i^\text{int})^k + (\boldsymbol{f}_i^\text{ext})^k + (\boldsymbol{f}_i^\text{drag})^k + (\boldsymbol{f}_i^\text{buoy})^k + 
	{(\boldsymbol{f}_i^{\boldsymbol{\tau}})^k,}
	 \quad \forall i \in [1,N_n].
	\end{equation}
	
	\item Use an explicit time integrator to obtain the updated solid phase velocity coefficients associated with the finite element grid nodes; treat the grid as though it were Lagrangian:
	\begin{equation}
	\boldsymbol{v}_{si}^{k+1'} = \boldsymbol{v}_{si}^k + \Delta t\  \boldsymbol{a}_{si}^k, \quad \forall i \in [1,N_n].
	\end{equation}
	
	\item Update the solid phase material state vector, $\boldsymbol{\bar{\xi}}_p$, and effective granular stress, $\boldsymbol{\tilde{\sigma}}_p$, according to the relevant constitutive update procedure. For stability, $\boldsymbol{L}_p$, the average material point velocity gradient is often used:
	\begin{equation}
	\boldsymbol{L}_p^{k+1} = \sum_{i=1}^{N_n} \boldsymbol{v}_{si}^{k+1'} \otimes \mathcal{G}^k_{ip}, \quad \boldsymbol{\tilde{\sigma}}_p^{k+1} = \boldsymbol{T}\big( \boldsymbol{L}_p^{k+1}, \boldsymbol{\bar{\xi}}_p^{k+1} \big), \quad \forall p \in [1,N_m].
	\end{equation}
	
	\item Map the solid phase nodal accelerations and velocities to the material points to update their velocity approximations, positions, and densities:
	\begin{equation}
	\begin{aligned}
	\boldsymbol{v}_{sp}^{*k+1} &= \boldsymbol{v}_{sp}^{*k} + \Delta t \sum_{i=1}^{N_n} \mathcal{S}_{ip}^k \boldsymbol{a}_{si}^{k}, \quad \forall p \in [1,N_m],\\
	\boldsymbol{x}_{p}^{k+1} &= \boldsymbol{x}_p^k + \Delta t \sum_{i=1}^{N_n} \mathcal{S}_{ip}^k \boldsymbol{v}_{si}^{k+1'}, \quad \forall p \in [1,N_m],\\
	v_{p}^{k+1} &= v_p^k \bigg( 1 + \Delta t \sum_{i=1}^{N_n} \mathcal{G}_{ip}^k \cdot \boldsymbol{v}_{si}^{k+1'}\bigg), \quad \forall p \in [1,N_m],\\[1em]
	\bar{\rho}_{sp}^{k+1} &= m_p / v_p^{k+1}, \quad \forall p \in [1,N_m].
	\end{aligned}
	\end{equation}
	
	\item Update the diagonal mass matrix, $[\mathcal{M}_D]^{k+1}$, and mapping matrices, $[\mathcal{S}]^{k+1}$ and $[\mathcal{G}]^{k+1}$, according to their definitions in \eqref{eqn:mapping_matrices}. Numerical quadrature can be particularly helpful in these calculations, but care must be taken in updating the location and weights of quadrature points at this step (see \cite{bardenhagen2004,sadeghirad2011,sadeghirad2013} for further discussion).
	
	\item Increment $k$ to $k+1$; go to (1).
	
\end{enumerate}

This algorithm is consistent with the governing equations in \eqref{eqn:mixture_equations} and is used to generate the results in sections \ref{sec:porous} and \ref{sec:terzaghi}. An alternative numerical algorithm is described in Appendix \ref{sec:alternative_algorithm} and is used to generate the results in sections \ref{sec:collapse}, \ref{sec:2d_erosion}, \ref{sec:3d_erosion}, and \ref{sec:rocket}. This alternative algorithm is more complex but provides better accuracy and stability of the simulated fluid.
Both numerical algorithms have two important conditions that must be satisfied to ensure their stability. First, the Forward Euler time-integration method has explicit expressions for each coefficient update; this requires consideration of the Courant--Friedrichs--Lewy (CFL; see \cite{courant1928}) condition for numerical methods (i.e.\ the time increment $\Delta t$ must be small enough that the numerical stencil for any explicit equation captures the motion of physical waves in the real system during that increment). In practice, we find the most limiting wave-speed to be the acoustic waves in the solid or fluid domains (depending on the specific elastic moduli and material properties of the problem). Second, the numerical evaluation of the fluid phase shear stresses and heat fluxes at the finite volume boundaries as well as the calculation of the inter-phase drag terms, $(\boldsymbol{f}_i^{\text{drag}})^k$ and $(\boldsymbol{F}_\alpha^{\text{drag}})^k$, can be unstable if the chosen time increment is not small enough to capture the decay rates associated with each of these terms; these instabilities are not unique to this framework and are common to many computational codes for transient problems. Full discussion of these stability conditions, an analysis of the overall accuracy of this simulation approach, and alternative numerical procedures can be found in Appendices \ref{sec:consistency}, \ref{sec:stability}, and \ref{sec:alternative_algorithm}.

\section{Order of Accuracy}
Suppose the numerical algorithm described above is implemented for a particular problem using time increment $\Delta t$, a finite volume mesh with characteristic length $h_\alpha$ (e.g.\ the average distance between cell centers), a finite element grid with characteristic length $h_i$ (e.g.\ the average distance between grid nodes), and a set of material points with characteristic length $h_p$ (e.g.\ the average distance between the initial positions of the material point centroids). If the stability criteria mentioned above are satisfied and equations \eqref{eqn:density_evolution}, \eqref{eqn:momentum_evolution}, \eqref{eqn:fluid_state_vector_equation}, and \eqref{eqn:weak_porosity} were evaluated exactly, the results of the algorithm would have numerical errors that were proportional to the truncation errors in the time-integration method (e.g.\ Forward Euler, $\mathcal{O}(\Delta t)$) and the discretization errors determined by the choices of finite volume mesh, nodal basis functions, and material point characteristic functions (e.g.\ Barth \& Jesperson reconstructions, $\mathcal{O}(h_\alpha^2)$; non-overlapping material point characteristics, $\mathcal{O}(h_p)$; $l$th-order nodal bases, $\mathcal{O}(h_i^{l})$). In practice, however, the use of common MPM quadrature schemes (e.g.\ uGIMP, CPDI, etc.), diagonal mapping matrices $[\mathcal{M}_D]$ and $[\mathcal{B}_D]$, and the approximations of the inter-phase forces described in Appendix \ref{sec:numerical_approximation}, limits the order of convergence of the overall method to the order of the truncation and quadrature errors associated with each of these terms: $\mathcal{O}(h_p) + \mathcal{O}(h_p/h_i) + \mathcal{O}(h_i) + \mathcal{O}(h_\alpha) + \mathcal{O}(\Delta t)$. The $\mathcal{O}(h_p/h_i)$ term is common to particle-in-cell (PIC) methods and is usually reduced by increasing the number of \textit{material points per grid cell}. (Note that in the special case of a regular Cartesian grid with B-spline basis functions, the convergence rate becomes $\mathcal{O}(h_p) + \mathcal{O}(h_p^2/h_i^2) + \mathcal{O}(h_i^2) + \mathcal{O}(h_\alpha)  + \mathcal{O}(\Delta t)$; see \cite{steffen2008analysis}.)  A full analysis of these convergence rates can be found in Appendix \ref{sec:consistency}.

\section{Numerical Examples}

\subsection{Flow through porous media test}\label{sec:porous}
Models for flow through porous media have practical importance for a number of engineering applications, including predicting groundwater movement and gas seepage through packed sediments. Laws for the motion of fluids through porous solids have been developed and generalized for more than a century (see \cite{darcy1856, dupuit1863, carman1937, vanderhoef2005, beetstra2007}) and have been implemented in many numerical frameworks (e.g.\ \cite{geiger2004}). In the case of uniaxial flow or for flows along a streamline, the apparent fluid velocity $u$ (i.e.\ the fluid volume flux divided by the total flow area, $u = q/A$) can be expressed as a function of the soil permeability $K$, fluid pressure change $\Delta p_f$, and the length of the soil section $L$:
\begin{equation}
	u = K \frac{\Delta p_f}{L}.
\end{equation}
Here $K$ is an empirical measure of resistance due to drag on the fluid as it flows through the pores of the solid.

\begin{figure}[!h]
	\centering
	$$
	\begin{matrix}
		\quad\hbox{\includegraphics[scale=0.35]{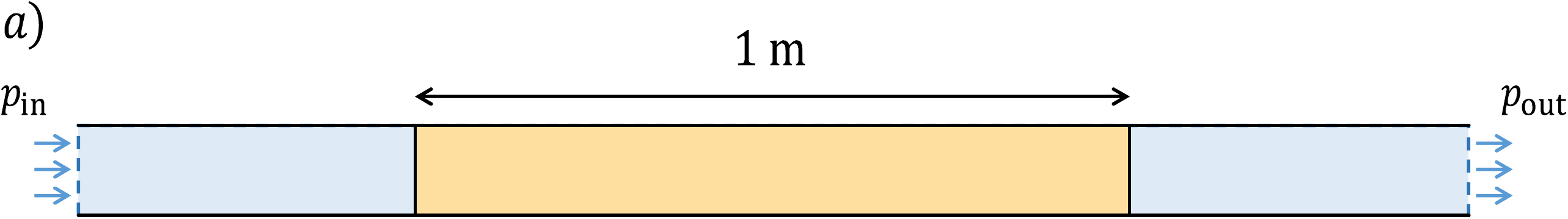}}\\[1em]
		\includegraphics[scale=0.35]{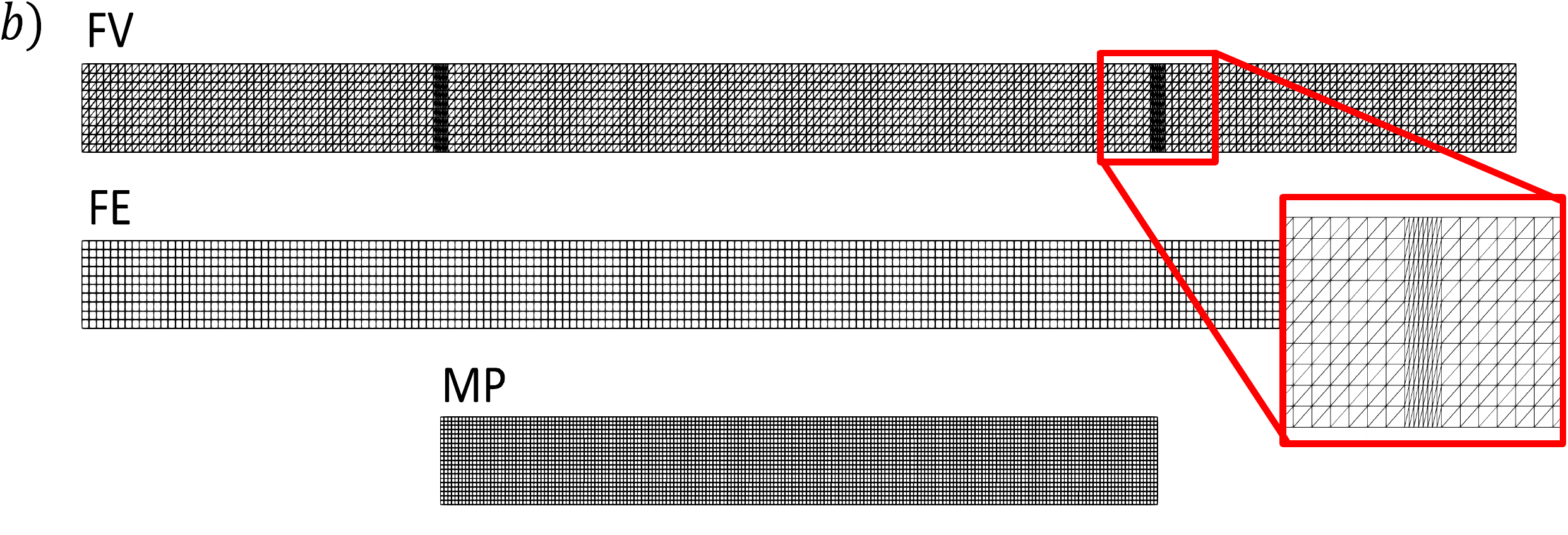}\\[0em]
		\includegraphics[scale=0.5]{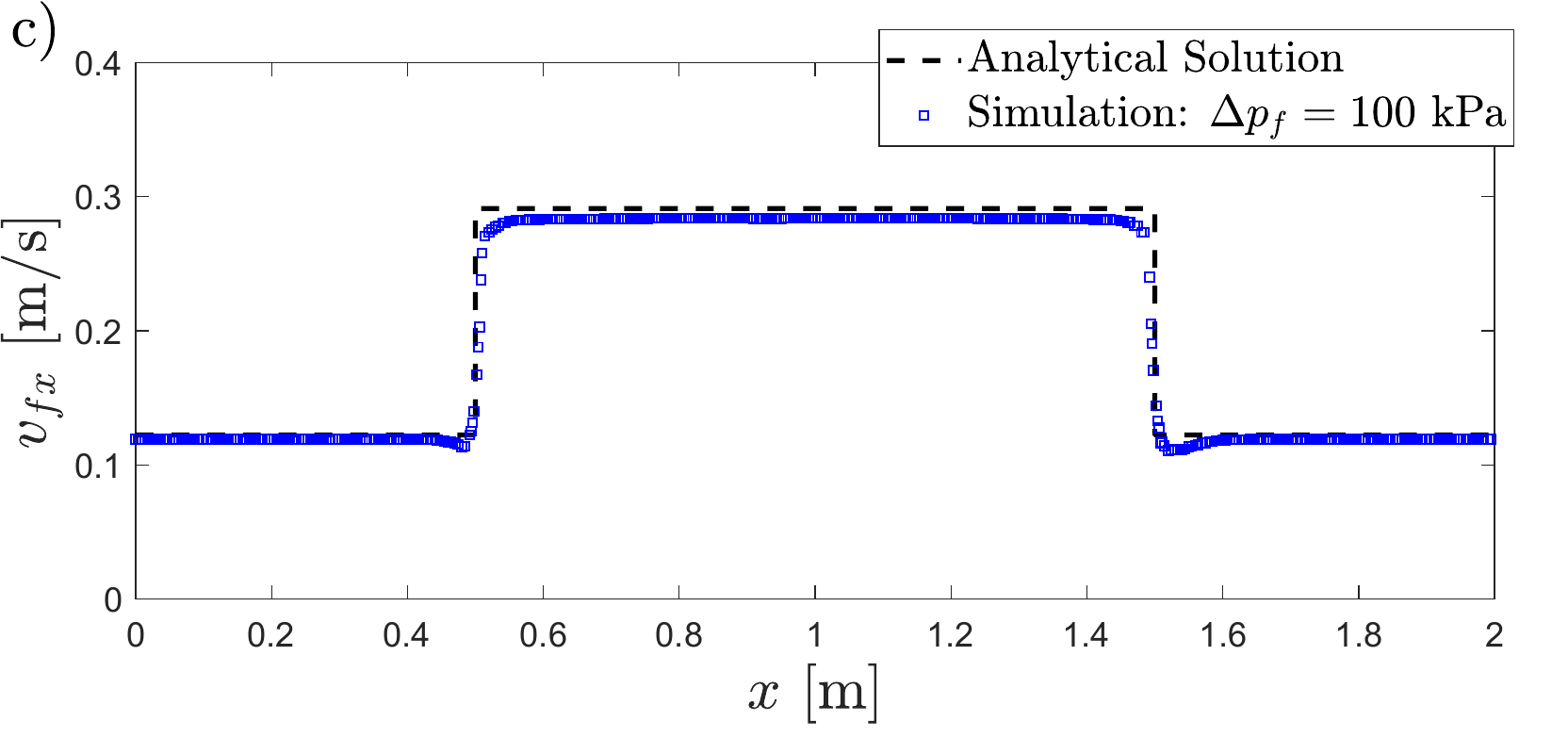}
	\end{matrix}
	\quad
	\vline
	\quad
	\begin{matrix}
		\includegraphics[scale=0.55]{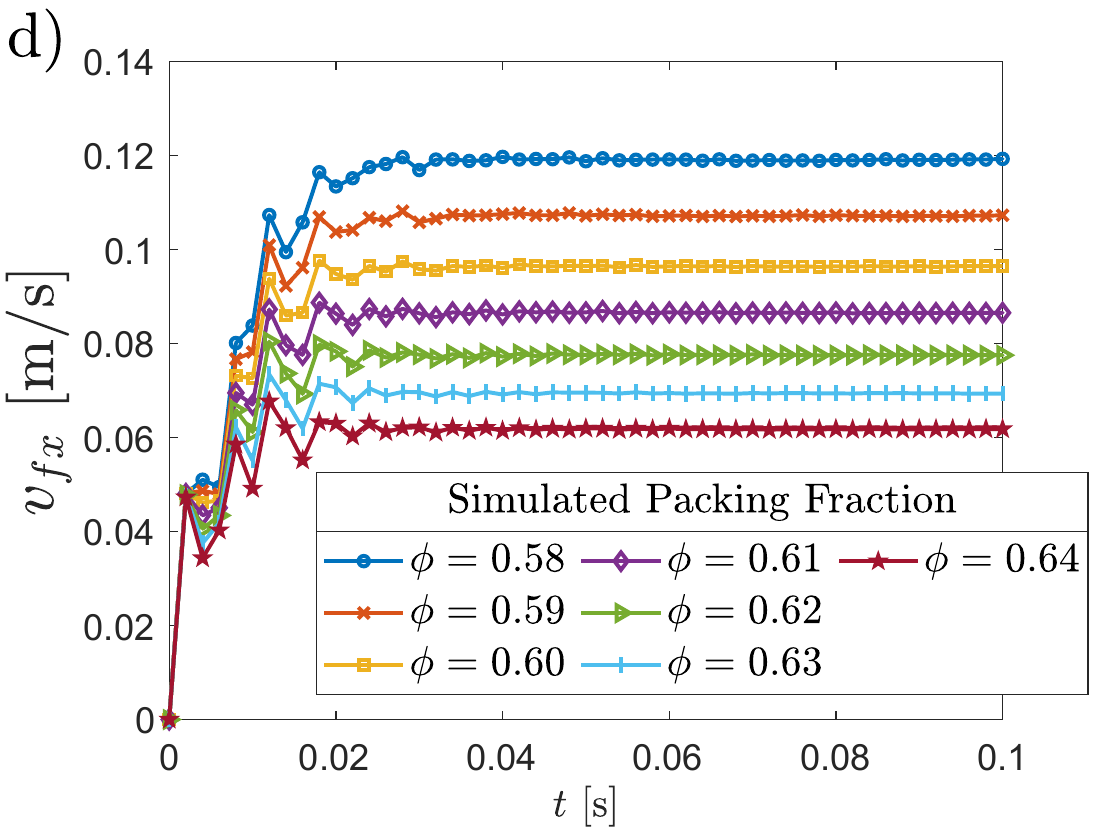}\\
		\includegraphics[scale=0.55]{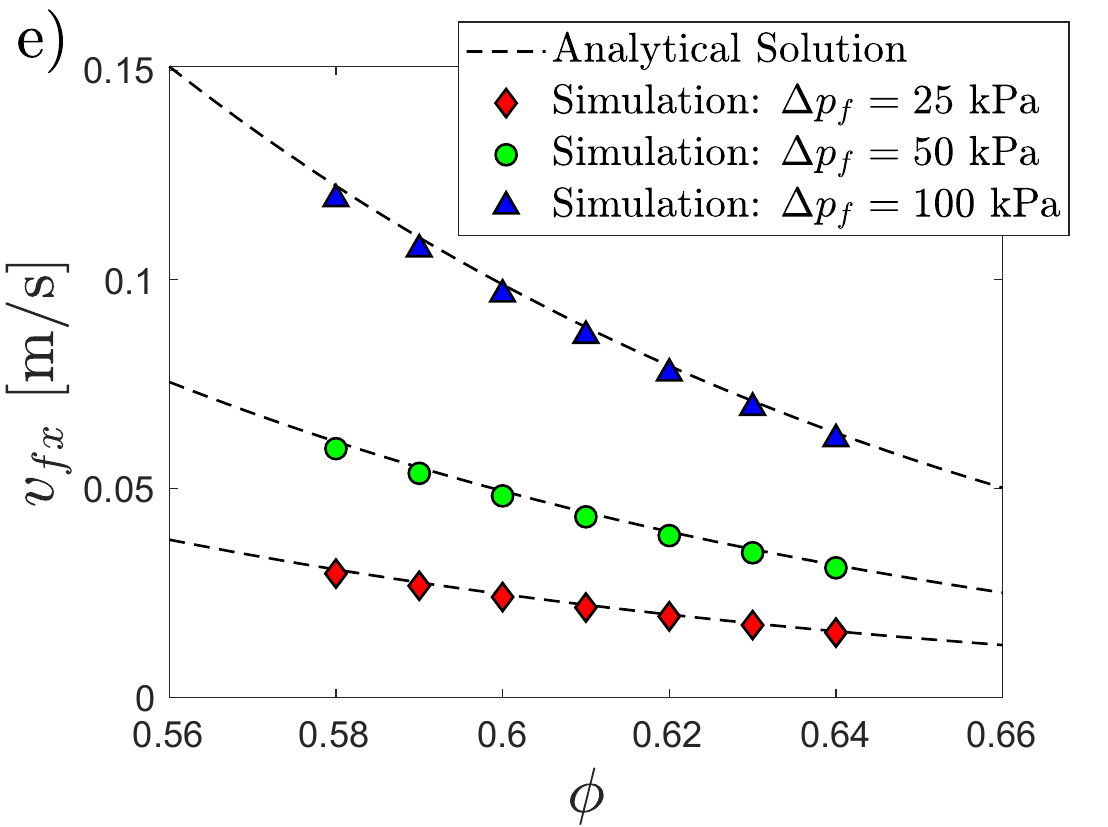}
	\end{matrix}
	$$
	\caption{Numerical test of flow through a porous solid: a) schematic diagram of test; b) finite volume (FV) grid, finite element (FE) grid, and material points (MP) used in numerical test; c) a representative analytical (dashed line) and simulated (blue squares) flow solution for the $x$-component of the fluid phase velocity, $\boldsymbol{v}_f$, for $\phi = 0.58$; d) time-history of the flow solution at $x=0.2$ m for several solid volume fractions; and e) comparison of analytical (dashed lines) and simulated (colored symbols) flow solutions at $x=0.2$ m for three sets of $\Delta p_f = p_{\text{in}} - p_{\text{out}}$.}
	\label{fig:darcy}
\end{figure}

We test our numerical framework on an example problem of this type. Consider a 2 m long, 10 cm tall pipe filled with water ($\kappa = 2.2$ GPa, $\rho_f = 1000$ kg/m$^3$, and $\eta_0 = 1$ mPa$\cdot$s) and a $L = 1$ m section of packed granular solid as in Figure \ref{fig:darcy}a. The porous solid in this problem can be modeled as a linear elastic material with effective Young's modulus $E = 10$ MPa, effective Poisson's ratio $\nu = 0.3$, true density $\rho_s = 2650$ kg/m$^3$, solid volume fraction $\phi \in [0.58,0.64]$, and average grain diameter $d = 1$ mm. (Note that the porous solid can be compressed without violating the assumption that the grains themselves are incompressible.) If we model the drag function from equation \eqref{eqn:drag_force} using the Carman--Kozeny formula, $F(\phi, \Reyn) = 10 \phi / (1 - \phi)^2$, then the permeability of this solid section can be expressed as	$K = d^2 (1-\phi)^3 / 180 \eta_0 \phi^2$. With this determined, we can express the true fluid velocity in the pipe, $v_{fx}(x)$, analytically in terms of the volume fraction of the solid, $\phi$, and the local porosity in the pipe, $n(x)$:
\begin{equation}\label{eqn:darcy_exact}
	v_{fx}(x) = \frac{1}{n(x)} \frac{d^2}{180 \eta_0} \frac{(1-\phi)^3}{\phi^2} \frac{\Delta p_f}{L}.
\end{equation}
Given this exact solution to the governing equations in \eqref{eqn:mixture_equations}, our numerical framework can be evaluated by comparing its predicted solutions with those from \eqref{eqn:darcy_exact}.

To simulate this problem, we generate a regular Cartesian grid with 200$\times$10 linear elements for the nodal (FE) basis functions --- with bi-linear (or ``tent''; see \cite{bardenhagen2004}) basis functions --- and discretize the porous solid using 4,000 material points (MP) as in Figure \ref{fig:darcy}b. To ensure accuracy of the fluid flow solution, we generate triangular finite volumes (FV) from this same Cartesian grid with additional refinement near the ends of the porous solid (where $\partial n/\partial x$ is large). We then assign static pressure boundary conditions, $p_{\text{in}}$ and $p_{\text{out}}$, at the ends of the pipe for three different pressure drops, $\Delta p_f \in \{25,50,100\}$ kPa, and seven different granular packings, $\phi \in \{0.58,0.59,0.60,0.61,0.62,0.63,0.64\}$. After reaching a steady flow, as in Figure \ref{fig:darcy}d, we compare the numerical value of $v_{fx}(x)$ with the analytical solution along length of the pipe. A representative flow solution for $\Delta p_f = 100$ kPa and $\phi = 0.58$ is shown in Figure \ref{fig:darcy}c, and a comparison of the flow solutions at $x=0.2$ m along the pipe for each set of pressure drops and volume fractions is shown in Figure \ref{fig:darcy}e.
{(Note that, in this example problem, the material points do not cross the cell boundaries of the FE grid, avoiding a known source of error in MPM.)}

\subsection{Consolidation test}\label{sec:terzaghi}
A common numerical test for mixture simulation frameworks is the one-dimensional consolidation of a fully-saturated, elastic, porous solid under external loading (see \cite{terzaghi1943,bandara2015}). Here we consider a $H=1$ m tall, infinitely wide, densely packed granular column with average grain diameter $d=0.58$ mm and volume fraction $\phi = 0.7$. This column is subjected to an external load, $\sigma_0$, of 10 kPa. The granular solid is assumed to be fully saturated by water ($\kappa = 2.2$ GPa, $\eta_0 = 1$ mPa$\cdot$s, $\rho_f = 1000$ kg/m$^3$) and is treated as a linear elastic material with Young's modulus $E$ ($E=10$ MPa), Poisson's ratio $\nu$ ($\nu = 0.3$), and density $\rho_s$ ($\rho_s = 2650$ kg/m$^3$). The fluid is assumed to have zero pressure at $z=1$ m, and the base ($z=0$ m) of the granular column is assumed to be impermeable. Here we again use the Carman--Kozeny formula for drag.

For this consolidation problem, the pore fluid pressure, $p_f(z,t)$, has the following analytical form,
\begin{equation}
	\label{eqn:terzaghi}
	p_f(z,t) = \sum_{m=0}^{\infty} \frac{2 \sigma_0}{M} \sin(Mz/H) e^{-M^2 T_v}, \text{\quad with\quad} M = \frac{\pi}{2} (2m+1), \quad T_v = \frac{c_v t}{H^2},
\end{equation}
for $c_v = E_v n^3 d^2/ (180 (1-n)^2 \eta_0)$ and $E_v = E (1 - \nu) / (1 - \nu - 2\nu^2)$ (see \cite{terzaghi1943, terzaghi1996}). As in the previous section, given this exact solution to the governing equations in \eqref{eqn:mixture_equations}, our numerical framework can be evaluated by comparing its predictions to those in \eqref{eqn:terzaghi}.

\begin{figure}[!h]
	\centering
	$\vcenter{\hbox{\includegraphics[scale=0.47]{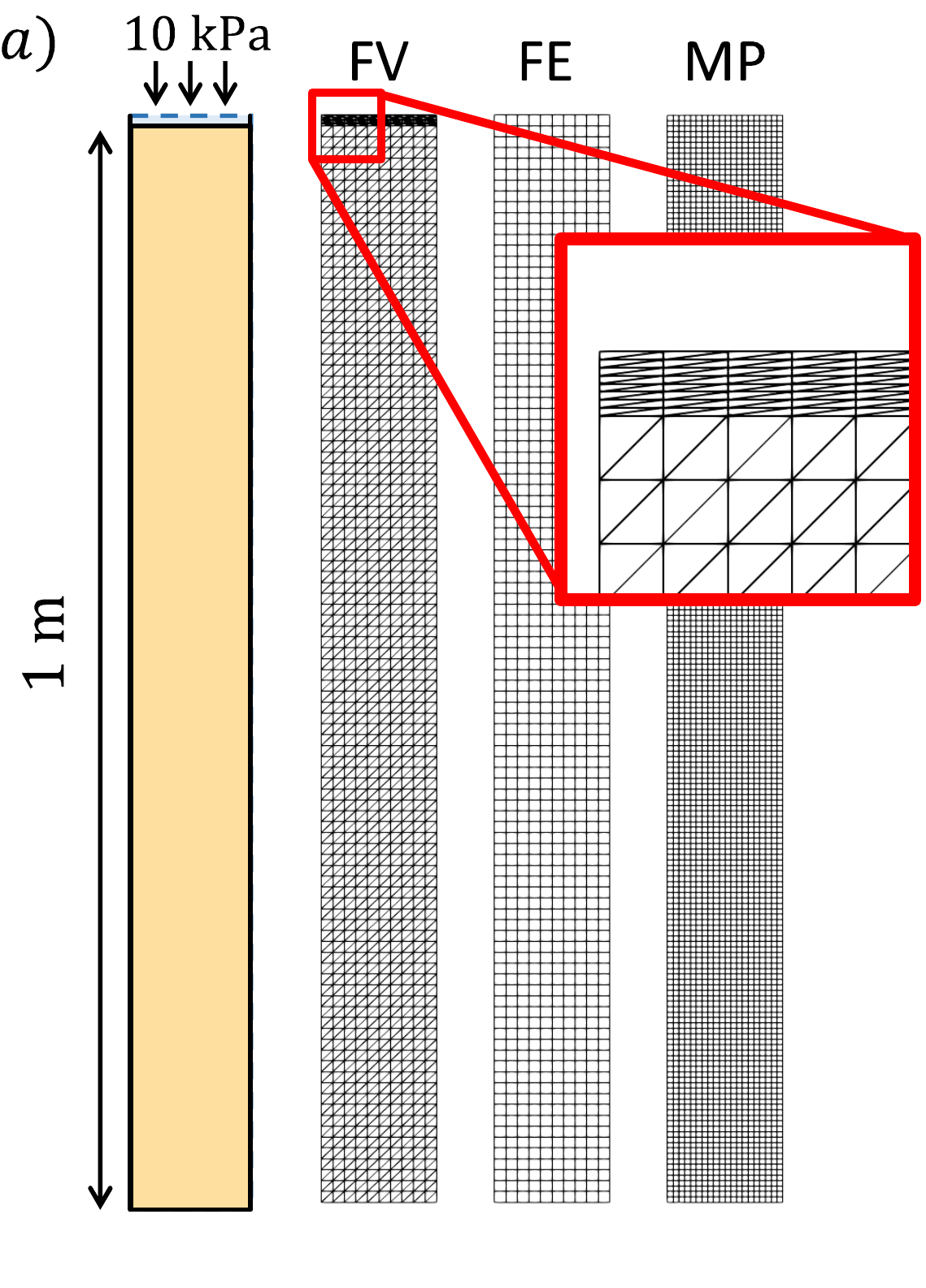}}}$
	$\vcenter{\hbox{\includegraphics[scale=0.65]{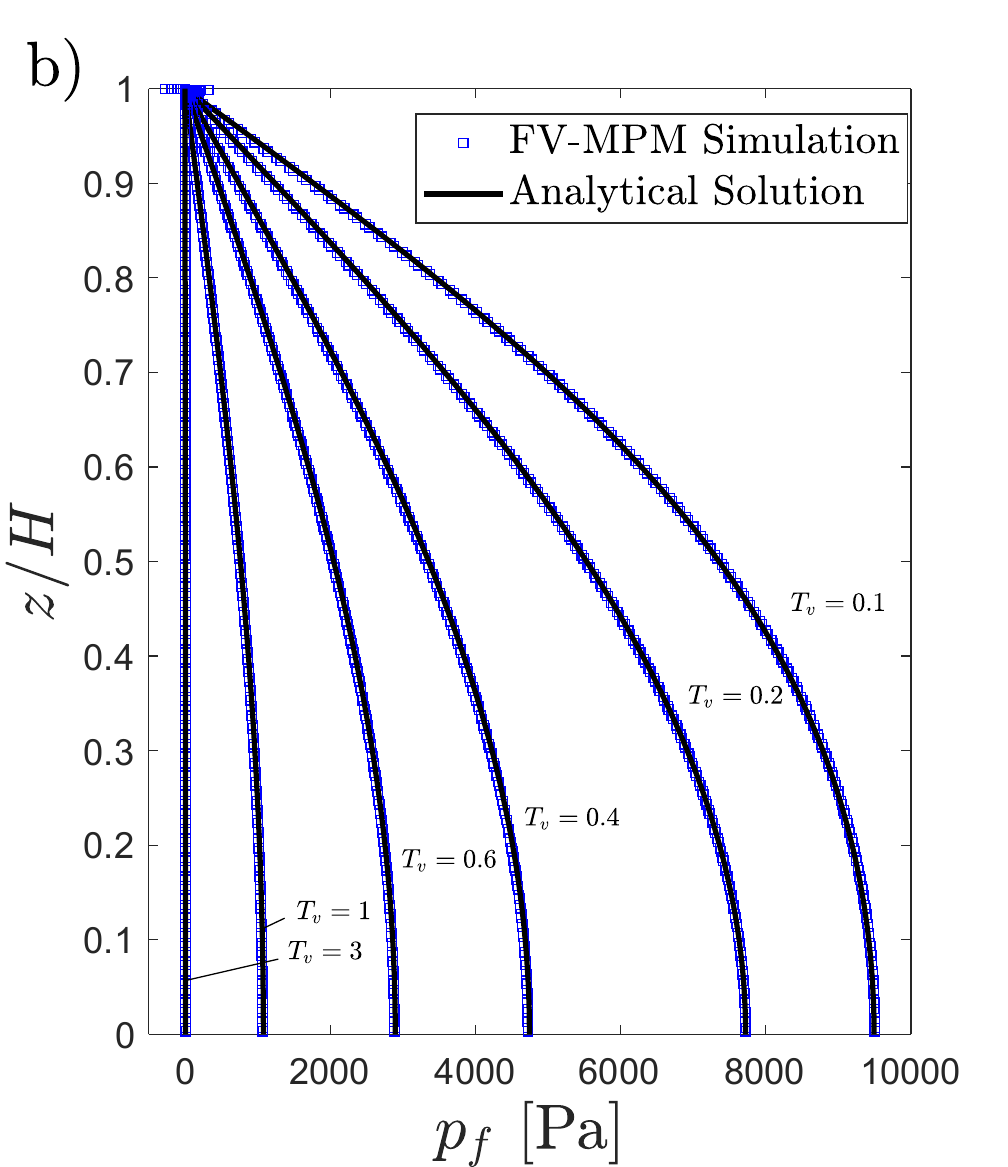}}}$
	\caption{Numerical consolidation: a) schematic diagram of test alongside finite volume (FV) grid, finite element (FE) grid, and material points (MP) used in numerical test; b) comparison of analytical (solid lines) and simulated (blue squares) pore pressure solutions at $T_v = 0.1,\ 0.2,\ 0.4,\ 0.6,\ 1,\text{ and }3$.}
	\label{fig:terzaghi}
\end{figure}

To evaluate our numerical framework as applied to this problem, we generate a 10$\times$100 element Cartesian FE grid with 4 material points per grid cell (4,000 total). Standard bi-linear (or ``tent'') basis functions are used in this discretization. To improve accuracy near the top surface, we generate the triangular FV mesh shown in Figure \ref{fig:terzaghi}a (with additional refinement at $z=1$ m). Impermeable, smooth boundary conditions are applied to the two walls and lower boundary of the simulated domain, and two traction boundary conditions are imposed at the top surface: the vertical load $\sigma_0$ is applied to the top layer of material points and a zero pressure condition is assigned to the fluid. To avoid non-physical pressure oscillations in the fluid at the onset of these loading conditions ($t=0$), we apply numerical damping of the form described in \cite{bandara2015} at the very beginning of the simulation.
After 2.5 s of simulated time, we compare the fluid pressure solution, $p_f(z,t)$, to the analytical solution at $T_v = $ 0.1, 0.2, 0.4, 0.6, 1.0, and 3.0 in Figure \ref{fig:terzaghi}b. Although there are small oscillations in the pressure solution near $z=1$m, the overall solution profiles are in very close agreement.
{(Note that, in this example problem, the material points do not cross the cell boundaries of the FE grid, avoiding a known source of error in MPM.)}

\subsection{Collapse of a submerged granular column: comparison to two-phase MPM}\label{sec:collapse}
\begin{figure}[!h]
	\centering
	$\vcenter{\hbox{\includegraphics[scale=0.37]{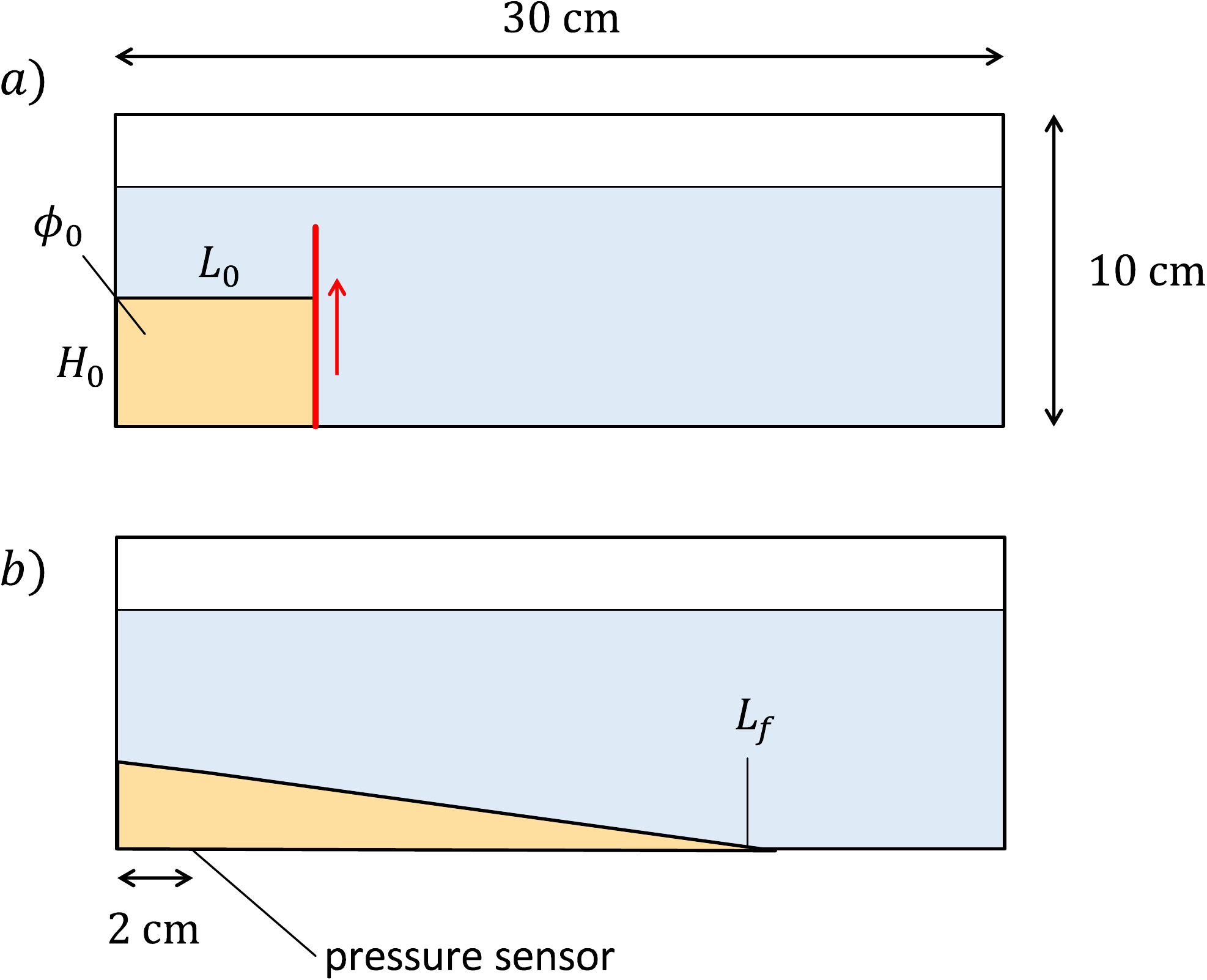}}}$
	\vline
	\ 
	$
	\begin{matrix}
		\includegraphics[scale=0.52]{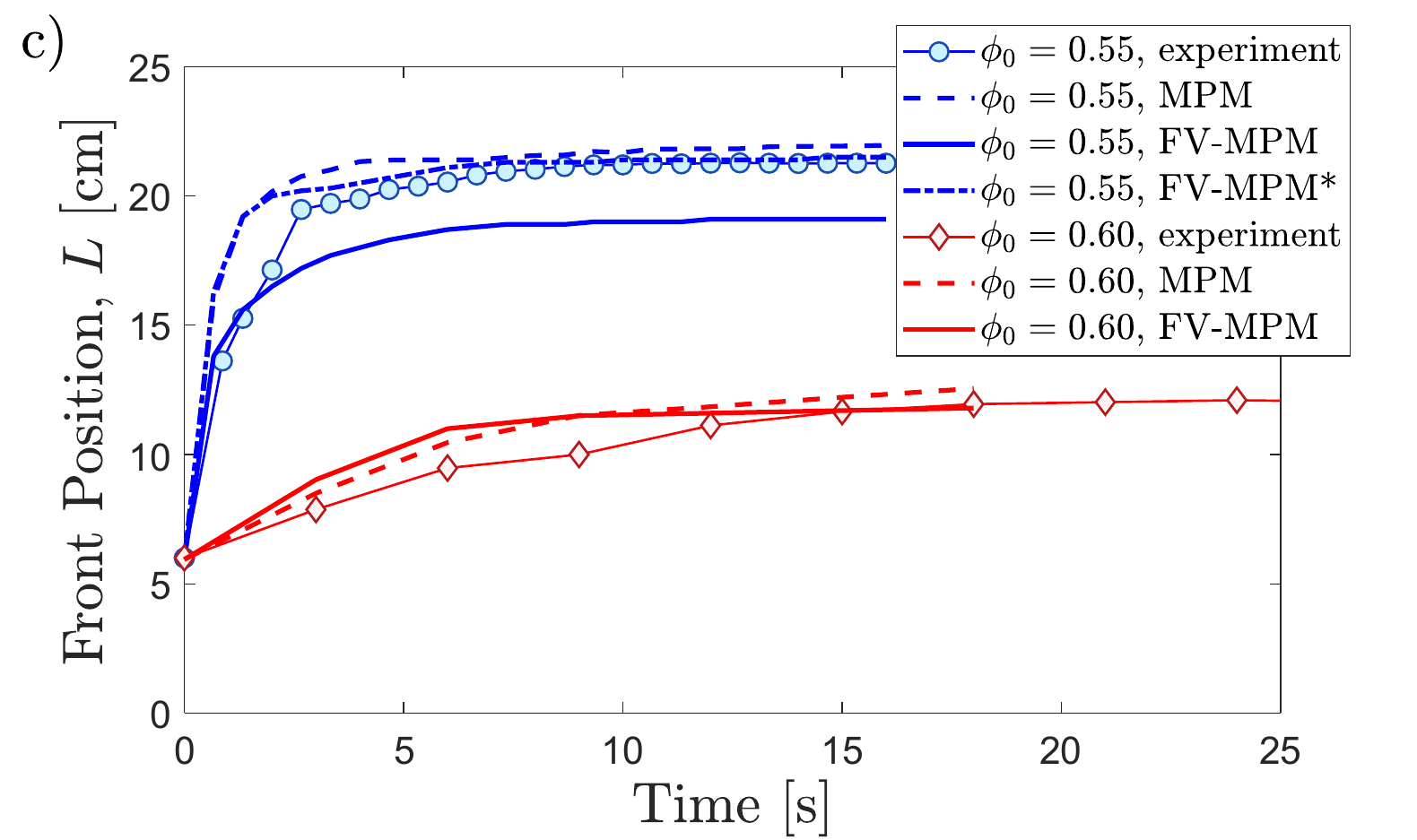}\\
		\includegraphics[scale=0.52]{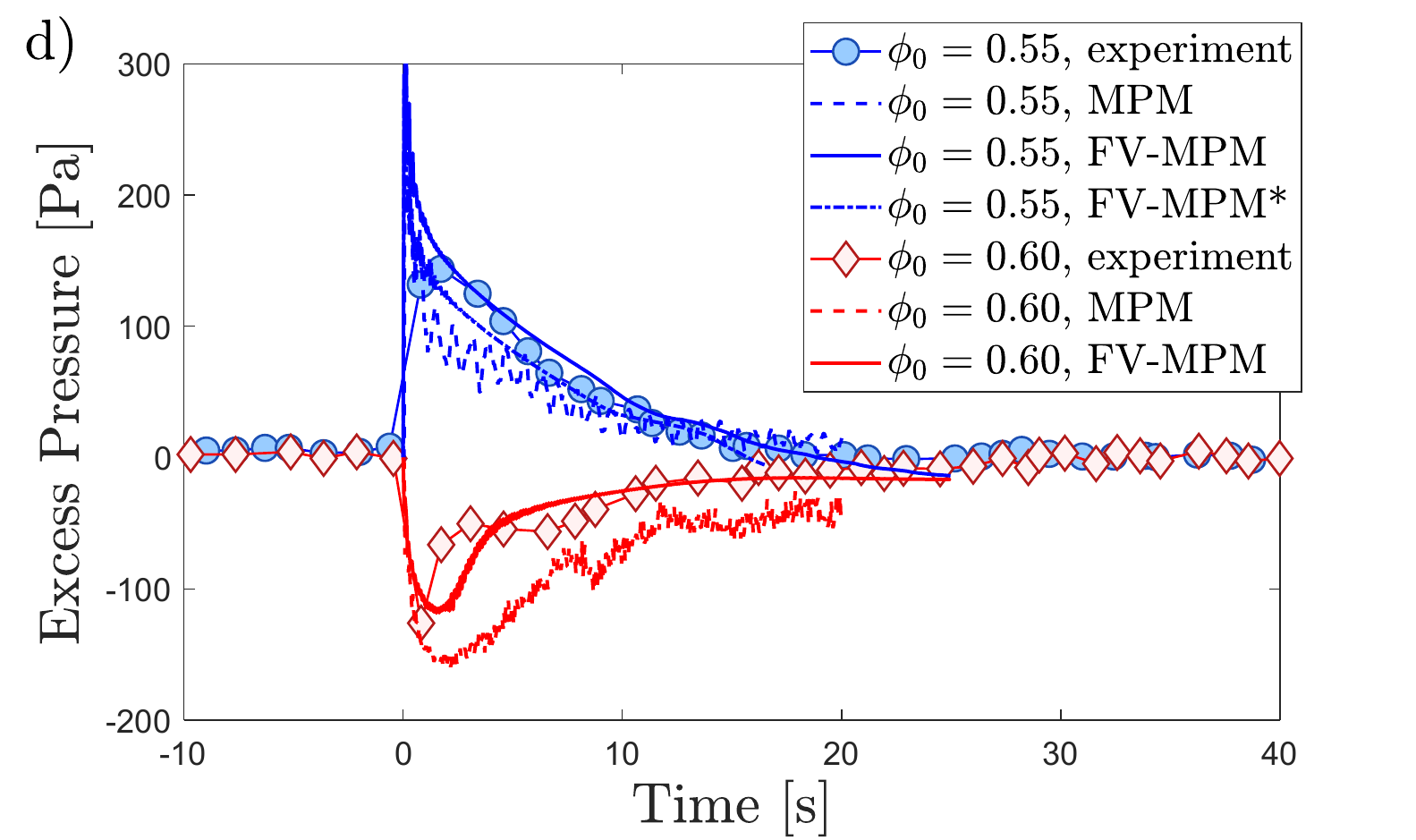}\\
	\end{matrix}
	$
	\caption{Collapse of a submerged column of glass beads: a) initial configuration of granular column with height $H_0$, length $L_0$, and solid volume fraction $\phi_0$; b) final configuration of granular column and position of pressure sensor; c) comparison of front position, $L$, found experimentally (symbols; see \cite{rondon2011}), with MPM (dashed lines; see \cite{baumgarten2019a}), and with FV-MPM (solid and dash-dotted lines); d) comparison of excess fluid pressures found experimentally (symbols), with MPM (dashed lines), and with FV-MPM (solid and dash-dotted lines). * indicates FV-MPM simulation run with zero pressure BC (see Figure \ref{fig:collapse_results}).}
	\label{fig:collapse_data}
\end{figure}

The third case that we examine using this numerical framework is the collapse of two columns of glass beads submerged in a viscous fluid. In particular, we seek to recreate the numerical simulations reported in \cite{baumgarten2019a} for two of the experiments in \cite{rondon2011}. The two columns have the same total mass and are composed of many small, spherical glass beads ($d = 255\ \mu$m and $\rho_s = 2500$kg/m$^3$). Both columns are held behind a retaining wall and submerged in a mixture of water and Ucon oil ($\eta_0=12$ mPa$\cdot$s and $\rho_f = 1000$ kg/m$^3$), which fills an open tank measuring 70 cm$\times$15 cm$\times$15 cm. (For the numerical simulations performed in \cite{baumgarten2019a}, it was assumed that the length and height of the tank could be shortened to 30 cm$\times$10 cm and that the whole problem could be modeled as quasi-two-dimensional.) The only difference between the two columns is their initial preparation: one column is densely packed with $\phi_0 = 0.60$, $H_0=0.42$ cm, and $L_0=6$ cm and the other is loosely packed with $\phi_0 = 0.55$, $H_0=0.45$ cm, and $L_0=6$ cm. At the beginning of the experiments, the retaining walls are removed and the columns collapse under their own weight. See Figure \ref{fig:collapse_data}a and \ref{fig:collapse_data}b for schematic of the initial and final configurations of the columns, respectively, along with the location of the pressure sensor used to measure the fluid pressure, $p_f$, at the base of the columns.

To simulate these two experiments, we use the same elasto-plastic effective granular stress model for $\boldsymbol{\tilde{\sigma}}$ as is reported in \cite{baumgarten2019a}, which accounts for frictional granular flow ($\mu_1 = 0.35$), rate-dependence ($\mu_2 = 1.39$ and $b = 0.31$), shear induced dilation ($\phi_m = 0.585$, $a = 1.23$, and $K_3 = 4.72$), granular elasticity ($E = 10$ kPa, $\nu = 0.3$), and free granular separation. We model the fluid phase of the mixture as a weakly compressible viscous fluid with bulk modulus $\kappa = 10$ kPa, baseline density $\rho_{f0} = 1000$ kg/m$^3$, and effective viscosity $\eta_r = \eta_0 (1 + \tfrac{5}{2}\phi)$. The drag interaction is modeled using the form of  $\hat{F}(\phi,\Reyn)$ from \cite{beetstra2007}.

\begin{figure}[!h]
	\centering
	\includegraphics[scale=0.43]{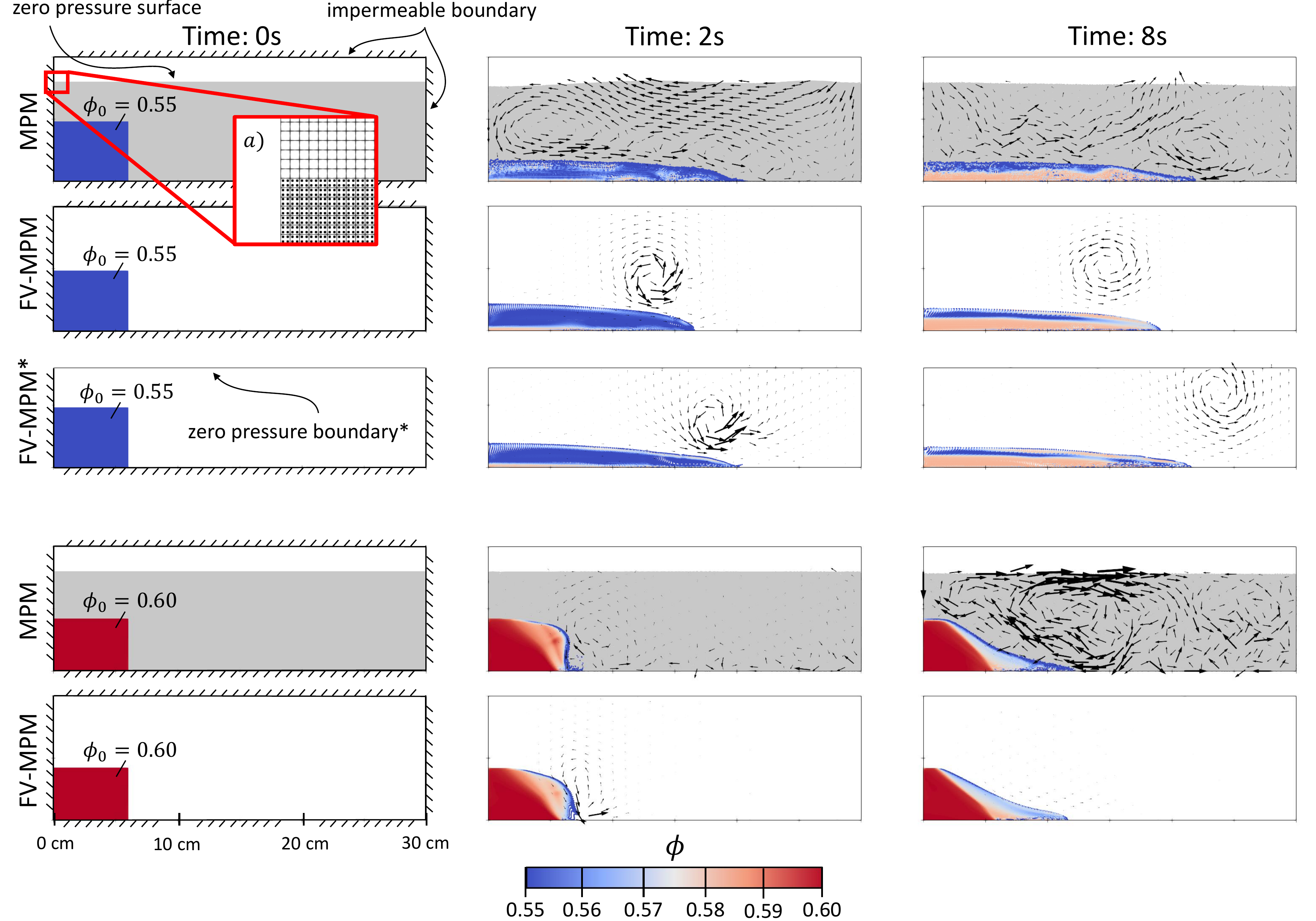}
	\caption{Simulation snapshots for collapse of a submerged column of glass beads. The images in each column represent the flow solution at 0 s, 2 s, and 8 s, respectively, and are found using the simulation frameworks and initial conditions labeled at the left of the figure. The first column images indicate the relevant boundary conditions applied in each case (* indicates the application of a { zero} pressure BC at the upper surface), and the arrows in the second and third columns indicate the strength and direction of the fluid velocity field, $\boldsymbol{v}_f$. The length scale of these arrows is 0.25 s $\times\ \boldsymbol{v}_f$. The material points associated with the granular phase are colored by the local granular volume fraction, $\phi$, and the material points associated with the fluid phase are colored gray. Inset a) highlights the regular Cartesian grid used in the MPM and FV-MPM simulations as well as markers for the initial material point discretization. }
	\label{fig:collapse_results}
\end{figure}

To discretize the simulation domain, we use the 300$\times$100 element Cartesian FE grid from \cite{baumgarten2019a} --- with the bi-cubic-spline basis functions from \cite{steffen2008analysis} --- and represent both granular columns with 4 material point tracers per FE grid cell. As in \cite{baumgarten2019a}, we model the boundaries of the domain as frictional walls.
The primary difference between our simulation approach and that used in \cite{baumgarten2019a} is the treatment of the fluid phase of the mixture. In \cite{baumgarten2019a}, the fluid fills the tank to a height of 8 cm and is modeled using its own set of material point tracers, which are shown in gray in Figure \ref{fig:collapse_results} and allow for direct simulation of the fluid free surface. In this work, however, we let the tank be fully filled so that the fluid phase can be modeled on a set of finite volumes, which are defined by the same 300$\times$100 element Cartesian grid and have an impermeable upper boundary.

In addition to the two simulations run in this manner (labeled `FV-MPM' in Figures \ref{fig:collapse_data} and \ref{fig:collapse_results}), we also assess the potential impact of the fluid free surface on these results by running a third simulation on a 30 cm $\times$ 8 cm (300$\times$80 element) Cartesian grid using a zero pressure, static upper boundary (labeled `FV-MPM*' in Figures \ref{fig:collapse_data} and \ref{fig:collapse_results}). This zero pressure condition does not represent a physical boundary condition but is a closer numerical approximation to the boundary conditions used in \cite{baumgarten2019a}. The various boundary conditions that are applied to the fluid phase of these mixtures are shown in the first column of Figure \ref{fig:collapse_results}.

A comparison of the predictions made by MPM and FV-MPM with the experimental measurements reported in \cite{rondon2011} is shown in Figures \ref{fig:collapse_data}c and \ref{fig:collapse_data}d, and snapshots of these simulated flow solutions (at 0 s, 2 s, and 8 s) are shown in Figure \ref{fig:collapse_results}. Both simulation approaches show close agreement with the front positions (i.e.\ the leading edge of the collapsing columns) and excess pore pressures (i.e.\ the fluid pressure, $p_f$, inside the columns minus any hydrostatic component) reported from the experiments, and the dynamics of the collapsing columns are remarkably similar. However, there are four differences apparent in Figures \ref{fig:collapse_data} and \ref{fig:collapse_results} that should be discussed.

The first two differences highlight the relative advantages of the FV-MPM approach over MPM. First, in Figure \ref{fig:collapse_data}d the excess pore pressure found in \cite{baumgarten2019a} for the densely packed column is about 50 Pa lower than the experimental measurements and the results found in this work; this is likely an artifact of the ringing instability in MPM (see \cite{gritton2014,gritton2017}) and is avoided by evaluating the fluid pressure with static finite volumes. Second, the snapshots of the MPM flow solutions at 8 s of simulated time in Figure \ref{fig:collapse_results} --- after the columns have essentially settled --- show the growth of noticeable secondary fluid flows that do not dissipate and are absent from the FV-MPM simulations; these spurious motions are indicative of quadrature errors accumulating in the MPM solutions (see \cite{zhang2018}) and are examples of another numerical artifact that is avoided with FV-MPM.

The other two differences that are apparent in these results highlight a short-coming of the proposed method: an inability to model the fluid free surface. First, in Figure \ref{fig:collapse_data}c, the final front position in the `FV-MPM' simulations of the loosely packed column is about 2 cm less than the experimental measurements, the MPM results, and the `FV-MPM*' results; this is likely due the lack of a nearby zero pressure boundary condition, which is associated with this fluid surface. Second, in Figure \ref{fig:collapse_results}, the fluid flow solution predicted by MPM for the loosely packed column shows substantial sloshing of the mixture surface: something markedly absent from both the `FV-MPM' and `FV-MPM*' solutions. To capture the behavior of the free fluid surface in this framework, without using Lagrangian material tracers, requires augmentation of the method with some form of surface tracking scheme (e.g.\ the volume of fluid method or the level set method; see \cite{hirt1981,sethian2003}).

\subsection{Erosion of sand by air}\label{sec:2d_erosion}
\begin{figure}[H]
	\centering
	\includegraphics[scale=0.4]{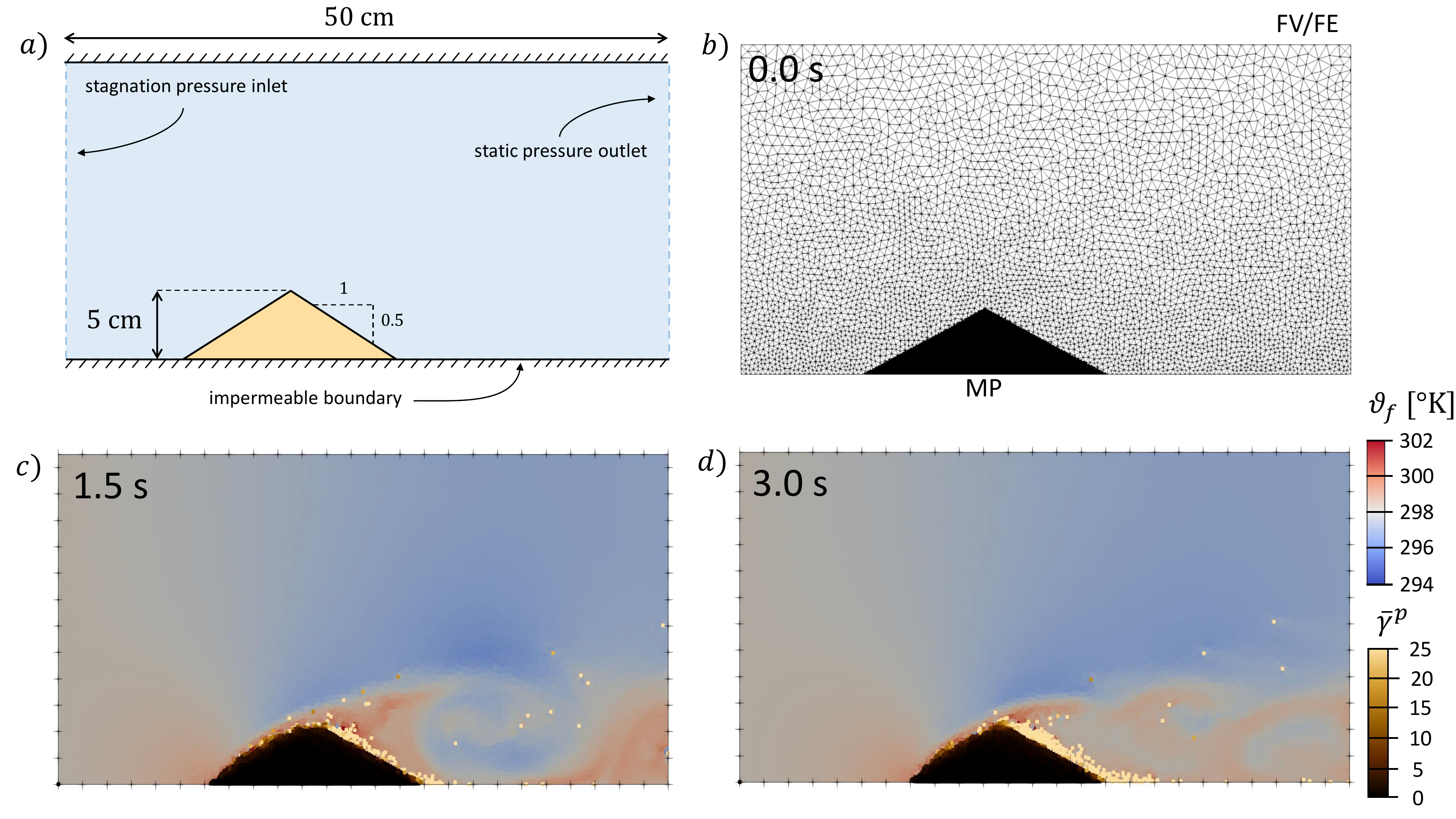}
	\caption{Numerical simulation sand pile erosion by air in two dimensions: a) schematic diagram of numerical test showing boundary conditions and initial dimensions of sand pile; b) finite volume (FV) and finite element (FE) triangular grids along with initial material point (MP) distribution (representing the sand pile at 0.0 s); c) mixed flow solution at 1.5 s with air colored by effective temperature, $\vartheta_f$, and material points (sand) colored by the effective shear strain measure, $\bar{\gamma}^{p}$; d) mixed flow solution at 3.0 s. Tick marks in c) and d) mark 2 cm intervals.
	}
	\label{fig:erosion_2D}
\end{figure}

We now look at problems that FV-MPM is uniquely suited to simulate and model. Consider a two-dimensional channel measuring 25 cm tall and 50 cm long filled with an air-like fluid ($\rho_f = 1.18$ kg/m$^3$ and $\vartheta_f = 298$ K). Suppose that a pile of sand-like material ($\rho_s = 2700$ kg/m$^3$, $d=0.25$ mm, and $\phi = 0.6$) measuring 5 cm tall and 20 cm wide is situated near the front of the channel as in Figure \ref{fig:erosion_2D}a. In this numerical example, we subject this pile of sand to a steady flow of air and model its erosion using FV-MPM.

To simulate this problem, we use the identical FE and FV grids shown in Figure \ref{fig:erosion_2D}b --- with linear FE basis functions --- and represent the pile of sand with 1,600 material point tracers. The sand is modeled using the effective granular stress model for $\boldsymbol{\tilde{\sigma}}$ from \cite{dunatunga2015} that accounts for simple frictional granular flow ($\mu_1 = \mu_2 = 0.67$), free granular separation ($\bar{\rho}_c = 1620$ kg/m$^3$), and granular elasticity ($E = 16$ MPa and $\nu = 0.3$). We treat the air-like material as an ideal gas with heat capacity ratio $\gamma_r = 1.4$, specific gas constant $R = 24$ J/kg$\cdot$K, viscosity $\eta_0 = 18\ \mu$Pa$\cdot$s, and coefficient of thermal conductivity $k_f = 0.026$ W/m$\cdot$K. The drag model from \cite{beetstra2007} is again used to determine $\hat{F}(\phi,\Reyn)$. To capture the impact of unresolved turbulence on our flow solution, we add the Smagorinsky eddy viscosity ($\mu_v = \bar{\rho}_f \alpha_h^2 h_\alpha^2 \sqrt{2} \|\boldsymbol{D}_{f0}\|$ with $\alpha_h \approx 0.16$) to our fluid model as in \cite{smagorinsky1963,wilcox1998}.

To induce air flow through the channel, we assign pressure boundary conditions that correspond to an unimpeded air speed of 17 m/s (i.e.\ a stagnation pressure and temperature at the channel inlet, $p_f^* = 8700$ Pa and $\vartheta_f^* = 299.7$ K, and a static pressure and temperature at the channel outlet, $p_f = 8408$ Pa and $\vartheta_f = 298$ K). Figure \ref{fig:erosion_2D}c and \ref{fig:erosion_2D}d show snapshots of the FV-MPM flow solution at 1.5 s and 3 s, respectively, including the effective air temperature, $\vartheta_f$, and equivalent granular shear strain measure, $\bar{\gamma}^p$, from \cite{dunatunga2015}. Here $\bar{\gamma}^p$ represents the accumulated shear deformation that the granular material has undergone. The high value of $\bar{\gamma}^p$ on the leeward slope of the sand pile is indicative of material that has undergone substantial shearing; in this case, it indicates that this material has been dragged along the windward face and over the crest of the dune.

\subsection{Formation of a barchan dune}\label{sec:3d_erosion}
\begin{figure}[!h]
	\centering
	\includegraphics[scale=0.4]{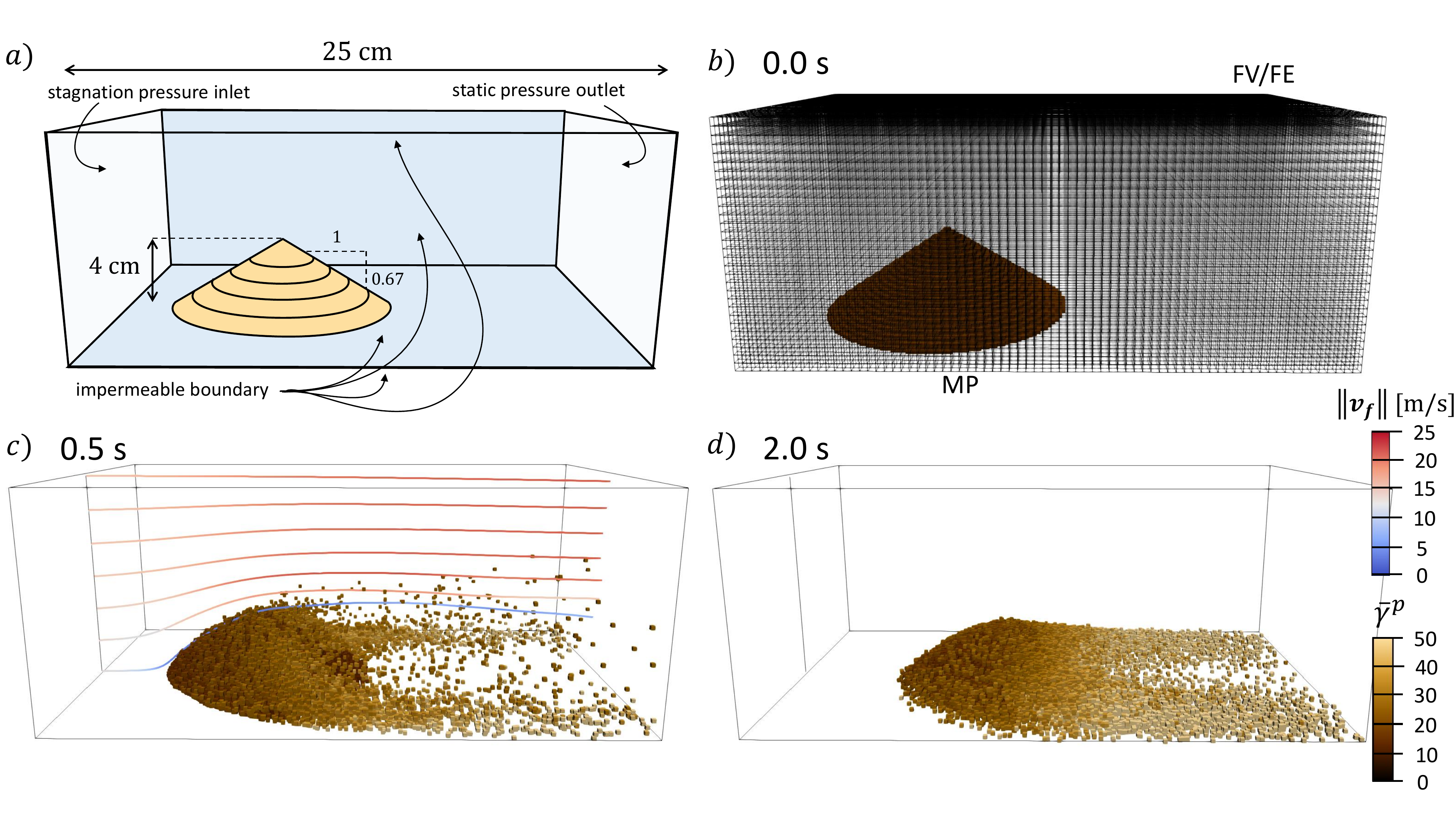}
	\caption{Numerical simulation of sand pile erosion by gust of wind in three dimensions: a) schematic diagram of numerical test showing boundary conditions and initial dimensions of sand pile; b) finite volume (FV) and finite element (FE) Cartesian grids along with initial material point (MP) distribution (representing the sand pile at 0.0 s); c) mixed flow solution at 0.5 s with air streamlines (originating along center-line of inlet) colored by velocity magnitude, $||\boldsymbol{v}_f||$, and material points (sand) colored by the effective shear strain measure, $\bar{\gamma}^{p}$; d) mixed flow solution at 2.0 s, after wind gust has stopped.}
	\label{fig:erosion_3D}
\end{figure}

We continue the assessment of our method by extending the previous example into three dimensions. Consider a channel measuring 10 cm tall, 15 cm wide, and 25 cm long filled with an air-like fluid ($\rho_f = 1.18$ kg/m$^3$ and $\vartheta_f = 298$ K). Suppose a conical pile of sand ($\rho_s = 2700$ kg/m$^3$, $d=0.25$ mm, and $\phi = 0.6$) measuring 4 cm tall and 12 cm wide is situated near the front of the channel as in Figure \ref{fig:erosion_3D}a. In this numerical example, we subject this pile of sand to a 1 s gust of air and model its erosion using FV-MPM.

To simulate this problem, we use the identical 75$\times$45$\times$30 element Cartesian FE and FV grids shown in Figure \ref{fig:erosion_3D}b --- with tri-linear ``tent'' FE basis functions --- and represent the sand pile with 32,224 material point tracers. The sand is again modeled using the effective granular stress model for $\boldsymbol{\tilde{\sigma}}$ from \cite{dunatunga2015} that accounts for simple frictional granular flow ($\mu_1 = \mu_2 = 0.67$), granular separation ($\bar{\rho}_c = 1620$ kg/m$^3$), and granular elasticity ($E = 16$ MPa and $\nu = 0.3$). We treat the air-like material as an ideal gas with heat capacity ratio $\gamma_r = 1.4$, specific gas constant $R = 24$ J/kg$\cdot$K, viscosity $\eta_0 = 18\ \mu$Pa$\cdot$s, and coefficient of thermal conductivity $k_f = 0.026$ W/m$\cdot$K. The drag model from \cite{beetstra2007} is used to determine $\hat{F}(\phi,\Reyn)$, and the Smagorinsky eddy viscosity correction from \cite{smagorinsky1963,wilcox1998} is applied to the fluid.

The channel boundary conditions are identical to those used in the two-dimensional erosion problem and correspond to an unimpeded air speed of 17 m/s. At the beginning of the simulation, the air is allowed to flow freely; however, after 1 s of simulated time, the boundary conditions are reset to stop the flow (i.e.\ $p_f^* = p_f = 8408$ Pa and $\vartheta_f^* = \vartheta_f = 298$ K). With more degrees of freedom in three dimensions, a more complex fluid flow occurs. Figures \ref{fig:erosion_3D}c and \ref{fig:erosion_3D}d show snapshots of the FV-MPM flow solutions at 0.5 s and 2 s, respectively. In the 0.5 s snapshot, streamlines are traced out beginning from the center-line of the channel inlet and colored by the local air speed, $\|\boldsymbol{v}_f\|$; at 2 s, the gust of air has stopped and there is no longer any airflow. The material points in both snapshots are colored by the accumulated equivalent shear strain measure, $\bar{\gamma}^p$.

\begin{figure}[H]
	\centering
	\includegraphics[scale=0.36]{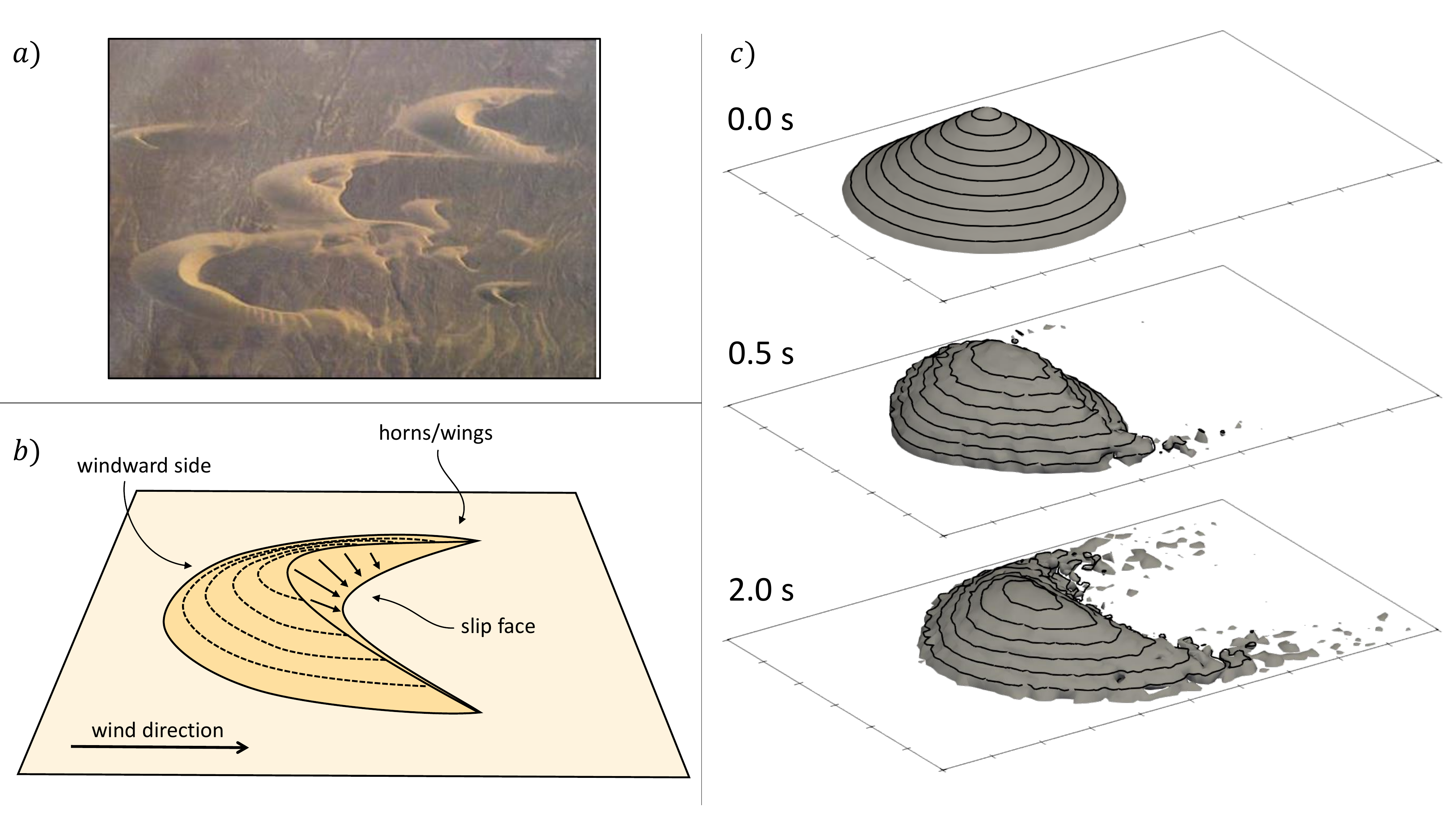}
	\caption{Numerical simulation of sand pile erosion by gust of wind in three dimensions: a) a field of crescent-shaped barchan sand dunes in the desert between Chimbote and Casma on the coast of Peru (image source: \cite{schwammle2003}); b) schematic diagram of characteristic barchan dune shape (see \cite{sauermann2000}); b) $\phi = 0.3$ surface contours for flow solutions at 0 s, 0.5 s, and 2 s. Black contour lines in b) are marked at 4 mm vertical increments, and the tick marks along the boundary denote 2.5 cm increments.}
	\label{fig:barchan}
\end{figure}

An interesting feature of this three-dimensional erosion process is the change in shape of the initially conical pile of sand. As the simulation progresses, the sand forms a crescent shape with two long horns extending in the direction of the airflow. When the material has settled, the final shape of the pile is reminiscent of a barchan dune: a common dune type observed in deserts around the world. An example image of several barchan dunes in Peru (from \cite{schwammle2003}) is shown in Figure \ref{fig:barchan}a alongside a schematic of the classic barchan dune features in Figure \ref{fig:barchan}b. To highlight the similarity between our simulated sand pile and this characteristic shape, we show several of the $\phi = 0.3$ surface contours from our FV-MPM simulation in Figure \ref{fig:barchan}c. Although this problem is an example of erosion over a much shorter time-scale, the similarity of these dune shapes suggests that FV-MPM is capable of modeling real world processes.

\subsection{Rocket exhaust impinging on Martian soil}\label{sec:rocket}
In this final numerical example, we examine a problem that is of growing interest to the aerospace community: the effect of rocket exhaust plumes on the dusts and sediments found on extraterrestrial bodies. In the last decade many authors have investigated this effect using numerical techniques that focus on the entrained granular material (i.e.\ the grains that are kicked up from the granular surface; see \cite{morris2015,ejtehadi2020}). These approaches generally rely on empirical erosion rates, which are based on laboratory experiments and analyses of previous extraterrestrial landings (e.g.\ subsonic cratering experiments \cite{metzger2010} and analyses of the Apollo 12 landing \cite{immer2011}), and do not account for the changing topography of the granular surface as it craters. This suggests that FV-MPM may be of use for modeling these types of problems, as it is capable of simulating the complex interaction between the impinging exhaust gases, the entrained granular material, and the deforming, solid-like granular surface.

Suppose we intend to use the Apollo Lunar Module Descent Engine (LMDE; see \cite{farrow1967,cherne1967}) in the design of a Martian lander and are interested in how the exhaust gases from this engine will interact with the dusty Martian surface. In particular, we want to predict the surface cratering that will occur as the lander comes to rest. In this numerical example, we simulate a rough approximation of this problem using FV-MPM.

\begin{figure}[!h]
	\centering
	\includegraphics[scale=0.36]{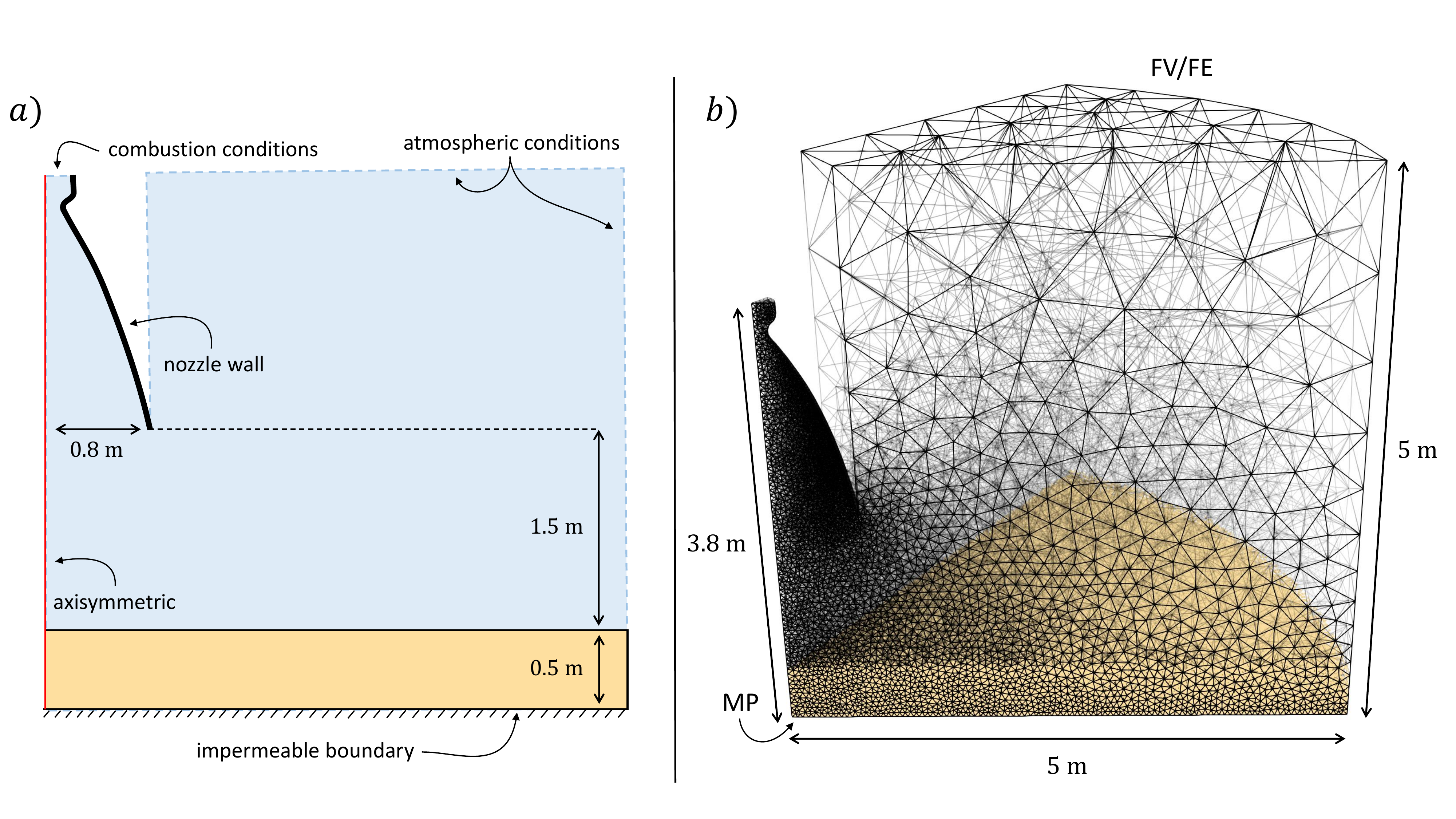}
	\caption{Numerical simulation of rocket exhaust impinging on Martian soil: a) schematic diagram of numerical test showing boundary conditions, dimensions, and initial configuration of soil; b) finite volume (FV) and finite element (FE) tetrahedral grids along with initial material point (MP) distribution. Simulation is performed on a 60$^\circ$ wedge of a cylindrical domain with symmetric boundary conditions applied to reflected boundaries of the domain. Martian gravity (3.7 m/s$^2$) is used for $\boldsymbol{g}$.}
	\label{fig:rocket_mesh}
\end{figure}

\begin{figure}[!h]
	\centering
	\includegraphics[scale=0.8]{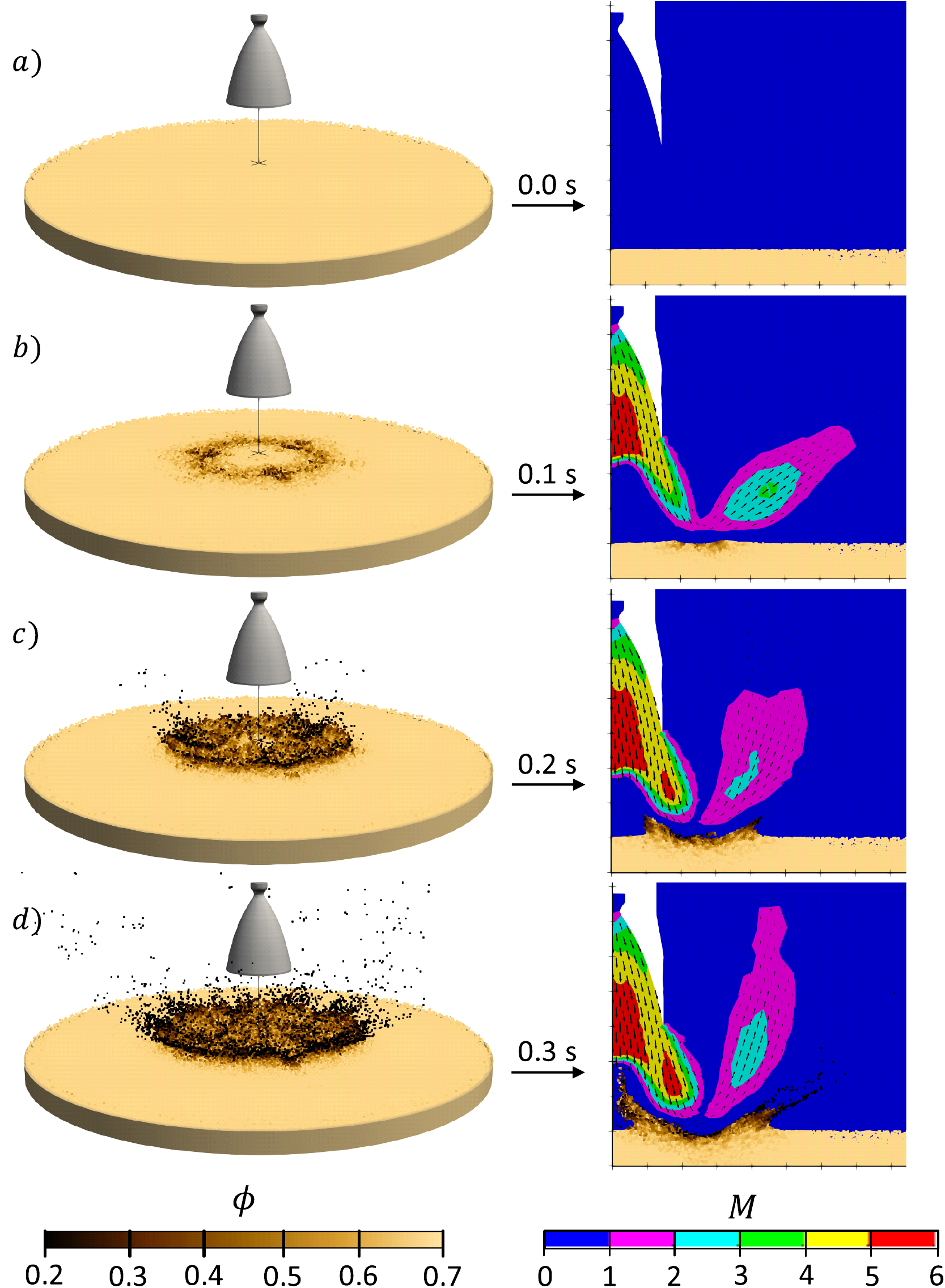}
	\caption{Numerical simulation of rocket exhaust impinging on Martian soil: a) flow solution at 0.0 s, b) flow solution at 0.1 s, c) flow solution at 0.2 s, and d) flow solution at 0.3 s. Left column: renderings of Apollo Lunar Module Descent Engine (LMDE) and material point tracers colored by the local volume fraction $\phi$. Right column: slice of simulated domain extending radially outward from the center-line highlighting the overlapping fluid and granular domains; the fluid domain is colored by the local mach number $M = \|\boldsymbol{v}_f\|/\sqrt{\gamma_r R \vartheta_f}$.}
	\label{fig:rocket_exhaust}
\end{figure}

To perform this simulation, we consider the axisymmetric domain shown in the schematic in Figure \ref{fig:rocket_mesh}a. The base of this domain is impermeable and covered in a 0.5 m thick bed of Martian soil, and the top and outside boundaries of this domain are open to the Martian atmosphere ($p_f = 610$ Pa and $\vartheta_f = 210$ K). The simulated LMDE is fixed in position with the nozzle exit 1.5 m above the soil surface. To model the rocket exhaust accurately, we treat the nozzle wall as impermeable and friction-less, and assign a chamber pressure and temperature consistent with the operating conditions of the LMDE ($p_f^*=710$ kPa and $\vartheta_f^* = 3000$ K).

For our numerical implementation, we use the identical tetrahedral FE and FV wedge grids shown in Figure \ref{fig:rocket_mesh}b --- with linear FE basis functions --- and represent the Martian soil with 498,563 material point tracers. 
{Note that simulation of the axisymmetric governing equations on a planar grid is also possible but requires additional considerations that are not included in this work.}
We model the soil using the effective granular stress model for $\boldsymbol{\tilde{\sigma}}$ from \cite{baumgarten2019a}, which accounts for frictional granular flow ($\mu_1 = \mu_2 = 0.47$), shear induced dilation ($\phi_m = 0.68$, $a = 1.23$, and $K_3 = 4.72$), granular elasticity ($E = 10$ MPa, $\nu = 0.3$), and free granular separation. The material properties for the Martian soil are approximated from \cite{perko2006}, ($\rho_f = 2400$ kg/m$^3$ and $d = 0.5$ mm). We treat the rocket exhaust and Martian atmosphere as a single ideal gas with heat capacity ratio $\gamma_r = 1.29$, specific gas constant $R = 189$ J/kg$\cdot$K, viscosity $\eta_0 = 14\ \mu$Pa$\cdot$s, and coefficient of thermal conductivity $k_f = 0.026$ W/m$\cdot$K. The drag model from \cite{beetstra2007} is used to determine $\hat{F}(\phi,\Reyn)$.

There is one final consideration that must be made before we can simulate this problem in our numerical framework: it is necessary to include additional numerical dissipation to suppress oscillations near strong shocks and avoid the carbuncle phenomenon near the central axis (see \cite{mccorquodale2011,pandolfi2001}). This requirement is not unique to our numerical method and has been studied thoroughly in the literature (e.g.\ see \cite{wilkins1980, barter2010}). Here we extend the artificial viscosity approach from \cite{mccorquodale2011}, which adds artificial diffusion to regions near strong shocks but does not affect smoothly varying flows. Discussion of this approach can be found in Appendix \ref{sec:artificial_viscosity}.

With this problem correctly implemented, we begin our simulation by ramping up the chamber pressure and temperature over several milliseconds. Once the simulated engine reaches the desired operating condition, we allow the flow to develop according to the equations of motion in \eqref{eqn:mixture_equations}. Figure \ref{fig:rocket_exhaust} shows snapshots of the mixed flow solution at 0.1 s intervals for this rocket impingement and cratering problem. The left column of Figure \ref{fig:rocket_exhaust} shows the reflections of our 60$^\circ$ wedge domain around the central axis and allows us to visualize the deformation of the soil bed using the material point tracers (colored by the local volume fraction, $\phi$). The right column of Figure \ref{fig:rocket_exhaust} shows slices through the simulated domain and allows us to visualize the shape of the exhaust plume predicted by the finite volumes (colored by the local mach number, $M=\|\boldsymbol{v}_f\|/\sqrt{\gamma_r R \vartheta_f}$).

Several features of this predicted flow solution are similar to those found in previous studies, including the mach contours inside the nozzle and the recirculation bubble just above the granular surface (see \cite{morris2015}); however there are several new features that have not been previously simulated. In particular, we observe that the flow direction of the exhaust gases changes by about 45$^\circ$ vertically as the crater is forming. All together, our results highlight a strength of FV-MPM in comparison with more traditional approaches: the ability to accurately simulate the coupling between surface deformations, entrained granular material, and impinging fluids.

\section{Conclusion}
We have introduced a novel numerical simulation approach for a special class of engineering problems: large deformation flows of mixtures of fluids and porous solids. In this work, we have focused our attention on fluid--sediment mixtures. These types of problems are unique in that they have features that require high accuracy numerical advection (e.g.\ granular stresses, history dependent variables, porosity, etc.), suggesting a Lagrangian simulation approach, as well as features that make such Lagrangian approaches difficult (e.g.\ highly turbulent regions, inlet and outlet conditions, etc.).

To overcome these challenges, we have proposed the FV-MPM simulation framework, which uses material point tracers and an Eulerian finite element grid to solve the equations of motion of the granular phase of these mixtures, and an Eulerian finite volume grid to solve the equations of motion of the fluid phase of these mixtures. By tracking the motion of the granular phase on the set of material point tracers, the advection of masses, stresses, and material properties occurs without the numerical diffusion associated with purely Eulerian approaches. Additionally, by using finite volumes to calculate the motion of the fluid, the numerical integration of the fluid equations is not affected by the amount of deformation that occurs over the course of a long simulation.

This numerical framework builds on recent successes using two-phase MPM approaches (see \cite{abe2013, bandara2015, baumgarten2019a}) by incorporating thoroughly studied aspects of more common FVM approaches (see \cite{roe1981,barth1989,geiger2004}). In our analysis, we have shown that the proposed framework is robust and capable of addressing several classes of time-dependent mixture problems, from soil compaction to submerged granular flows to sediment erosion by air. Additional discussion of the method, its accuracy and assumptions, and useful approximations can be found in the Appendix.

\section*{Acknowledgments}
AB and KK acknowledge support from Army Research Office Grants W911NF-16-1-0440 and W911NF-19-1-0431.

\appendix

\gdef\thesection{\Alph{section}} 
\makeatletter
\renewcommand\@seccntformat[1]{Appendix \csname the#1\endcsname.\hspace{0.5em}}
\makeatother

\section*{Appendices}

\setcounter{figure}{0}
\section{Mixture Theory and Simplifying Assumptions}\label{sec:mixture_theory}
The governing equations summarized in \eqref{eqn:mixture_equations} derive from the mixture theories of \cite{jackson2000, drumheller2000, coussy2004}. In this section, we briefly review the specialization of these theories to fluid--sediment mixtures and explicate the simplifying assumptions that we've made.

\begin{figure}[!h]
	\centering
	\includegraphics[scale=0.5]{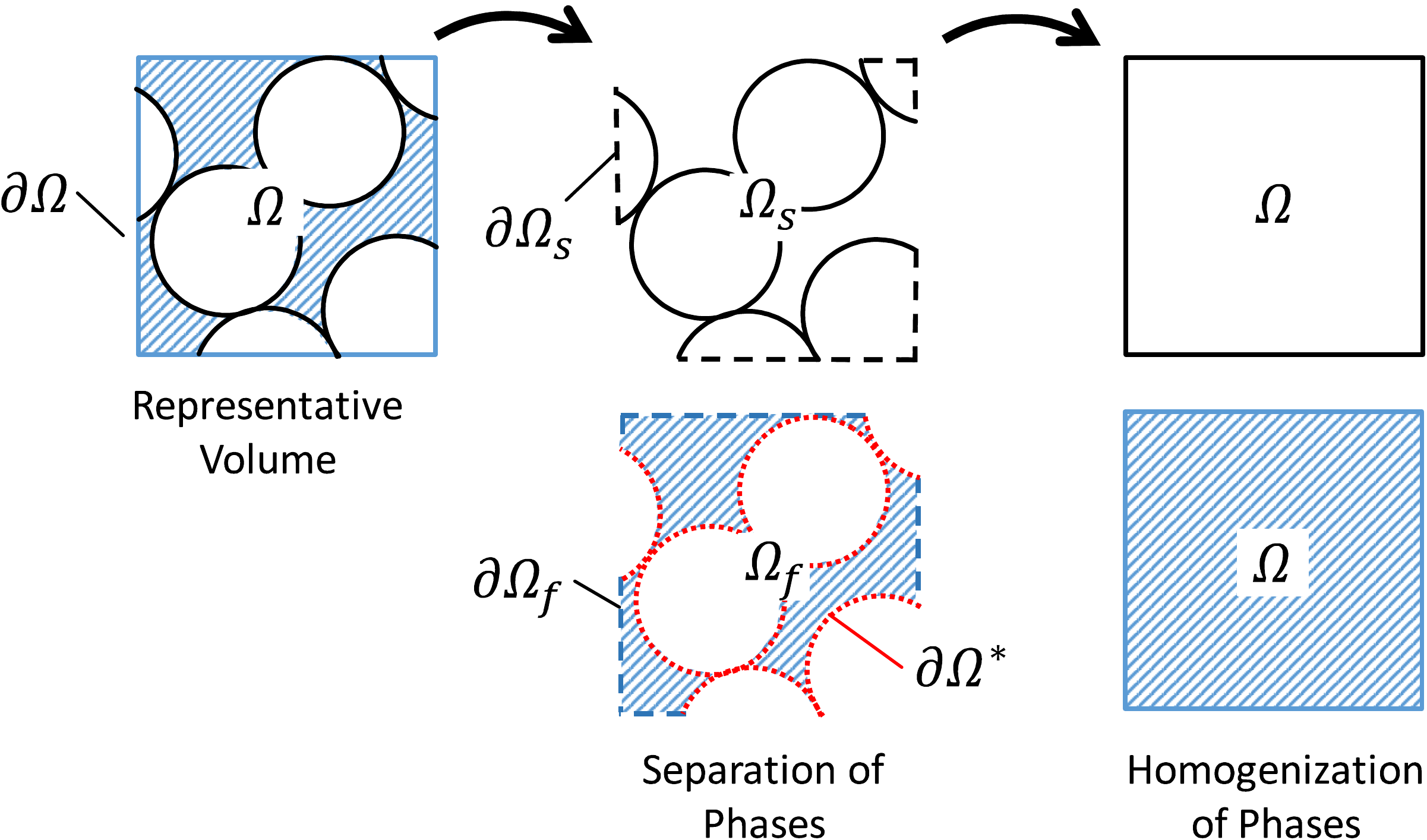}
	\caption{Pictorial description of the representative volume $\Omega$ and boundary $\partial \Omega$, the decomposition of the domain into fluid and solid volumes, and the homogenization of the two phases. Here $\Omega = \Omega_s \cup \Omega_f$ and $\partial \Omega = \partial \Omega_s \cup \partial \Omega_f$ with $\partial \Omega^*$ defining the interior surface separating the solid and fluid domains.}
	\label{fig:homogenization}
\end{figure}

As stated above, this formulation assumes that the grains are quasi-mono-disperse, rough, incompressible with density $\rho_s$, and fully immersed in a compressible fluid with density $\rho_f$. A representative volume of material, $\Omega$, can therefore be decomposed into a solid volume, $\Omega_s$, and a fluid volume, $\Omega_f$, such that $\Omega = \Omega_s \cup \Omega_f$. Figure \ref{fig:homogenization} shows how this volume is decomposed and the important step of homogenizing the solid and fluid volumes into two, overlapping continua.

\subsection{Homogenization}
Following from the works of \cite{jackson2000, drumheller2000, baumgarten2019a, coussy2004, batchelor1970}, we define the effective densities ($\bar{\rho}_s$ and $\bar{\rho}_f$), continuum velocities ($\boldsymbol{v}_s$ and $\boldsymbol{v}_f$), and effective internal energies ($\varepsilon_s$ and $\varepsilon_f$) such that conservation of mass, momentum, and energy in the continuum correspond to conservation of mass, momentum, and energy in the real mixture. Toward this end, we consider a representative volume of material, $\Omega$, that contains a \textit{large} number of individual grains: for the continuum approximation to be valid, the volume must be large enough to smooth out grain-scale phenomena and capture the average bulk behavior of the mixture. For such a volume, we define:
\begin{equation}
	\begin{aligned}
		\int_\Omega{\bar{\rho}_s dv} &\equiv \int_{\Omega_s}{\rho_s dv},
		\\ \int_\Omega{\bar{\rho}_f dv} &\equiv \int_{\Omega_f}{\rho_f dv},
		\\ \int_\Omega{\bar{\rho}_s \boldsymbol{v}_s dv} &\equiv \int_{\Omega_s}{\rho_s \boldsymbol{v} dv},
		\\ \int_\Omega{\bar{\rho}_f \boldsymbol{v}_f dv} &\equiv \int_{\Omega_f}{\rho_f \boldsymbol{v} dv},
		\\ \int_\Omega{\bar{\rho}_s (\varepsilon_s + \tfrac{1}{2}\boldsymbol{v}_s \cdot \boldsymbol{v}_s) dv} &\equiv \int_{\Omega_s}{\rho_s (\varepsilon + \tfrac{1}{2} \boldsymbol{v} \cdot \boldsymbol{v}) dv},
		\\ \int_\Omega{\bar{\rho}_f (\varepsilon_f + \tfrac{1}{2}\boldsymbol{v}_f \cdot \boldsymbol{v}_f) dv} &\equiv \int_{\Omega_f}{\rho_f (\varepsilon + \tfrac{1}{2} \boldsymbol{v} \cdot \boldsymbol{v}) dv},
	\end{aligned}
	\label{eqn:homogenization}
\end{equation}
with $\rho_s$ and $\rho_f$ the \textit{true} material densities, $\boldsymbol{v}$ the \textit{true} material velocity, $\varepsilon$ the \textit{true} internal energy, and $\Omega_s$ and $\Omega_f$ defined as in Figure \ref{fig:homogenization}. Following this procedure, the local solid volume fraction ($\phi$) can be defined as the ratio of the volume of solid grains ($\Omega_s$) to the volume of mixture ($\Omega$) and the local porosity ($n$) can be defined similarly. If many such volumes are chosen over which  variations in these local fields are negligible (i.e.\ relatively \textit{small} volumes), we can then relate the effective fields to the true mixture densities as in \eqref{eqn:porosity}.

\subsection{Mass Conservation}
Conservation of mass within an arbitrary volume of real mixture $\Omega = \Omega_s \cup \Omega_f$ with boundary $\partial \Omega = \partial \Omega_s \cup \partial \Omega_f$ has the following form:
\begin{equation}\label{eqn:mass_conservation_specific}
	\begin{aligned}
		\frac{\partial}{\partial t} \int_{\Omega_s}{\rho_s dv} + \int_{\partial \Omega_s}{\rho_s \boldsymbol{v} \cdot \hat{\boldsymbol{n}} da} &= 0, \\
		\frac{\partial}{\partial t} \int_{\Omega_f}{\rho_f dv} + \int_{\partial \Omega_f}{\rho_f \boldsymbol{v} \cdot \hat{\boldsymbol{n}} da} &= 0,
	\end{aligned}
\end{equation}
with $\hat{\boldsymbol{n}}$ the outward facing surface normal vector, and $\partial\Omega_s$ and $\partial\Omega_f$ defining the exterior solid and fluid surfaces as in Figure \ref{fig:homogenization}. (Note that $\partial \Omega^*$, the interior surface within the domain $\Omega$, moves with the particle boundaries, therefore there is no mass flux across it.) Using the identities in \eqref{eqn:homogenization} for $\Omega$ encompassing relatively \textit{small} volumes, \eqref{eqn:mass_conservation_specific} can be re-expressed as,
\begin{equation}\label{eqn:mass_conservation}
	\begin{aligned}
		\frac{\partial}{\partial t} \int_\Omega{\bar{\rho}_s dv} + \int_{\partial \Omega}{\bar{\rho}_s \boldsymbol{v}_s \cdot \hat{\boldsymbol{n}} da} &= 0, \\
		\frac{\partial}{\partial t} \int_\Omega{\bar{\rho}_f dv} + \int_{\partial \Omega}{\bar{\rho}_f \boldsymbol{v}_f \cdot \hat{\boldsymbol{n}} da} &= 0,
	\end{aligned}
\end{equation}
and since \eqref{eqn:mass_conservation} must hold for \textit{any} representative volume, we require,
\begin{equation}\label{eqn:mass_strong}
	\begin{aligned}
		\frac{\partial \bar{\rho}_s}{\partial t} + \divr (\bar{\rho}_s \boldsymbol{v}_s) &= 0,\\
		\frac{\partial \bar{\rho}_f}{\partial t} + \divr (\bar{\rho}_f \boldsymbol{v}_f) &= 0.
	\end{aligned}
\end{equation}

\subsection{Momentum Conservation}
Momentum conservation within an arbitrary volume of real mixture $\Omega = \Omega_s \cup \Omega_f$ with exterior boundary $\partial \Omega = \partial \Omega_s \cup \partial \Omega_f$ has the following form:
\begin{equation}\label{eqn:momentum_conservation_specific}
	\begin{aligned}
		\frac{\partial}{\partial t} \int_{\Omega_s}{\rho_s \boldsymbol{v} dv} + \int_{\partial \Omega_s}{\rho_s \boldsymbol{v} (\boldsymbol{v} \cdot \hat{\boldsymbol{n}}) da} &= \int_{\Omega_s}{\rho_s \boldsymbol{g} dv} - \int_{\partial\Omega^*}\boldsymbol{t}(\hat{\boldsymbol{n}}) da + \int_{\partial\Omega_s}\boldsymbol{t}(\hat{\boldsymbol{n}}) da,\\
		\frac{\partial}{\partial t} \int_{\Omega_f}{\rho_f \boldsymbol{v} dv} + \int_{\partial \Omega_f}{\rho_f \boldsymbol{v} (\boldsymbol{v} \cdot \hat{\boldsymbol{n}}) da} &= \int_{\Omega_s}{\rho_f \boldsymbol{g} dv} + \int_{\partial\Omega^*}\boldsymbol{t}(\hat{\boldsymbol{n}})da + \int_{\partial\Omega_f}\boldsymbol{t}(\hat{\boldsymbol{n}})da,
	\end{aligned}
\end{equation}
with $\boldsymbol{t}(\hat{\boldsymbol{n}})$ the surface traction vector, a function of the outward pointing surface normal vector $\hat{\boldsymbol{n}}$; $\boldsymbol{g}$ the gravitational acceleration vector; and $\partial\Omega^*$ the interior surface separating the solid and fluid domains of the mixture. (Note that $\partial \Omega^*$ moves with the particle boundaries, therefore there is no momentum flux across it.) A decomposition of the \textit{true} velocity ($\boldsymbol{v}$) in the solid and fluid phases of the mixture into a homogenized component ($\boldsymbol{v}_s$ and $\boldsymbol{v}_f$) and fluctuational component ($\boldsymbol{\delta v}_s$ and $\boldsymbol{\delta v}_f$) is common in poromechanics texts (see \cite{batchelor1970, coussy2004}) and necessary to account for the effects of tortuosity (see \cite{wilmanski2005, kosinski2002}). Here $\boldsymbol{\delta v}_s \equiv \boldsymbol{v} - \boldsymbol{v}_s$ in the solid domain ($\Omega_s$) and $\boldsymbol{\delta v}_f \equiv \boldsymbol{v} - \boldsymbol{v}_f$ in the fluid domain ($\Omega_f$). Adding this decomposition and the definitions in \eqref{eqn:homogenization} to \eqref{eqn:momentum_conservation_specific} yields,
\begin{equation}\label{eqn:momentum_conservation}
	\begin{aligned}
		\frac{\partial}{\partial t} \int_\Omega{\bar{\rho}_s \boldsymbol{v}_s dv} + \int_{\partial \Omega}{\bar{\rho}_s \boldsymbol{v}_s (\boldsymbol{v}_s \cdot \hat{\boldsymbol{n}}) da} &= \int_\Omega{\bar{\rho}_s \boldsymbol{g} dv} - \int_{\partial\Omega^*}\boldsymbol{t}(\hat{\boldsymbol{n}})da + \int_{\partial\Omega_s}\boldsymbol{t}(\hat{\boldsymbol{n}}) - \rho_s \boldsymbol{\delta v}_s (\boldsymbol{\delta v}_s \cdot \hat{\boldsymbol{n}}) da,\\
		\frac{\partial}{\partial t} \int_\Omega{\bar{\rho}_f \boldsymbol{v}_f dv} + \int_{\partial \Omega}{\bar{\rho}_f \boldsymbol{v}_f (\boldsymbol{v}_f \cdot \hat{\boldsymbol{n}}) da} &= \int_\Omega{\bar{\rho}_f \boldsymbol{g} dv} + \int_{\partial\Omega^*}\boldsymbol{t}(\hat{\boldsymbol{n}})da + \int_{\partial\Omega_f}\boldsymbol{t}(\hat{\boldsymbol{n}})  - \rho_f \boldsymbol{\delta v}_f (\boldsymbol{\delta v}_f \cdot \hat{\boldsymbol{n}}) da.
	\end{aligned}
\end{equation}
We can evaluate the exterior surface integrals in \eqref{eqn:momentum_conservation} by defining an effective Cauchy stress for each phase of the continuum mixture ($\boldsymbol{\sigma}_s$ and $\boldsymbol{\sigma}_f$) according to the integral form of Cauchy's Theorem (and including of the fluctuational stresses described in \cite{coussy2004, jackson2000}) for a sufficiently large, arbitrary domain ($\Omega$) with boundary $\partial \Omega = \partial \Omega_s \cup \partial \Omega_f$,
\begin{equation}
	\begin{aligned}
		\int_{\partial \Omega}{\boldsymbol{\sigma}_s\hat{\boldsymbol{n}} da} &\equiv \int_{\partial \Omega_s}{\boldsymbol{t}(\hat{\boldsymbol{n}}) - \rho_s \boldsymbol{\delta v}_s (\boldsymbol{\delta v}_s \cdot \hat{\boldsymbol{n}}) da},
		\\ \int_{\partial \Omega}{\boldsymbol{\sigma}_f\hat{\boldsymbol{n}} da} &\equiv \int_{\partial \Omega_f}{\boldsymbol{t}(\hat{\boldsymbol{n}}) - \rho_f \boldsymbol{\delta v}_f (\boldsymbol{\delta v}_f \cdot \hat{\boldsymbol{n}}) da}.
	\end{aligned}
	\label{eqn:cauchy_theorem}
\end{equation}
Similarly, we can evaluate the interior surface integrals in \eqref{eqn:momentum_conservation} by homogenizing the tractions along the internal boundary ($\partial \Omega^*$) into a set of body forces describing the buoyant ($\boldsymbol{f_b}$) and drag ($\boldsymbol{f_d}$) interactions occurring between the phases,
\begin{equation} \label{eqn:interior_traction_integral}
	\int_\Omega (\boldsymbol{f_b} + \boldsymbol{f_d}) dv \equiv \int_{\partial \Omega^*} \boldsymbol{t}(\hat{\boldsymbol{n}}) da.
\end{equation}
Together \eqref{eqn:momentum_conservation}, \eqref{eqn:cauchy_theorem}, and \eqref{eqn:interior_traction_integral} must hold for \textit{any} volume $\Omega$ such that momentum conservation can be expressed locally as,
\begin{equation}
	\begin{aligned}
		\frac{\partial \bar{\rho}_s \boldsymbol{v}_s}{\partial t} + \divr\big(\bar{\rho}_s\boldsymbol{v}_s \otimes \boldsymbol{v}_s \big) &= \bar{\rho}_s \boldsymbol{g} - \boldsymbol{f_b} - \boldsymbol{f_d} + \divr \boldsymbol{\sigma}_s,
		\\ \frac{\partial \bar{\rho}_f \boldsymbol{v}_f}{\partial t} + \divr \big(\bar{\rho}_f\boldsymbol{v}_f \otimes \boldsymbol{v}_f \big) &= \bar{\rho}_f \boldsymbol{g} + \boldsymbol{f_b} + \boldsymbol{f_d} + \divr \boldsymbol{\sigma}_f.
		\label{eqn:momentum_strong}
	\end{aligned}
\end{equation}

\subsection{Energy Conservation}
Conservation of energy over an arbitrary volume of real mixture $\Omega = \Omega_s \cup \Omega_f$ with boundary $\partial \Omega = \partial \Omega_s \cup \partial \Omega_f$ has the following form:
\begin{equation}
	\begin{aligned}
		\frac{\partial}{\partial t}\int_{\Omega_s} \rho_s \big(\varepsilon + \tfrac{1}{2} \boldsymbol{v} \cdot \boldsymbol{v} \big) dv =& - \int_{\partial \Omega_s}{\rho_s \big(\varepsilon + \tfrac{1}{2} \boldsymbol{v} \cdot \boldsymbol{v} \big)(\boldsymbol{v}\cdot\hat{\boldsymbol{n}})da}\\
		& -\int_{\partial\Omega_s}(\boldsymbol{q} \cdot \hat{\boldsymbol{n}}) da + \int_{\partial \Omega^*} (\boldsymbol{q} \cdot \hat{\boldsymbol{n}}) da\\
		& + \int_{\Omega_s} q\ dv \\
		& + \int_{\partial \Omega_s}(\boldsymbol{t}(\hat{\boldsymbol{n}}) \cdot \boldsymbol{v}) da -\int_{\partial \Omega^*}(\boldsymbol{t}(\hat{\boldsymbol{n}}) \cdot \boldsymbol{v}) da\\
		&+ \int_{\Omega_s}(\rho_s \boldsymbol{g} \cdot \boldsymbol{v}) dv,
	\end{aligned}
	\label{eqn:first_law_solid_specific}
\end{equation}
and,
\begin{equation}
	\begin{aligned}
		\frac{\partial}{\partial t}\int_{\Omega_f} \rho_f \big(\varepsilon + \tfrac{1}{2} \boldsymbol{v} \cdot \boldsymbol{v} \big) dv =& - \int_{\partial \Omega_f}{\rho_f \big(\varepsilon + \tfrac{1}{2} \boldsymbol{v} \cdot \boldsymbol{v} \big)(\boldsymbol{v}\cdot\hat{\boldsymbol{n}})da}\\
		& -\int_{\partial\Omega_f}(\boldsymbol{q} \cdot \hat{\boldsymbol{n}}) da - \int_{\partial \Omega^*} (\boldsymbol{q} \cdot \hat{\boldsymbol{n}}) da\\
		& + \int_{\Omega_f} q\ dv \\
		& + \int_{\partial \Omega_f}(\boldsymbol{t}(\hat{\boldsymbol{n}}) \cdot \boldsymbol{v}) da + \int_{\partial \Omega^*}(\boldsymbol{t}(\hat{\boldsymbol{n}}) \cdot \boldsymbol{v}) da\\
		&+ \int_{\Omega_f}(\rho_f \boldsymbol{g} \cdot \boldsymbol{v}) dv,
	\end{aligned}
	\label{eqn:first_law_fluid_specific}
\end{equation}
where $\boldsymbol{q}$ is the \textit{true} heat flow vector and $q$ is the \textit{true} rate of internal heat generation. (Note that although there is no energy flux across $\partial \Omega^*$, heat flow through this boundary is still possible.) To evaluate the exterior boundary integrals in \eqref{eqn:first_law_solid_specific} and \eqref{eqn:first_law_fluid_specific}, we first define a set of effective heat flow vectors ($\boldsymbol{q}_s$ and $\boldsymbol{q}_f$) \textit{that include the extra mixing of internal energies due to local fluctuations in velocity} as,
\begin{equation}\label{eqn:heat_flow}
	\begin{aligned}
		\int_{\partial \Omega}{(\boldsymbol{q}_s \cdot \hat{\boldsymbol{n}}) da} &\equiv \int_{\partial \Omega_s}{(\boldsymbol{q} \cdot \hat{\boldsymbol{n}}) + \rho_s \varepsilon (\boldsymbol{\delta v}_s \cdot \hat{\boldsymbol{n}}) da}, \\
		\int_{\partial \Omega}{(\boldsymbol{q}_f \cdot \hat{\boldsymbol{n}}) da} &\equiv \int_{\partial \Omega_f}{(\boldsymbol{q} \cdot \hat{\boldsymbol{n}}) + \rho_f \varepsilon (\boldsymbol{\delta v}_f \cdot \hat{\boldsymbol{n}}) da},
	\end{aligned}
\end{equation}
and a set of effective phase-wise rates of internal heat generation ($q_s$ and $q_f$) as,
\begin{equation}\label{eqn:heat_generation}
	\begin{aligned}
		\int_{\Omega}{q_s\ dv} &\equiv \int_{\Omega_s}{q\ dv}, \\
		\int_{\Omega}{q_f\ dv} &\equiv \int_{\Omega_f}{q\ dv}.
	\end{aligned}
\end{equation}
Similarly, we can evaluate the heat flow integrals over the internal boundary in \eqref{eqn:first_law_solid_specific} and \eqref{eqn:first_law_fluid_specific} by defining an effective inter-phase heat flow rate ($q_i$) that represents the flow from the solid domain to the fluid domain as,
\begin{equation}\label{eqn:interphase_heat_flow}
	\int_{\Omega}{q_i\ dv} \equiv -\int_{\partial \Omega^*}{(\boldsymbol{q} \cdot \hat{\boldsymbol{n}})da}.
\end{equation}
{The remaining integrals over the internal boundary describe the inter-phase work associated with tractions along particle surfaces. To homogenize these integrals, we return to the definition of the buoyant and drag forces ($\boldsymbol{f_b}$ and $\boldsymbol{f_d}$) from \eqref{eqn:interior_traction_integral}. There, these surface tractions were homogenized into effective body forces; here, we homogenize the work associated with these tractions into components associated with these body forces ($\boldsymbol{f_b} \cdot \boldsymbol{v}_s$ and $\boldsymbol{f_d}\cdot\boldsymbol{v}_s$) and a component associated with expansion or compaction of the solid phase:
\begin{equation}\label{eqn:interphase_work_homogenized}
	\int_{\partial \Omega^*}{\big(\boldsymbol{t}(\hat{\boldsymbol{n}})\cdot\boldsymbol{v}\big)\ da} \approx \int_{\partial \Omega^*}{\big(\boldsymbol{t}(\hat{\boldsymbol{n}})\cdot\boldsymbol{\delta v}_s\big)\ da} + \int_{\Omega}{(\boldsymbol{f_b} + \boldsymbol{f_d})\cdot \boldsymbol{v}_s + \sigma_n \divr(\boldsymbol{v}_s)\ da}.
\end{equation}
Here $\sigma_n$ represents the effective normal force on the particle surfaces. (Note that this approximation assumes that the accumulated deformation of any individual grain will be substantially smaller than the associated deformation of the fluid in the pores between grains; hence, the inter-phase work is homogenized as forces conjugate to the solid phase velocity, $\boldsymbol{v}_s$.)}

Substituting these definitions along with those from \eqref{eqn:homogenization} and  \eqref{eqn:cauchy_theorem} into \eqref{eqn:first_law_solid_specific} and \eqref{eqn:first_law_fluid_specific} allows us to express the evolution of the effective phase-wise energies as follows,
\begin{equation}
	\begin{aligned}
		\frac{\partial}{\partial t}\int_{\Omega} \bar{\rho}_s \big(\varepsilon_s + \tfrac{1}{2} \boldsymbol{v}_s \cdot \boldsymbol{v}_s \big) dv =& - \int_{\partial \Omega}{\bar{\rho}_s \big(\varepsilon_s + \tfrac{1}{2} \boldsymbol{v}_s \cdot \boldsymbol{v}_s \big)(\boldsymbol{v}_s\cdot\hat{\boldsymbol{n}})da}\\
		& +\int_{\partial\Omega} (\boldsymbol{\sigma}_s \boldsymbol{v}_s - \boldsymbol{q}_s) \cdot \hat{\boldsymbol{n}}\ da\\
		& + \int_{\Omega} q_s - q_i + (\bar{\rho}_s \boldsymbol{g}) \cdot \boldsymbol{v}_s - (\boldsymbol{f_b} + \boldsymbol{f_d}) \cdot \boldsymbol{v}_s - {\sigma_n \divr(\boldsymbol{v}_s)}\ dv\\
		&+ \int_{\partial \Omega_s}{(\boldsymbol{t}(\hat{\boldsymbol{n}}) \cdot \boldsymbol{\delta v}_s) + \tfrac{1}{2} \rho_s (\boldsymbol{\delta v}_s\cdot\boldsymbol{\delta v}_s) (\boldsymbol{\delta v}_s \cdot \hat{\boldsymbol{n}})\ da}\\
		&- \int_{\partial \Omega^*}{(\boldsymbol{t}(\hat{\boldsymbol{n}}) \cdot \boldsymbol{\delta v}_s)da},
	\end{aligned}
	\label{eqn:first_law_solid}
\end{equation}
and,
\begin{equation}
	\begin{aligned}
		\frac{\partial}{\partial t}\int_{\Omega} \bar{\rho}_f \big(\varepsilon_f + \tfrac{1}{2} \boldsymbol{v}_f \cdot \boldsymbol{v}_f \big) dv =& - \int_{\partial \Omega}{\bar{\rho}_f \big(\varepsilon_f + \tfrac{1}{2} \boldsymbol{v}_f \cdot \boldsymbol{v}_f \big)(\boldsymbol{v}_f\cdot\hat{\boldsymbol{n}})da}\\
		& +\int_{\partial\Omega} (\boldsymbol{\sigma}_f \boldsymbol{v}_f - \boldsymbol{q}_f) \cdot \hat{\boldsymbol{n}}\ da\\
		& + \int_{\Omega} q_f + q_i + (\bar{\rho}_f \boldsymbol{g}) \cdot \boldsymbol{v}_f + (\boldsymbol{f_b} + \boldsymbol{f_d}) \cdot \boldsymbol{v}_s + {\sigma_n \divr(\boldsymbol{v}_s)}\ dv\\
		&+ \int_{\partial \Omega_f}{(\boldsymbol{t}(\hat{\boldsymbol{n}}) \cdot \boldsymbol{\delta v}_f) + \tfrac{1}{2} \rho_f (\boldsymbol{\delta v}_f\cdot\boldsymbol{\delta v}_f) (\boldsymbol{\delta v}_f \cdot \hat{\boldsymbol{n}})\ da}\\
		& + \int_{\partial \Omega^*}{(\boldsymbol{t}(\hat{\boldsymbol{n}}) \cdot \boldsymbol{\delta v}_s)da}.
	\end{aligned}
	\label{eqn:first_law_fluid}
\end{equation}
The remaining integrals over the boundaries $\partial \Omega_s$, $\partial \Omega_f$, and $\partial \Omega^*$ in \eqref{eqn:first_law_solid} and \eqref{eqn:first_law_fluid} capture the extra dissipation and diffusion of energy that occurs to do the fluctuations in the \textit{true} velocity field. To account for this in our equations of motion, we introduce measures of this extra work in the solid phase ($\delta w_s$), in the fluid phase ($\delta w_f$), and between the phases ($\delta w_i$) as follows,
\begin{equation}\label{eqn:fluctuational_work}
	\begin{aligned}
		\int_{\Omega}{\delta w_s} &\equiv \int_{\partial \Omega_s}{(\boldsymbol{t}(\hat{\boldsymbol{n}}) \cdot \boldsymbol{\delta v}_s) + \tfrac{1}{2} \rho_s (\boldsymbol{\delta v}_s\cdot\boldsymbol{\delta v}_s) (\boldsymbol{\delta v}_s \cdot \hat{\boldsymbol{n}})\ da},\\
		\int_{\Omega}{\delta w_f} &\equiv \int_{\partial \Omega_f}{(\boldsymbol{t}(\hat{\boldsymbol{n}}) \cdot \boldsymbol{\delta v}_f) + \tfrac{1}{2} \rho_f (\boldsymbol{\delta v}_f\cdot\boldsymbol{\delta v}_f) (\boldsymbol{\delta v}_f \cdot \hat{\boldsymbol{n}})\ da},\\
		\int_{\Omega}{\delta w_i} &\equiv \int_{\partial \Omega^*}{(\boldsymbol{t}(\hat{\boldsymbol{n}}) \cdot \boldsymbol{\delta v}_s) da}.\\
	\end{aligned}
\end{equation}
For \textit{large} domains $\Omega$ over which $\boldsymbol{\delta v}_s$ and $\boldsymbol{\delta v}_f$ have no inherent directionality, these integrals will vanish; however, if the local fluctuations in velocity have directionality that is correlated with their magnitudes or with the local traction vector, $\boldsymbol{t}(\hat{\boldsymbol{n}})$, these extra contributions to the evolution of internal energy must be taken into account. Together \eqref{eqn:first_law_solid}, \eqref{eqn:first_law_fluid}, and \eqref{eqn:fluctuational_work} must hold for \textit{any} volume $\Omega$ such that energy conservation can be expressed locally as,
\begin{equation}\label{eqn:energy_strong}
	\begin{aligned}
		\frac{\partial \bar{\rho}_s E_s}{\partial t} &= -\divr(\bar{\rho}_s E_s \boldsymbol{v}_s) + \divr(\boldsymbol{\sigma}_s \boldsymbol{v}_s - \boldsymbol{q}_s) + q_s - q_i + (\bar{\rho}_s \boldsymbol{g}) \cdot \boldsymbol{v}_s - (\boldsymbol{f_b}+\boldsymbol{f_d}) \cdot \boldsymbol{v}_s - {\sigma_n \divr(\boldsymbol{v}_s)} + \delta w_s - \delta w_i,\\
		\frac{\partial \bar{\rho}_f E_f}{\partial t} &= -\divr(\bar{\rho}_f E_f \boldsymbol{v}_f) + \divr(\boldsymbol{\sigma}_f \boldsymbol{v}_f - \boldsymbol{q}_f) + q_f + q_i + (\bar{\rho}_f \boldsymbol{g}) \cdot \boldsymbol{v}_f + (\boldsymbol{f_b}+\boldsymbol{f_d}) \cdot \boldsymbol{v}_s + {\sigma_n \divr(\boldsymbol{v}_s)} + \delta w_f + \delta w_i,
	\end{aligned}
\end{equation}
with $E_s \equiv \varepsilon_s + \tfrac{1}{2} \boldsymbol{v}_s \cdot \boldsymbol{v}_s$ and $E_f \equiv \varepsilon_f + \tfrac{1}{2} \boldsymbol{v}_f \cdot \boldsymbol{v}_f$ the specific total energies for each phase.

\subsection{Common Expressions for the Cauchy Stresses and Buoyant Force}
The effective Cauchy stresses ($\boldsymbol{\sigma}_s$ and $\boldsymbol{\sigma}_f$) defined in \eqref{eqn:cauchy_theorem} are evaluated over many \textit{large} domains $\Omega$ (or, as in \cite{batchelor1970}, averages over an ensemble of realizations), we therefore expect these stresses contain contributions from both phases of the real mixture and can be decomposed as follows. The solid phase stress, $\boldsymbol{\sigma}_s$, takes the following form:
\begin{equation}
	\boldsymbol{\sigma}_s = \boldsymbol{\tilde{\sigma}} - (1-n) p_f \boldsymbol{I},
	\label{eqn:solid_stress}
\end{equation}
with $\boldsymbol{\tilde{\sigma}}$ the portion of the solid phase stress resulting from \textit{true} granular contacts and microscopic viscous interactions between grains due to immersion in a fluid medium (e.g.\ lubrication forces), and $p_f$ the \textit{true} fluid pore pressure. The fluid phase stress, $\boldsymbol{\sigma}_f$, takes the following form:
\begin{equation}
	\boldsymbol{\sigma}_f = \boldsymbol{\tau_f} - n p_f \boldsymbol{I},
	\label{eqn:fluid_stress}
\end{equation}
with $\boldsymbol{\tau_f}$ a deviatoric shear stress tensor and $n p_f \boldsymbol{I}$ the spherical stress given by the porosity of the mixture and \textit{true} fluid pore pressure.

Additionally, we decompose the interior traction integral in  \eqref{eqn:interior_traction_integral} into normal components (i.e.\ buoyant forces; $\boldsymbol{f_b}$) and shear components (i.e.\ drag forces; $\boldsymbol{f_d}$). Following the works of \cite{drumheller2000, jackson2000}, we let the buoyant force take the following form:
\begin{equation}
	\boldsymbol{f_b} = p_f \nabla n,
	\label{eqn:buoyant_force}
\end{equation}
which accounts for the distribution of pressures on particle surfaces that arises due to the presence of a pore fluid. 
{
An analogous assessment of the inter-phase work integral in \eqref{eqn:interphase_work_homogenized} yields the following expression for $\sigma_n$:
\begin{equation}\label{eqn:sigma_n}
	\sigma_n = -\phi p_f,
\end{equation}
which is associated with the volumetric work of the fluid pressing on a dilating granular material. (Note that this formulation of $\sigma_n$ neglects contributions from the drag force $\boldsymbol{f}_d$, which may arise from lifting-drag models; see \cite{chauchat2018}.)}

\subsection{Closure of Equations and Simplifying Assumptions}
As stated in \cite{jackson2000}, closure of this system of conservation equations requires definitions of following quantities in terms of the homogenized fields in \eqref{eqn:homogenization}:
\begin{equation}\label{eqn:constitutive_fields}
	\boldsymbol{\tilde{\sigma}}, \quad p_f, \quad \boldsymbol{\tau_f}, \quad \boldsymbol{f_d}, \quad \boldsymbol{q}_s, \quad \boldsymbol{q}_f, \quad q_s, \quad q_f, \quad q_i, \quad \delta w_s, \quad \delta w_f, \quad \delta w_i.
\end{equation}
In the main text, we have reduced this set by making the following simplifying assumptions.

Since we are primarily interested in fluid--sediment flows where the absolute temperature of the grains does not considerably affect the solution, we model the granular phase as an isothermal solid: the granular temperature, $\vartheta_s$, is constant; the solid phase heat generation and heat flux, $q_s$ and $\boldsymbol{q}_s$, are implicitly defined;  and the effective granular stress $\boldsymbol{\tilde{\sigma}}$ evolves according to an admissible isothermal equation of state (see \cite{baumgarten2019a}). If we further assume that there is no inter-phase heat flow,
\begin{equation}
	q_i \to 0,
\end{equation}
then there is no need to explicitly track $\varepsilon_s$ in the mixture.

In addition to modeling the granular phase as isothermal, we will also limit ourselves to considering chemically stable mixtures ($q_f \to 0$) and neglect the extra work terms ($\delta w_s$, $\delta w_f$, and $\delta w_i$), which derive from higher order correlations of velocity fluctuations $\boldsymbol{\delta v}_s$ and $\boldsymbol{\delta v}_f$. Although there are constitutive models for the average magnitudes of $\boldsymbol{\delta v}_s$ and $\boldsymbol{\delta v}_f$ (see \cite{kosinski2002,zhang2017microscopic,kim2020}), there are few descriptions of how the direction and magnitude of these local velocity fluctuations are correlated. 

{
With these assumptions, \eqref{eqn:solid_stress}, \eqref{eqn:fluid_stress}, \eqref{eqn:buoyant_force}, and \eqref{eqn:sigma_n} can be substituted into \eqref{eqn:mass_strong}, \eqref{eqn:momentum_strong}, and \eqref{eqn:energy_strong} to yield the simplified system of equations presented in \eqref{eqn:mixture_equations}.
}

\setcounter{figure}{0}
\section{Useful Numerical Approximations}\label{sec:numerical_approximation}
In this section, we present numerical approximations that we have found useful when implementing the numerical algorithm described in the main text. These approximations of the buoyant force vectors ($\boldsymbol{f}_i^{\text{buoy}}$ and $\boldsymbol{F}_\alpha^{\text{buoy}}$), the drag force vectors ($\boldsymbol{f}_i^{\text{drag}}$ and $\boldsymbol{F}_\alpha^{\text{drag}}$), and the fluid phase gradients ($\langle \nabla \rho_f \rangle_\alpha$, $\langle \nabla \rho_f \boldsymbol{v}_f \rangle_\alpha$, and $\langle \nabla \rho_f E_f \rangle_\alpha$) improve the stability of the numerical algorithm without adding significant approximation errors.

\subsection{Buoyant Force Vectors}
The buoyant force vectors, $\boldsymbol{f}_i^{\text{buoy}}$ and $\boldsymbol{F}_\alpha^{\text{buoy}}$, are presented in \eqref{eqn:momentum_evolution} and \eqref{eqn:fluid_state_vector_equation}, respectively. Although the expression for $\boldsymbol{F}_\alpha^\text{buoy}$ can be evaluated explicitly using numerical quadrature (i.e.\ sampling the solid and fluid phase fields within each grid cell), the exact expression for $\boldsymbol{f}_i^{\text{buoy}}$ includes a gradient that can be tricky to evaluate on non-uniform grids:
\begin{equation*}
\boldsymbol{f}_i^{\text{buoy}} = \sum_{j=1}^{N_n} (1 - n_j) \int_{\Omega}{\nabla \big( \mathcal{N}_i(\boldsymbol{x}) \mathcal{N}_j(\boldsymbol{x}) \big) p_f\ dv}.
\end{equation*}
Additionally, since this expression reflects an exact evaluation of the weak-form equation in \eqref{eqn:weak_momentum_conservation_solid}, it may produce instabilities in the method when the diagonal matrices, $[\mathcal{M}_D]$ and $[\mathcal{B}_D]$, are used in place of the consistent matrices, $[\mathcal{M}]$ and $[\mathcal{B}]$, (see \cite{mackenzie2010}).

To avoid these issues, we introduce the following approximation of $\boldsymbol{f}_i^{\text{buoy}}$:
\begin{equation}
	\label{eqn:numerical_mpm_buoyancy}
	{\boldsymbol{f}_i^{\text{buoy}}}^{*} = (1 - n_i) \int_{\Omega}{\nabla \mathcal{N}_i(\boldsymbol{x})\ p_f\ dv}.
\end{equation}
Here, the integral expression is easily evaluated with numerical quadrature, and we can ensure that the coupling force is numerically stable (i.e.\  ${\boldsymbol{f}_i^{\text{buoy}}}^* = \boldsymbol{0}$ when $n_i = 1$).
If we assume that for a particular choice of finite element basis functions $\{\mathcal{N}_i(\boldsymbol{x})\}$ there exists a unique set of coefficients $\{\psi_i\}$ that produce the node values $\{\psi(\boldsymbol{x}_i)\}$ (with $\boldsymbol{x}_i$ the position of the $i$th node) and that these functions are only non-zero over a limited region of the simulated domain $\Omega$ (e.g.\ ``tent'' functions or cubic splines; see \cite{bardenhagen2004,steffen2008analysis}), then it is possible to show that,
\begin{equation}
	\|\boldsymbol{f}_i^{\text{buoy}} - {\boldsymbol{f}_i^{\text{buoy}}}^*\| \leq c_0 h_i \|\nabla n\|_\infty \|\nabla p_f\|_\infty \int_{\Omega}{\mathcal{N}_i(\boldsymbol{x}) dv},
\end{equation}
with $c_0$ a basis specific constant, $h_i$ the characteristic length of the finite element grid (e.g.\ the average distance between grid nodes), and $\|\psi\|_\infty$ the maximum component of $\psi$ across the entire simulated domain.

\subsection{Drag Force Vectors}
The drag force vectors, $\boldsymbol{f}_i^{\text{drag}}$ and $\boldsymbol{F}_\alpha^{\text{drag}}$, are presented in \eqref{eqn:momentum_evolution} and \eqref{eqn:fluid_state_vector_equation}, respectively. Although both expressions can be evaluated using numerical quadrature, we find that this approach can develop sudden instabilities. One source of these instabilities is the consistency of the exact expression for $\boldsymbol{f}_i^{\text{drag}}$ with the weak-form equation in \eqref{eqn:weak_momentum_conservation_solid}: in numerical algorithms that use the diagonal matrices, $[\mathcal{M}_D]$ and $[\mathcal{B}_D]$, exact calculation of these force vectors can produce spurious accelerations.

To help avoid these instabilities we approximate these drag force vectors by defining a resistance field, $K$, as follows:
\begin{equation}
	\label{eqn:inverse_permeability}
	K \equiv \frac{\|\boldsymbol{f_d}\|}{\|\boldsymbol{v}_s - \boldsymbol{v}_f\|}.
\end{equation}
(Note that this definition of resistance is only valid when the drag model for $\boldsymbol{f_d}$ is `co-directional' with the velocity difference, $\boldsymbol{v}_s - \boldsymbol{v}_f$.) We then project this field onto the finite element basis and call this projection $K^*$:
\begin{equation}\label{eqn:drag_approximation}
	K^* = \sum_{i=1}^{N_n} K^*_i(t) \mathcal{N}_i(\boldsymbol{x}).
\end{equation}
We let the coefficient $K^*_i$ be approximated as,
\begin{equation}
	\label{eqn:numerical_inverse_permeability}
	K^*_i \approx \frac{18 n_i (1 - n_i) \eta_0}{d^2} \hat{F}(1-n_i, \Reyn_i^*),
\end{equation} 
with $\Reyn_i^*$ an approximation of the Reynolds number ($\Reyn_i^* \equiv (\bar{\rho}_{fi}^* d\ \|\boldsymbol{v}_{si} - \boldsymbol{v}_{fi}^*\|)/\eta_0$), $\bar{\rho}_{fi}^*$ an approximation of the effective fluid density ($\bar{\rho}_{fi}^* \equiv \sum_{\alpha=1}^{N_v} (V_\alpha \mathcal{A}_{i\alpha} \langle \bar{\rho}_f \rangle_\alpha)/V_i$), $\boldsymbol{v}_{fi}^*$ an approximation of the fluid velocity ($\boldsymbol{v}_{fi}^* \equiv \sum_{\alpha=1}^{N_v}(V_\alpha \mathcal{A}_{i\alpha} \langle \bar{\rho}_f \boldsymbol{v}_f \rangle_\alpha / \langle \bar{\rho}_f \rangle_\alpha) / V_i$), and
\begin{equation*}
	V_i = \int_\Omega \mathcal{N}_i(\boldsymbol{x}) dv.
\end{equation*}
The drag vectors can then be approximated as,
\begin{equation}
	\label{eqn:numerical_mpm_drag}
	{\boldsymbol{f}_i^{\text{drag}}}^* = -V_i K_i^* ( \boldsymbol{v}_{si} - \boldsymbol{v}_{fi}^*),
\end{equation}
and,
\begin{equation}
	\label{eqn:numerical_fvm_drag}
	{\boldsymbol{F}_\alpha^{\text{drag}}}^* = \sum_{i=1}^{N_n} \mathcal{A}_{i\alpha} K^*_i \big(\boldsymbol{v}_{si} - \langle \bar{\rho}_f \boldsymbol{v}_{f} \rangle_\alpha / \langle \bar{\rho}_f \rangle_\alpha \big) \cdot
	\begin{pmatrix}
		0\\
		1\\
		\boldsymbol{v}_{si}
	\end{pmatrix}.
\end{equation}

If we assume that for a particular choice of finite element basis functions $\{\mathcal{N}_i(\boldsymbol{x})\}$ there exists a unique set of coefficients $\{\psi_i\}$ that produce the node values $\{\psi(\boldsymbol{x}_i)\}$ and that these functions are only non-zero over a limited region of the simulated domain $\Omega$, then it is possible to show that,
\begin{equation}
	\begin{aligned}
	\|\boldsymbol{f}_i^{\text{drag}} - {\boldsymbol{f}_i^{\text{drag}}}^*\| \leq&\ c_0 h_i \|\nabla \boldsymbol{f_d}\|_\infty V_i\\
	&+ c_1 h_\alpha \|\nabla \boldsymbol{v}_f\|_\infty  \|\partial \boldsymbol{f_d}/ \partial \boldsymbol{v}_f \|_\infty V_i\\
	&+ c_1 h_\alpha \|\nabla \bar{\rho}_f\|_\infty  \|\partial \boldsymbol{f_d}/ \partial \bar{\rho}_f \|_\infty V_i,
	\end{aligned}
\end{equation}
and,
\begin{equation}
	\begin{aligned}
		\|\boldsymbol{F}_\alpha^{\text{drag}} - {\boldsymbol{F}_\alpha^{\text{drag}}}^*\| \leq&\ c_0 h_i \|\nabla (K \boldsymbol{v}_s)\|_\infty \\
		&+ c_2 h_\alpha^2 \|\partial^2 \boldsymbol{v}_f/\partial \boldsymbol{x}^2\|_\infty  \|\partial \boldsymbol{f_d}/\partial \boldsymbol{v}_f \|_\infty\\
		&+ c_2 h_\alpha^2 \|\partial^2 \bar{\rho}_f/\partial \boldsymbol{x}^2\|_\infty  \|\partial \boldsymbol{f_d}/ \partial \bar{\rho}_f \|_\infty\\
		&+ c_3 h_i h_\alpha \|\nabla \boldsymbol{v}_f\|_\infty  \|\partial \boldsymbol{f_d}/\partial \boldsymbol{v}_f \|_\infty\\
		&+ c_3 h_i h_\alpha \|\nabla \bar{\rho}_f\|_\infty  \|\partial \boldsymbol{f_d}/\partial \bar{\rho}_f \|_\infty,
	\end{aligned}
\end{equation}
with $c_0$, $c_1$, $c_2$, and $c_3$, grid specific constants; $h_i$ the characteristic length of the finite element grid; and $h_\alpha$ the characteristic length of the finite volume grid. If a finite element basis is chosen that has the property $\mathcal{N}_i(\boldsymbol{x}_i) = 1$, then the leading error term above becomes $c_l h_i^l \|\partial^{(l+1)} (K \boldsymbol{v}_s)/\partial \boldsymbol{x}^{(l+1)}\|_\infty$ with $l$ the order of the basis functions.

\subsection{Fluid Phase Gradients}
The numerical reconstruction of the fluid phase fields ($\bar{\rho}_f$, $\rho_f$, $\rho_f \boldsymbol{v}_f$, and $\rho_f E_f$) shown in \eqref{eqn:fluid_phase_reconstruction} uses local approximations of the fluid phase gradients ($\langle \nabla \rho_f \rangle_\alpha$, $\langle \nabla \rho_f \boldsymbol{v}_f \rangle_\alpha$, and $\langle \nabla \rho_f E_f \rangle_\alpha$) computed from surrounding centroid data. Although these reconstructions are accurate to second order, such an approach can produce significant errors when the porosity field has unresolved fluctuations (i.e.\ fluctuations with a characteristic length that is the same magnitude as the finite volume length scale, $h_\alpha$). In this section, we propose an alternative approach for gradient calculations which will have higher accuracy in steady, adiabatic flows, but may introduce small errors in transient problems. This alternative approach is used in the simulations reported in sections \ref{sec:porous}, \ref{sec:terzaghi}, \ref{sec:2d_erosion}, and \ref{sec:3d_erosion}.

\begin{figure}[!h]
	\centering
	\includegraphics[scale=0.5]{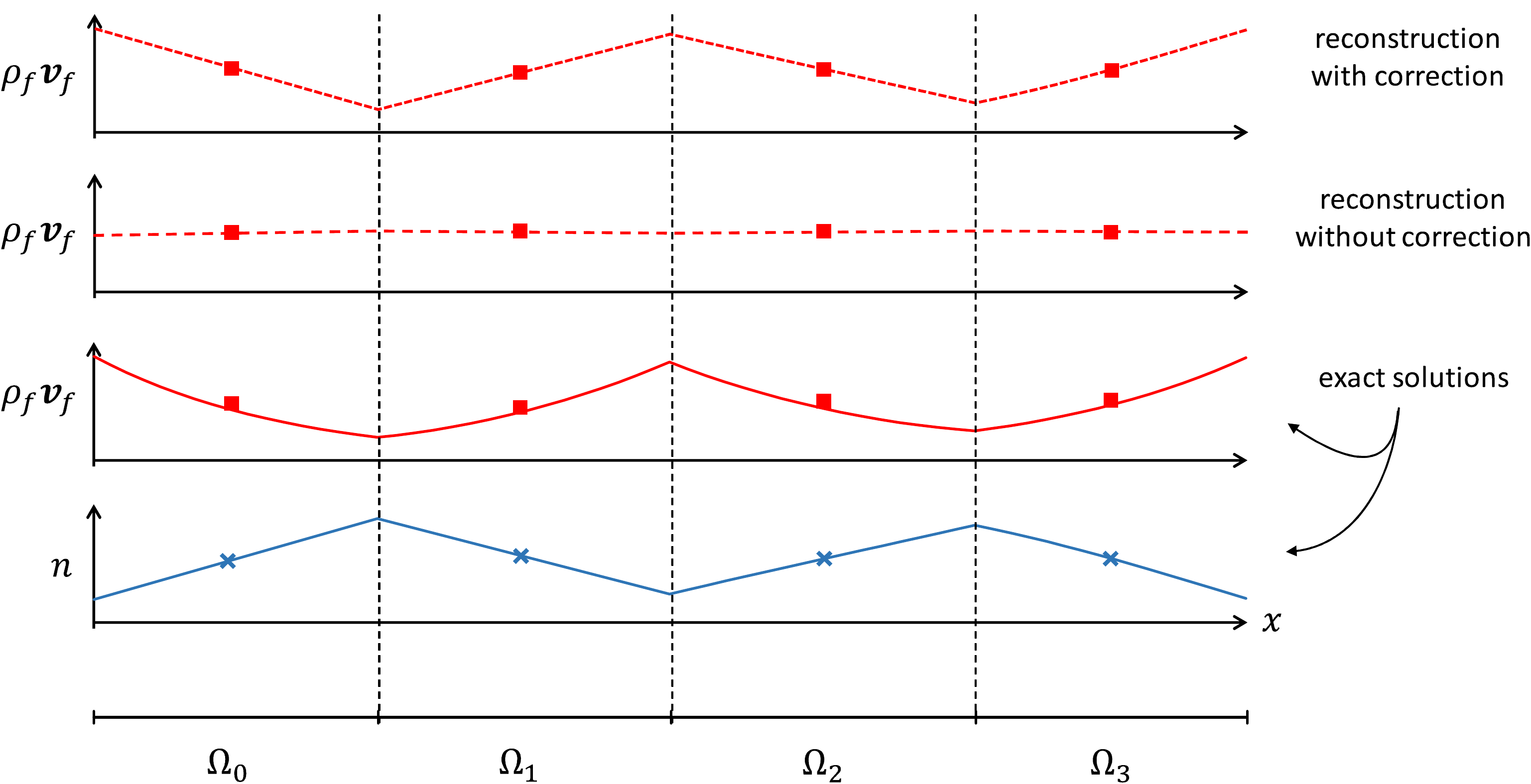}
	\caption{Pictorial representation of a one-dimensional, steady-state flow solution. The blue solid line shows a representative porosity field, $n$, with fluctuations that are not resolved on the finite volume grid, and the blue `$\times$'s denote the cell-wise average values. The red solid line shows a steady-state \textit{true} momentum field ($\rho_f \boldsymbol{v}_f$) solution that is consistent with the porosity field below, and the red `$\square$'s denote the cell-wise average values. The red long-dashed line shows the reconstructed \textit{true} momentum field found using the gradient approximations from \eqref{eqn:barth_and_jespersen_flux_limiter}. The red short-dashed line shows the reconstructed \textit{true} momentum field found using the corrected gradient approximations from \eqref{eqn:corrected_flux_limiter}.}
	\label{fig:reconstruction}
\end{figure}

Consider, the one-dimensional finite volume grid and porosity field shown in Figure \ref{fig:reconstruction}. In steady flow, mass conservation dictates that the \textit{true} momentum field must fluctuate with the porosity field, which here corresponds to linear flow area (see the solid red line in the figure); as the porosity decreases, there should be an inversely proportional change in \textit{true} fluid velocity. This relationship can be expressed for adiabatic, inviscid flows as follows:
\begin{equation}
	\label{eqn:analytical_streamwise_fluctuations}
	\begin{aligned}
		\frac{\partial \rho_f}{\partial s} &= \frac{\rho_f}{n} \bigg(\frac{M^2}{1 - M^2}\bigg) \frac{\partial n}{\partial s}, \\
		\frac{\partial  \rho_f \boldsymbol{v}_f}{\partial s} &= - \frac{\rho_f \boldsymbol{v}_f}{n} \frac{\partial n}{\partial s},\\
		\frac{\partial \rho_f E_f}{\partial s} &= \frac{\rho_f E_f + p_f - \rho_f a^2}{n} \bigg( \frac{M^2}{1 - M^2}\bigg) \frac{\partial n}{\partial s},
	\end{aligned}
\end{equation}
with $M$ the local Mach number, $a$ the local speed of sound in the fluid, and $\partial s$ denoting infinitesimal increments along a streamline.

Applying the reconstruction procedure described in the main text to this one-dimensional problem would produce a uniform momentum field, inconsistent with the grid-scale fluctuations expected in the true solution (see the long-dashed red line in Figure \ref{fig:reconstruction}). Although refining the finite volume grid would improve the accuracy of the reconstruction, it is also possible to directly incorporate the analytical expressions from \eqref{eqn:analytical_streamwise_fluctuations} into our reconstruction procedure. To do this, we calculate the cell-wise average porosity gradient, $\langle \nabla n \rangle_\alpha$, as follows,
\begin{equation}
	\langle \nabla n \rangle_\alpha = \frac{1}{V_\alpha} \sum_{i=1}^{N_i} n_i \int_{\Omega_\alpha}{ \nabla \mathcal{N}_i(\boldsymbol{x}) dv},
\end{equation}
and calculate the associated best estimate gradient $[\nabla n]_\alpha$ and flux limiter $\Phi_n$ from the surrounding centroid data. We can then quantify the amount of grid-scale porosity fluctuation with $\delta \nabla n$, the porosity gradient correction vector:
\begin{equation}
	\delta \nabla n = \langle \nabla n \rangle_\alpha - \Phi_n [\nabla n]_\alpha.
\end{equation}
Next, we estimate the relative motion of the solid and fluid phases by calculating the average solid phase velocity in the finite volume cell, $\langle \boldsymbol{v}_s \rangle_\alpha$, as,
\begin{equation}
	\langle \boldsymbol{v}_s \rangle_\alpha = \sum_{i=1}^{N_i} \mathcal{A}_{i\alpha} \boldsymbol{v}_{si},
\end{equation}
and defining $\boldsymbol{v}^*$, the relative fluid velocity vector, as follows,
\begin{equation}
	\boldsymbol{v}^* = \frac{\langle \rho_f \boldsymbol{v}_f \rangle_\alpha}{\langle \rho_f \rangle_\alpha} - \langle \boldsymbol{v}_s \rangle_\alpha.
\end{equation}
For subsonic flows, we then apply the following correction to \eqref{eqn:barth_and_jespersen_flux_limiter}:
\begin{equation}
	\label{eqn:corrected_flux_limiter}
	\begin{aligned}
		\langle \nabla \rho_f \rangle_\alpha &= \Phi_{\rho_f} [\nabla \rho_f]_\alpha + \frac{\langle \rho_f \rangle_\alpha}{\langle n \rangle_\alpha}\bigg(\frac{{M^*}^2}{1 - {M^*}^2} \bigg) \frac{\boldsymbol{v}^* (\boldsymbol{v}^* \cdot \delta \nabla n)}{\|\boldsymbol{v}^*\|^2} ,\\
		\langle \nabla \rho_f \boldsymbol{v}_f \rangle_\alpha &= \Phi_{\boldsymbol{v}_f} [\nabla \rho_f \boldsymbol{v}_f]_\alpha - \frac{\langle \rho_f \boldsymbol{v}_f \rangle_\alpha}{\langle n \rangle_\alpha} \bigg( \frac{\boldsymbol{v}^* (\boldsymbol{v}^* \cdot \delta \nabla n)}{\|\boldsymbol{v}^*\|^2} \bigg) ,\\
		\langle \nabla \rho_f E_f \rangle_\alpha &= \Phi_{E_f} [\nabla \rho_f E_f]_\alpha + \frac{(\langle \rho_f E_f \rangle_\alpha + p_f - \langle \rho_f \rangle_\alpha\ a^{2})}{\langle n \rangle_\alpha} \bigg(\frac{{M^*}^2}{1 - {M^*}^2} \bigg) \frac{\boldsymbol{v}^* (\boldsymbol{v}^* \cdot \delta \nabla n)}{\|\boldsymbol{v}^*\|^2} ,
	\end{aligned}
\end{equation}
with $p_f$ and $a$ calculated at the centroid of $\Omega_\alpha$ and $M^*$ the relative fluid mach number, $M^* = \|\boldsymbol{v}^*\|/a$.

Applying the corrected reconstruction from \eqref{eqn:corrected_flux_limiter} for the steady flow problem described in Figure \ref{fig:reconstruction} produces the field shown by the short-dashed red line in the figure: a much closer match to the analytical solution. For some fluid flows, particularly those near steady-state, this corrected reconstruction approach can improve the overall accuracy of the method without requiring additional refinement. Further, for general flows, any errors introduced by applying this correction will disappear when the porosity field is properly resolved:
\begin{equation}
	\begin{aligned}
		\big\| \nabla \rho_f \big|_{\boldsymbol{X}_\alpha} - \langle \nabla \rho_f \rangle_\alpha \big\| \leq&\ c_4 h_\alpha \|\partial^2 \rho_f / \partial \boldsymbol{x}^2\|_\infty\\
		& + A_1 \big( c_4 h_\alpha \|\partial^2 n / \partial \boldsymbol{x}^2\|_\infty + c_5 h_\alpha^2 \|\partial^3 n / \partial \boldsymbol{x}^3\|_\infty \big),\\[1em]
		\big\| \nabla \rho_f \boldsymbol{v}_f \big|_{\boldsymbol{X}_\alpha} - \langle \nabla \rho_f \boldsymbol{v}_f \rangle_\alpha \big\| \leq&\ c_4 h_\alpha \|\partial^2 (\rho_f \boldsymbol{v}_f) / \partial \boldsymbol{x}^2\|_\infty \\
		&+ A_2 \big( c_4 h_\alpha \|\partial^2 n / \partial \boldsymbol{x}^2\|_\infty + c_5 h_\alpha^2 \|\partial^3 n / \partial \boldsymbol{x}^3\|_\infty \big),\\[1em]
		\big\| \nabla \rho_f E_f \big|_{\boldsymbol{X}_\alpha} - \langle \nabla \rho_f E_f \rangle_\alpha \big\| \leq&\ c_4 h_\alpha \|\partial^2 (\rho_f E_f) / \partial \boldsymbol{x}^2\|_\infty\\
		& + A_3 \big( c_4 h_\alpha \|\partial^2 n / \partial \boldsymbol{x}^2\|_\infty + c_5 h_\alpha^2 \|\partial^3 n / \partial \boldsymbol{x}^3\|_\infty \big).
	\end{aligned}
\end{equation}
Here $c_4$ and $c_5$ represent grid specific constants, $h_\alpha$ represents the characteristic finite volume length scale, and $\{A_j\}$ represent the leading coefficients in \eqref{eqn:corrected_flux_limiter}.

\setcounter{figure}{0}
\section{Notes on Numerical Consistency}\label{sec:consistency}
In this section, we briefly discuss the accuracy of the numerical method described in the main text. We begin by evaluating the consistency of solutions to the weak- and integral-forms of the governing equations in \eqref{eqn:weak_mass_conservation_solid}, \eqref{eqn:weak_momentum_conservation_solid}, and \eqref{eqn:weak_mixture_equations_fluid} with solutions to the strong forms in \eqref{eqn:mixture_equations}. This analysis provides some basic intuition about the expected convergence rates of the numerical algorithm, which we then validate using the method of manufactured solutions (MMS; see \cite{sadeghirad2011,wallstedt2008}).

\subsection{Order of Accuracy}
We begin this analysis by considering the weak expression for solid phase mass conservation in \eqref{eqn:weak_mass_conservation_solid}. Using the definition of $w$ from \eqref{eqn:solid_test_function} and introducing the exact solution fields $\bar{\rho}_s^\dagger$, $\boldsymbol{v}_s^\dagger$, $n^\dagger$, which solve \eqref{eqn:mixture_equations}, we can express the rate of growth of density error as follows,
\begin{equation}
	\int_{\Omega_p} \frac{D^s (\bar{\rho}_s^\dagger - \bar{\rho}_s)}{Dt}\ dv = - \int_{\Omega_p} (\bar{\rho}_s^\dagger - \bar{\rho}_s) \divr(\boldsymbol{v}_s^\dagger) + \bar{\rho}_s \divr(\boldsymbol{v}_s^\dagger - \boldsymbol{v}_s) + (\boldsymbol{v}_s^\dagger - \boldsymbol{v}_s)\cdot \nabla \bar{\rho}_s^\dagger\ dv,
\end{equation}
with $\Omega_p$ the domain of the $p$th material point characteristic function. If we let $e_{\bar{\rho}_s} \equiv \bar{\rho}_s^\dagger - \bar{\rho}_s$ and $e_{\boldsymbol{v}_s} \equiv \boldsymbol{v}_s^\dagger - \boldsymbol{v}_s$ it is possible to show that this growth rate is bounded as,
\begin{equation}
	\label{eqn:solid_density_error}
	\begin{aligned}
		\frac{d}{dt} \|e_{\bar{\rho}_s} \|_\infty \leq&\
	 	\|e_{\bar{\rho}_s} \|_\infty \|\nabla \boldsymbol{v}_s^\dagger\|_\infty\\
	 	& +  \|\bar{\rho}_{s}\|_\infty \big( c_6 h_i^{-1} \|e_{\boldsymbol{v}_s}\|_\infty + c_7 h_i^{l} \|\partial^{(l+1)} \boldsymbol{v}_s^\dagger/\partial \boldsymbol{x}^{(l+1)}\|_\infty \big)\\
	 	& + \rho_s \|e_{\boldsymbol{v}_s}\|_\infty \|\nabla n^\dagger\|_\infty\\
	 	& + c_8 h_p \rho_s \big( \|\nabla n^\dagger\|_\infty \|\nabla \boldsymbol{v}_s\|_\infty + \|\nabla \partial n^\dagger/\partial t\|_\infty \big),
 	\end{aligned}
\end{equation}
with $c_6$, $c_7$, and $c_8$ discretization specific constants, $h_p$ the characteristic length scale of material points, $h_i$ the characteristic length of the finite element grid, and $l$ the order of the finite element basis functions. If we assume that $e_{\boldsymbol{v}_s}$ only accounts for discretization error (i.e.\ the error between the exact solution and the projection of the exact solution onto the finite element basis), and that $e_{\bar{\rho}_s}$ is initially zero, then we find that as the finite element grid and material points are refined, $e_{\bar{\rho}_s}$ will develop errors that are $\mathcal{O}(h_i^l) + \mathcal{O}(h_p)$.

Now consider the weak expression for solid phase momentum conservation in \eqref{eqn:weak_momentum_conservation_solid}. Using the definition of $\boldsymbol{w}$ from \eqref{eqn:test_function_and_porosity} and introducing the exact solution fields $\boldsymbol{\tilde{\sigma}}^\dagger$, $\boldsymbol{f_d}^\dagger$, and $p_f^\dagger$, which solve \eqref{eqn:mixture_equations}, we can express the rate of growth of velocity error as follows: 
\begin{equation}
	\label{eqn:error_weak_form}
	\begin{aligned}
		\int_{\Omega} \bar{\rho}_s^\dagger \boldsymbol{w} \cdot \frac{D^s (\boldsymbol{v}_s^\dagger - \boldsymbol{v}_s)}{Dt}\ dv = \int_{\Omega}& -(\bar{\rho}_s^\dagger - \bar{\rho}_s) \boldsymbol{a}_s \cdot \boldsymbol{w}  - (\boldsymbol{v}_s^\dagger - \boldsymbol{v}_s) \cdot (\nabla \boldsymbol{v}_s^\dagger) \boldsymbol{w}\\
		& - (\boldsymbol{\tilde{\sigma}}^\dagger - \boldsymbol{\tilde{\sigma}}) : \nabla \boldsymbol{w} + (\bar{\rho}_s^\dagger - \bar{\rho}_s) \boldsymbol{g}\cdot \boldsymbol{w}\\
		& - (\boldsymbol{f_d}^\dagger - \boldsymbol{f_d}) \cdot \boldsymbol{w}_s - (p_f^\dagger - p_f) \divr(n^\dagger \boldsymbol{w}) - p_f \divr\big((n^\dagger - n) \boldsymbol{w}\big)\ dv. 
	\end{aligned}
\end{equation}
Following a similar procedure to that described above, it is possible to convert \eqref{eqn:error_weak_form} into an inequality that relates the growth of solid velocity error ($e_{\boldsymbol{v}_s}$) to the finite element grid scale ($h_i$); the material point length scale ($h_p$); the maximum material stretch $\lambda_{\text{max}}$; the solution gradients (e.g.\ $\nabla n^\dagger$ and $\nabla \boldsymbol{v}_s^\dagger$); and the errors present in the density, stress, drag, and pressure calculations. Under ideal initial conditions, $e_{\boldsymbol{v}_s}$ can be shown to develop errors that are $\mathcal{O}(h_i^l) + \mathcal{O}(h_p^2/h_i) + \mathcal{O}(h_\alpha^2/h_i)$, with $h_\alpha$ the characteristic length of the finite volume grid.

Finally, we consider the integral expressions for fluid phase mass, momentum, and energy conservation in \eqref{eqn:weak_mixture_equations_fluid}. Introducing the exact solution fields $\bar{\rho}_f^\dagger$, $\boldsymbol{v}_f^\dagger$, $E_f^\dagger$, $\boldsymbol{\tau_f}^\dagger$, and $\boldsymbol{q}_f^\dagger$, which solve the strong form of the governing equations in \eqref{eqn:mixture_equations}, we can express the rates of change of the local density error as,
\begin{equation}\label{eqn:fluid_density_error}
	\begin{aligned}
		\frac{\partial}{\partial t}\bigg(\int_{\Omega_\alpha} \bar{\rho}_f^\dagger - \bar{\rho}_f\ dv\bigg) =& -\int_{\partial \Omega_\alpha} (\bar{\rho}_f^\dagger \boldsymbol{v}_f^\dagger - \bar{\rho}_f \boldsymbol{v}_f) \cdot \hat{\boldsymbol{n}}\ da,
	\end{aligned}
\end{equation}
the local momentum error as,
\begin{equation}\label{eqn:fluid_momentum_error}
	\begin{aligned}
		\frac{\partial}{\partial t}\bigg(\int_{\Omega_\alpha} (\bar{\rho}_f^\dagger \boldsymbol{v}_f^\dagger - \bar{\rho}_f \boldsymbol{v}_f)\ dv\bigg) =&- 
		\int_{\partial \Omega_\alpha}{ \big( (\bar{\rho}_f^\dagger \boldsymbol{v}_f^\dagger - \bar{\rho}_f \boldsymbol{v}_f) (\boldsymbol{v}_f \cdot \hat{\boldsymbol{n}}) + \bar{\rho}_f^\dagger \boldsymbol{v}_f^\dagger ( \boldsymbol{v}_f^\dagger - \boldsymbol{v}_f )\cdot \hat{\boldsymbol{n}} \big)\ da}\\
		& +\int_{\partial \Omega_\alpha}{ (\boldsymbol{\tau_f}^\dagger - \boldsymbol{\tau_f}) \hat{\boldsymbol{n}} - (n^\dagger - n) p_f \hat{\boldsymbol{n}} - n^\dagger (p_f^\dagger - p_f)\hat{\boldsymbol{n}}\ da}\\
		& + \int_{\Omega_\alpha}{ \big((\bar{\rho}_f^\dagger - \bar{\rho}_f) \boldsymbol{g} + (\boldsymbol{f_d}^\dagger - \boldsymbol{f_d}) + (p_f^\dagger - p_f) \nabla n + p_f^\dagger (\nabla n^\dagger - \nabla n) \big)\ dv},
	\end{aligned}
\end{equation}
and the local energy error as,
\begin{equation}\label{eqn:fluid_energy_error}
	\begin{aligned}
		\frac{\partial}{\partial t} \bigg(\int_{\Omega_\alpha} (\bar{\rho}_f^\dagger E_f^\dagger - \bar{\rho}_f E_f)\ dv \bigg) =&-\int_{\partial \Omega_\alpha} {\big((\bar{\rho}_f^\dagger E_f^\dagger - \bar{\rho}_f E_f) + (n^\dagger - n)p_f + n^\dagger (p_f^\dagger - p_f)\big) (\boldsymbol{v}_f \cdot \hat{\boldsymbol{n}})\ da}\\
		& - \int_{\partial \Omega_\alpha}{(\bar{\rho}_f^\dagger E_f^\dagger + n^\dagger p_f^\dagger)(\boldsymbol{v}_f^\dagger - \boldsymbol{v}_f)\cdot \hat{\boldsymbol{n}}\ da}\\
		& + \int_{\partial \Omega_\alpha}{\big((\boldsymbol{\tau_f}^\dagger - \boldsymbol{\tau_f})\boldsymbol{v}_f + \boldsymbol{\tau_f}^\dagger(\boldsymbol{v}_f^\dagger - \boldsymbol{v}_f) - (\boldsymbol{q}_f^\dagger - \boldsymbol{q}_f)\big) \cdot \hat{\boldsymbol{n}}\ da}\\
		& + \int_{\Omega_\alpha}{ (\bar{\rho}_f^\dagger \boldsymbol{v}_f^\dagger - \bar{\rho}_f \boldsymbol{v}_f) \cdot \boldsymbol{g}\ dv}\\
		& + \int_{\Omega_\alpha}{\big((p_f^\dagger - p_f) \nabla n + p_f^\dagger (\nabla n^\dagger - \nabla n) + (\boldsymbol{f_d}^\dagger - \boldsymbol{f}_d)\big) \cdot \boldsymbol{v}_s^\dagger\ dv}\\
		& + \int_{\Omega_\alpha}{\big((p_f^\dagger - p_f) \nabla n + p_f^\dagger (\nabla n^\dagger - \nabla n) + (\boldsymbol{f_d}^\dagger - \boldsymbol{f}_d)\big) \cdot (\boldsymbol{v}^\dagger_s-\boldsymbol{v}_s)\ dv}\\
		& {+ \int_{\Omega_\alpha}{(n^\dagger - n) p_f^\dagger \divr(\boldsymbol{v}_s^\dagger)\ dv}}\\
		& { - \int_{\Omega_\alpha}{(1-n) \big((p_f^\dagger - p_f) \divr(\boldsymbol{v}_s) + p_f^\dagger \divr(\boldsymbol{v}_s^\dagger - \boldsymbol{v}_s)\big)\ dv}.}
	\end{aligned}
\end{equation}
Again following the procedure used to calculate \eqref{eqn:solid_density_error}, it is possible to convert \eqref{eqn:fluid_density_error}, \eqref{eqn:fluid_momentum_error}, and \eqref{eqn:fluid_energy_error} into an inequality that relates the growth of the fluid density error ($e_{\bar{\rho}_f} \equiv \bar{\rho}_f^\dagger - \bar{\rho}_f$), velocity error ($e_{\boldsymbol{v}_f} \equiv \boldsymbol{v}_f^\dagger - \boldsymbol{v}_f$), and energy error ($e_{E_f} \equiv E_f^\dagger - E_f$) to the finite element grid scale ($h_i$); the solution gradients (e.g.\ $\nabla n^\dagger$); and the errors present in the density, stress, drag, pressure, and heat flux calculations. Under ideal conditions, the fluid fields can be shown to develop errors that are  $\mathcal{O}(h_i^l) + \mathcal{O}(h_\alpha^2)$.

All together, this simplified analysis provides insight into the potential sources of error and the convergence rates we might expect when the method is implemented. It is clear that the initial discretization errors in $\bar{\rho}_s$, $\boldsymbol{v}_s$, $\bar{\rho}_f$, $\boldsymbol{v}_f$, and $E_f$ --- as well as the propagation of these errors to the derived quantities $n$, $\boldsymbol{\tilde{\sigma}}$, $p_f$, $\boldsymbol{\tau_f}$, $\boldsymbol{f_d}$, and $\boldsymbol{q}_f$ --- play a crucial role determining the overall accuracy of the method. Additionally, error in \textit{any} of these terms will contribute to the growth of the error of \textit{all} of these terms. Therefore, assuming that an ideal initial discretization is implemented and that the weak- and integral-forms of the governing equations in \eqref{eqn:weak_mass_conservation_solid}, \eqref{eqn:weak_momentum_conservation_solid}, and \eqref{eqn:weak_mixture_equations_fluid} are evaluated exactly, the analysis above indicates that the leading error terms in our numerical solutions should be of the following order: $\mathcal{O}(h_i^{l}) + \mathcal{O}(h_\alpha^2) + \mathcal{O}(h_p) + \mathcal{O}(h_p^2/h_i) + \mathcal{O}(h_\alpha^2/h_i)$.

As mentioned in the main text, some of the numerical approximations that we use will introduce additional errors and reduce the order of accuracy of our method. Several of these approximations have been studied thoroughly in the literature (e.g.\ uGIMP integration; see \cite{bardenhagen2004,sadeghirad2011}) and a few were discussed in the previous section. With these approximations incorporated, the overall accuracy of the method should be of the following order: $\mathcal{O}(h_i) + \mathcal{O}(h_\alpha) + \mathcal{O}(h_p) + \mathcal{O}(h_p/h_i)$.

\subsection{Validation with Method of Manufactured Solutions}
In this section, we validate the convergence rates described above using the method of manufactured solutions (MMS). In the MMS, a solution to the governing equations is assumed \textit{a priori}; then the set of external forces and heats necessary to achieve this solution are determined analytically (see \cite{sadeghirad2011, wallstedt2008}). For the purposes of this numerical test, we adjust the governing equations in \eqref{eqn:mixture_equations} as follows:
\begin{equation}\label{eqn:mms_equations}
	\begin{aligned}
		\frac{D^s \bar{\rho}_s}{D t} &= -\bar{\rho}_s \divr(\boldsymbol{v}_s),\\
		\bar{\rho}_s \frac{D^s \boldsymbol{v}_s}{Dt} &= \divr(\boldsymbol{\tilde{\sigma}}) + \bar{\rho}_s \boldsymbol{b}_s - \boldsymbol{f_d} - \phi \nabla p_f,\\
		\frac{\partial \bar{\rho}_f}{\partial t} &= -\divr (\bar{\rho}_f \boldsymbol{v}_f),\\
		\frac{\partial \bar{\rho}_f \boldsymbol{v}_f}{\partial t} &= - \divr \big( \bar{\rho}_f\boldsymbol{v}_f \otimes \boldsymbol{v}_f + n p_f \boldsymbol{I} \big)  + \divr(\boldsymbol{\tau_f}) + \bar{\rho}_f \boldsymbol{b}_f + \boldsymbol{f_d} + p_f \nabla n,\\
		\frac{\partial \bar{\rho}_f E_f}{\partial t} &= -\divr \big((\bar{\rho}_f E_f + np_f) \boldsymbol{v}_f \big) + \divr(\boldsymbol{\tau_f} \boldsymbol{v}_f) + (\bar{\rho}_f \boldsymbol{b}_f) \cdot \boldsymbol{v}_f + \big(p_f \nabla n + \boldsymbol{f_d}\big) \cdot \boldsymbol{v}_s - {\phi p_f \divr(\boldsymbol{v}_s).} + q_f.
	\end{aligned}
\end{equation}
Here $\boldsymbol{b}_s$ and $\boldsymbol{b}_f$ represent the solid and fluid phase external force vectors, respectively, and $q_f$ represents the external heat flow into the fluid phase. We let the effective granular stress obey a pseudo-Neo-Hookean material model,
\begin{equation}
	\boldsymbol{\tilde{\sigma}} := G \boldsymbol{B}_0 + K \log(J) \boldsymbol{I},
\end{equation}
with $\boldsymbol{B}_0$ the deviator of the left Cauchy--Green tensor ($\boldsymbol{B} = \boldsymbol{F}\boldsymbol{F}^\top$ for the deformation gradient $\boldsymbol{F}$), $J$ a measure of volume change ($J^2 = \det(\boldsymbol{B})$), and $G$ and $K$ the shear and bulk moduli ($G = 3.8$ kPa and $K=8.3$ kPa); additionally, we let the fluid phase obey a standard ideal gas law,
\begin{equation}
	p_f := \rho_f R \vartheta_f, \quad \rho_f \varepsilon_f := \frac{p_f}{\gamma_r - 1}, \quad \text{and} \quad \boldsymbol{\tau_f} := \eta_0 (1 + \tfrac{5}{2}\phi) \big( \nabla \boldsymbol{v}_f + \nabla \boldsymbol{v}_f^\top - \tfrac{2}{3}\divr(\boldsymbol{v}_f)\boldsymbol{I} \big),
\end{equation}
with $R$ the specific gas constant ($R = 0.002871$ J/kg$\cdot$K), $\gamma_r$ the ratio of specific heats ($\gamma_r = 1.4$), and $\eta_0$ the dynamic fluid viscosity ($\eta_0 = 2$ Pa$\cdot$s). The final component of our mixture model is the inter-phase drag, $\boldsymbol{f}_d$, which we let take the form described in \cite{beetstra2007} for $d = 0.2$ m. 

To calculate the MMS, we first choose the desired solution fields for the solid and fluid phases of the mixture. For simplicity, we choose uniform density fields, $\rho_s^\dagger = 1000$ kg/m$^3$ and $\rho_f^\dagger = 117.7$ kg/m$^3$; a uniform porosity field, $n^\dagger=0.4$; and a uniform internal fluid energy field, $\varepsilon_f^\dagger = 21.4$ J/kg. We then choose two independent, divergence-free velocity fields:
\begin{equation}
	\label{eqn:mms_fluid_velocity}
	\boldsymbol{v}_f^\dagger = A(t)
	\begin{pmatrix}
		(x^2-1)^2 y (y^2-1)\\
		-x(x^2-1)(y^2-1)^2\\
		0
	\end{pmatrix}, \quad A(t) = -4t^2,
\end{equation}
and,
\begin{equation}
	\label{eqn:mms_solid_velocity}
	\boldsymbol{v}_s^\dagger = B(t)
	\begin{pmatrix}
		-y(x^2 + y^2 - 1)^2\\
		x(x^2 + y^2 - 1)^2\\
		0
	\end{pmatrix} \text{ if } x^2 + y^2 \leq 1, \quad B(t) = 4t^2,
\end{equation}
with $\boldsymbol{v}_s^\dagger = \boldsymbol{0}$ for $x^2 + y^2 > 1$. These velocity fields also have the useful property that $\boldsymbol{v}_s^\dagger = \boldsymbol{v}_f^\dagger = \boldsymbol{0}$ for $x = \pm 1$ or $y = \pm 1$.
Calculation of the necessary external forces ($\boldsymbol{b}_s$ and $\boldsymbol{b}_f$) and heat flow ($q_f$) to produce these solutions can be found by carefully inverting the system of equations in \eqref{eqn:mms_equations} and calculating the deformation gradient as follows,
\begin{equation}
	\boldsymbol{F} = \frac{\partial \boldsymbol{x}}{\partial \boldsymbol{X}}, \quad\text{and}\quad \frac{D^s \boldsymbol{x}}{Dt} = \boldsymbol{v}_s,
\end{equation}
with $\boldsymbol{x}$ a spatial point in the deformed solid phase material, which corresponds to a position $\boldsymbol{X}$ in the reference (or initial) material configuration.

\begin{figure}[!h]
	\centering
	\includegraphics[scale=0.4]{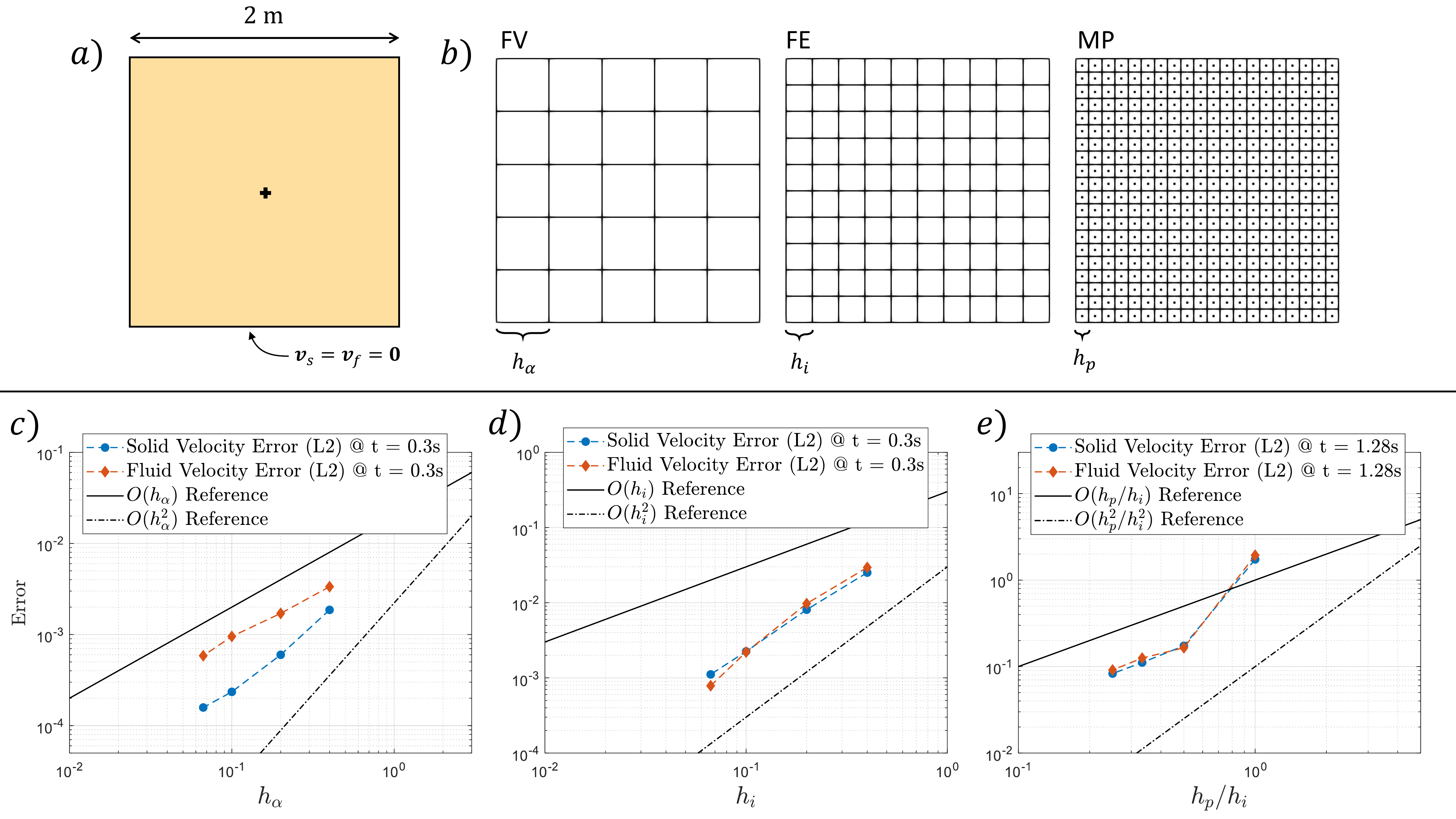}
	\caption{Method of manufactured solutions (MMS) results. a) Simulated domain and boundary conditions. b) FV, FE, and MP discretizations and associated characteristic lengths $h_\alpha$, $h_i$, and $h_p$. c) $L_2$-error vs. finite volume length scale, $h_\alpha$; $h_i$ and $h_p$ are constant. d) $L_2$-error vs. finite element length scale, $h_i$; $h_\alpha$ and $h_p/h_i$ are constant. e) $L_2$-error vs. material point refinement measure, $h_p/h_i$; $h_\alpha$ and $h_i$ are constant.}
	\label{fig:moms}
\end{figure}

With the analytical external forces and heat determined, we implement this problem in our FV-MPM framework. We use numerical quadrature to integrate the fluid body force and heat addition and --- as is standard in MPM --- apply the solid phase body force to the material points directly. This problem is simulated on a 2 m $\times$ 2 m domain centered at the origin and discretized using three overlapping, regular, Cartesian grids. The respective grid spacing is $h_p$ for the material points, $h_i$ for the finite element grid, and $h_\alpha$ for the finite volume grid. Standard bi-linear (or ``tent'') functions are used to define the finite element basis, $\mathcal{N}_i(\boldsymbol{x})$. Additionally, to enable larger time-increments ($\Delta t$), we use the alternative numerical algorithm detailed in Appendix \ref{sec:alternative_algorithm} to integrate the time derivatives of each field. The simulated domain, boundary conditions, and results are shown in Figure \ref{fig:moms}.

To analyze the convergence behavior of our method, we simulate this problem on twelve unique sets of meshes and calculate the $L_2$ norm of the velocity errors ($e_{\boldsymbol{v}_s}$ and $e_{\boldsymbol{v}_f}$) for each as follows,
\begin{equation}
	\|e_{\boldsymbol{v}_s}\|_{L_2} = \bigg[ \int_{\Omega} \|\boldsymbol{v}_s^\dagger - \boldsymbol{v}_s\|^2\ dv \bigg]^{\tfrac{1}{2}}, \quad \text{and} \quad \|e_{\boldsymbol{v}_f}\|_{L_2} = \bigg[ \int_{\Omega} \|\boldsymbol{v}_f^\dagger - \boldsymbol{v}_f\|^2\ dv \bigg]^{\tfrac{1}{2}}.
\end{equation}
The twelve sets of meshes are divided into three groups, one to test each of the convergence rates expected of our method: $\mathcal{O}(h_i) + \mathcal{O}(h_\alpha) + \mathcal{O}(h_p) + \mathcal{O}(h_p/h_i)$. (Note that the latter two convergence rates are grouped together and analyzed according to the more restrictive scale: $h_p/h_i$.) The $L_2$-errors calculated at 0.3 s simulated time for the first group of grids --- with $h_\alpha$ varying between $6.7$ cm and $40$ cm, $h_p/h_i$ fixed at 4, and $h_i$ fixed at 1.1 cm --- are shown in Figure \ref{fig:moms}c and have clear $\mathcal{O}(h_\alpha)$ convergence. The $L_2$-errors calculated at 0.3 s simulated time for the second group of grids --- with $h_i$ varying between $6.7$ cm and $40$ cm, $h_p/h_i$ fixed at 4, and $h_\alpha$ fixed at 3.3 cm --- are shown in Figure \ref{fig:moms}d and have clear $\mathcal{O}(h_i^2)$ convergence. This convergence rate exceeds our expectations, but is likely an artifact of the regular Cartesian grid we use in this particular simulation. And finally, the $L_2$-errors calculated at 1.28 s simulated time for the final group of grids --- with $h_p/h_i$ varying between 1 and 4, $h_i$ and $h_\alpha$ fixed at 3.3 cm --- are shown in Figure \ref{fig:moms}e and have clear $\mathcal{O}(h_p/h_i)$ convergence.

This validation of our method and implementation using the MMS shows that FV-MPM can reliably solve the governing equations in \eqref{eqn:mixture_equations} with at least first-order convergence in each discretization length scale. This low-order of convergence suggests that the method may be limited computationally for large complex flows; however, it does guarantee convergence for a wide range of problems which are extremely difficult to address with other numerical approaches (e.g.\ granular separation, solid-to-fluid and fluid-to-solid transition, long-duration simulations, etc.).

\setcounter{figure}{0}
\section{Notes on Numerical Stability}\label{sec:stability}
In this section, we briefly discuss the stability of the numerical method described in the main text. There are two primary factors that affect this stability for the classes of engineering problems we are interested in addressing: wave-propagation and dissipation. Assuming that the relevant features of a mixed flow solution are properly resolved, we can express the limitations on our discrete time increment ($\Delta t$) that are necessary to ensure stability when an explicit time integration approach is taken.

\subsection{Courant--Friedrichs--Lewy Condition}
The first condition is the Courant--Friedrichs--Lewy (CFL; see \cite{courant1928}) condition. As stated in the main text, this condition generally requires that the characteristic lines associated with a hyperbolic system of equations are appropriately resolved by the stencil of the numerical algorithm. In other words, the time increment must be small enough that the distance traveled by a physical wave (e.g.\ acoustic or elastic waves) does not exceed the relevant grid discretization length in a single time step. For the mixture problems we are generally interested in, this condition can be expressed as follows:
\begin{equation}
	\Delta t \leq \frac{c_{9} h_\alpha}{\big\|a_f + \|\boldsymbol{v}_f\|\big\|_\infty}, \quad \text{and} \quad \Delta t \leq \frac{c_{10} h_i}{\big\|a_s + \|\boldsymbol{v}_s\|\big\|_\infty},
\end{equation}
where $a_f$ and $a_s$ are the acoustic wave speeds in the fluid and solid phase, respectively; $c_{9}$ and $c_{10}$ are $\mathcal{O}(1)$ constants; $h_i$ is the characteristic length of the finite element grid; and $h_\alpha$ is the characteristic length of the finite volume grid. In our simulations, we generally assume $c_{9}$, $c_{10} \lesssim \tfrac{1}{5}$.

\subsection{Effective Fluid Shear Stress}
The second condition relates to the ability of our time integration procedure to resolve the rates of dissipation in a physical mixture. One contributing factor to dissipation in our mixed flow problems is the viscous shear stress in the fluid, $\boldsymbol{\tau_f}$. Although there many admissible forms of this stress, we will focus here on those that obey the relationship in \eqref{eqn:shear_stress_equation_fluid}:
\begin{equation*}
	\boldsymbol{\tau_f} = \eta_r \big( \nabla \boldsymbol{v}_f + \nabla \boldsymbol{v}_f^\top - \tfrac{2}{3}\divr(\boldsymbol{v}_f)\boldsymbol{I} \big).
\end{equation*}
When implemented numerically, as needed to solve \eqref{eqn:weak_mixture_equations_fluid}, the shear stress is approximated along the finite volume boundaries using either cell-centered or boundary-centered reconstructions. In either case, we can generally express the integrated shear stress along the boundary $\partial \Omega_{(\beta,\gamma)}$ (i.e.\ the cell boundary between volumes $\Omega_\beta$ and $\Omega_\gamma$) as follows,
\begin{equation}\label{eqn:shear_integral}
	\begin{aligned}
		\int_{\partial \Omega_{(\beta,\gamma)}}{ \boldsymbol{\tau_f}\boldsymbol{\hat{n}}_{\beta\to \gamma}\ da} \approx&\ \tfrac{1}{2} \eta_r dA_{(\beta,\gamma)} (C_{(\beta,\gamma)} - C_{(\gamma,\beta)})\bigg( \frac{\langle \bar{\rho}_f \boldsymbol{v}_f \rangle_\gamma}{\langle \bar{\rho}_f \rangle_\gamma} -  \frac{\langle \bar{\rho}_f \boldsymbol{v}_f \rangle_\beta}{\langle \bar{\rho}_f \rangle_\beta} \bigg)\\
		&+ \tfrac{1}{2} \eta_r dA_{(\beta,\gamma)} \sum_{\alpha\in S_\beta/\{\gamma\}} C_{(\beta,\alpha)} \bigg( \frac{\langle \bar{\rho}_f \boldsymbol{v}_f \rangle_\alpha}{\langle \bar{\rho}_f \rangle_\alpha} -  \frac{\langle \bar{\rho}_f \boldsymbol{v}_f \rangle_\beta}{\langle \bar{\rho}_f \rangle_\beta} \bigg)\\ 
		&+ \tfrac{1}{2} \eta_r dA_{(\beta,\gamma)} \sum_{\alpha\in S_\gamma/\{\beta\}} C_{(\gamma,\alpha)} \bigg( \frac{\langle \bar{\rho}_f \boldsymbol{v}_f \rangle_\alpha}{\langle \bar{\rho}_f \rangle_\alpha} -  \frac{\langle \bar{\rho}_f \boldsymbol{v}_f \rangle_\gamma}{\langle \bar{\rho}_f \rangle_\gamma} \bigg),
	\end{aligned}
\end{equation}
with $\boldsymbol{\hat{n}}_{\beta\to \gamma}$ the oriented face normal pointing from $\Omega_\beta$ to $\Omega_\gamma$; $dA_{(\beta,\gamma)}$ the area of the boundary; $C_{(\beta,\gamma)}$ a grid specific parameter used to calculate gradients between the centroids of $\Omega_\beta$ and $\Omega_\gamma$ (typically scales with $h_\alpha^{-1}$); and $S_\beta$ and $S_\gamma$ the set of neighbor indices for $\Omega_\beta$ and $\Omega_\gamma$, respectively. Here $(C_{(\beta,\gamma)} - C_{(\gamma,\beta)})$ is non-zero and positive.

When substituted into \eqref{eqn:weak_mixture_equations_fluid}, \eqref{eqn:shear_integral} represents one component of the momentum flux between the volumes $\Omega_\beta$ and $\Omega_\gamma$. For positive $(C_{(\beta,\gamma)} - C_{(\gamma,\beta)})$, this momentum flux will generally reduce the relative velocities between these neighboring cells; however, when an explicit time integrator (e.g.\ Forward Euler) is used with a large time increment ($\Delta t$), this diffusive flux can `over correct' velocity differences and become unstable. To ensure $\langle \bar{\rho}_f \boldsymbol{v}_f \rangle_\beta/\langle \bar{\rho}_f \rangle_\beta - \langle \bar{\rho}_f \boldsymbol{v}_f \rangle_\gamma/\langle \bar{\rho}_f \rangle_\gamma$ doesn't explode numerically, we require that,
\begin{equation}
	\Delta t \lesssim \frac{c_{11} h_\alpha^2}{\eta_r \big[ \langle \bar{\rho}_f \rangle_\beta^{-1} + \langle \bar{\rho}_f \rangle_\gamma^{-1} \big]}, \quad \forall \gamma \in S_\beta, \quad \forall \beta \in [1,N_v],
\end{equation}
with $c_{11}$ an $\mathcal{O}(2)$ constant and $h_\alpha$ the characteristic length of the finite volume grid. In our simulations, we generally assume $c_{11} \lesssim 1$.

\subsection{Effective Fluid Heat Flux}
Another contributing factor to dissipation in mixed flow problems is the diffusion of temperature through the heat flux $\boldsymbol{q}_f$. Although there many admissible forms of this heat flux, we will focus here on those that obey the relationship in \eqref{eqn:heat_flux_equation_fluid}:
\begin{equation*}
	\boldsymbol{q}_f = - n k_f \nabla \vartheta_f.
\end{equation*}
When implemented numerically, as needed to solve \eqref{eqn:weak_mixture_equations_fluid}, the heat flux is approximated along the finite volume boundaries using either cell-centered or boundary-centered reconstructions of $\nabla \vartheta_f$. For the ideal gas law considered in this work, we can express the reconstructed heat flux along the boundary $\partial \Omega_{(\beta,\gamma)}$ as follows,
\begin{equation}\label{eqn:heat_flux_integral}
	\begin{aligned}
		\int_{\partial \Omega_{(\beta,\gamma)}}{ -\boldsymbol{q}_f \cdot\boldsymbol{\hat{n}}_{\beta\to \gamma}\ da} \approx&\ \frac{nk_f}{2c_v} dA_{(\beta,\gamma)} (C_{(\beta,\gamma)} - C_{(\gamma,\beta)})\big(\langle \varepsilon_f \rangle_\gamma - \langle \varepsilon_f \rangle_\beta \big)\\
		&+ \frac{nk_f}{2c_v} dA_{(\beta,\gamma)} \sum_{\alpha\in S_\beta/\{\gamma\}} C_{(\beta,\alpha)} \big(\langle \varepsilon_f \rangle_\alpha - \langle \varepsilon_f \rangle_\beta \big)\\ 
		&+ \frac{nk_f}{2c_v} dA_{(\beta,\gamma)} \sum_{\alpha\in S_\gamma/\{\beta\}} C_{(\gamma,\alpha)} \big(\langle \varepsilon_f \rangle_\alpha - \langle \varepsilon_f \rangle_\gamma \big),\\[1em]
		\langle \varepsilon_f \rangle_\alpha \approx&\ \frac{\langle \bar{\rho}_f \rangle_\alpha \langle \bar{\rho}_f E_f \rangle_\alpha - \tfrac{1}{2}\langle \bar{\rho}_f \boldsymbol{v}_f \rangle_\alpha^2}{\langle \bar{\rho}_f \rangle_\alpha^2}, \quad \forall \alpha \in [1,N_v],
	\end{aligned}
\end{equation}
with $c_v$ the specific heat of the gas at constant volume and $\boldsymbol{\hat{n}}_{\beta\to \gamma}$, $dA_{(\beta,\gamma)}$, $C_{(\beta,\gamma)}$, and $S_\beta$ and $S_\gamma$ defined in the previous section.

When substituted into \eqref{eqn:weak_mixture_equations_fluid}, \eqref{eqn:heat_flux_integral} represents one component of the energy flux between volumes $\Omega_\beta$ and $\Omega_\gamma$. For positive $\big( C_{(\beta,\gamma)} - C_{(\gamma,\beta)}\big)$, this energy flux will generally reduce the difference between the specific internal energies of the neighboring cells; however, when an explicit time integrator (e.g.\ Forward Euler) is used with a large time increment, this diffusive term can `over correct' the energy difference and become unstable. To ensure $\langle \varepsilon_f \rangle_\beta - \langle \varepsilon_f \rangle_\gamma$ doesn't explode numerically, we require that,
\begin{equation}
	\Delta t \lesssim \frac{c_{11} c_v h_\alpha^2}{n k_f \big[ \langle \bar{\rho}_f \rangle_\beta^{-1} + \langle \bar{\rho}_f \rangle_\gamma^{-1} \big]}, \quad \forall \gamma \in S_\beta, \quad \forall \beta \in [1,N_v],
\end{equation}
with $c_{11}$ and $\mathcal{O}(2)$ constant and $h_\alpha$ the characteristic length of the finite volume grid.

\subsection{Inter-phase Drag Force}
The final form of dissipation that can affect the stability of our method is the inter-phase drag, $\boldsymbol{f_d}$, which tends to resist relative motion of the two mixed phases. An exact assessment of the stability condition relating to drag is beyond the scope of this Appendix; however, if we analyze the behavior of the approximate drag force vectors in \eqref{eqn:numerical_mpm_drag} and \eqref{eqn:numerical_fvm_drag},
\begin{equation*}
	{\boldsymbol{f}_i^{\text{drag}}}^* = -K_i^* \bigg( V_i \boldsymbol{v}_{si} - \sum_{\alpha=1}^{N_v} V_\alpha \mathcal{A}_{i\alpha} \langle \bar{\rho}_f \boldsymbol{v}_{f} \rangle_\alpha / \langle \bar{\rho}_f \rangle_\alpha \bigg),
\end{equation*}
\begin{equation*}
	{\boldsymbol{F}_\alpha^{\text{drag}}}^* = \sum_{i=1}^{N_n} \mathcal{A}_{i\alpha} K^*_i \big(\boldsymbol{v}_{si} - \langle \bar{\rho}_f \boldsymbol{v}_{f} \rangle_\alpha / \langle \bar{\rho}_f \rangle_\alpha \big) \cdot
	\begin{pmatrix}
		0\\
		1\\
		\boldsymbol{v}_{si}
	\end{pmatrix},
\end{equation*}
with $K_i^*$ defined in \eqref{eqn:numerical_inverse_permeability}, we can make a broad statements about when this term may become unstable and propose an adjustment to enhance its stability.

Consider the numerical expressions for momentum conservation in the solid and fluid phases from \eqref{eqn:momentum_evolution} and \eqref{eqn:fluid_state_vector_equation}. If we limit our analysis to methods that uses the diagonal matrices $[\mathcal{M}_D]$ and $[\mathcal{B}_D]$, then we have:
\begin{equation}\label{eqn:mpm_w_drag}
	{\mathcal{M}_D}_{ii} \frac{D^s \boldsymbol{v}_s}{Dt} = -K_i^* \bigg( V_i \boldsymbol{v}_{si} - \sum_{\alpha=1}^{N_v} V_\alpha \mathcal{A}_{i\alpha} \langle \bar{\rho}_f \boldsymbol{v}_{f} \rangle_\alpha / \langle \bar{\rho}_f \rangle_\alpha \bigg) + \boldsymbol{f}_i^{\text{int}} + \boldsymbol{f}_i^{\text{ext}} + \boldsymbol{f}_i^{\text{buoy}} + \boldsymbol{f}_i^{\boldsymbol{\tau}},
\end{equation}
and,
\begin{equation}\label{eqn:fvm_w_drag}
	\frac{d}{dt}
	{\small
		\begin{pmatrix}
			\langle \bar{\rho}_f \rangle_\alpha\\\
			\langle \bar{\rho}_f \boldsymbol{v}_f \rangle_\alpha\\
			\langle \bar{\rho}_f E_f \rangle_\alpha
		\end{pmatrix}
	}
	= \sum_{i=1}^{N_n} \mathcal{A}_{i\alpha} K^*_i \big(\boldsymbol{v}_{si} - \langle \bar{\rho}_f \boldsymbol{v}_{f} \rangle_\alpha / \langle \bar{\rho}_f \rangle_\alpha \big) \cdot
	{\small
	\begin{pmatrix}
		0\\
		1\\
		\boldsymbol{v}_{si}
	\end{pmatrix}} + \boldsymbol{F}_\alpha^{\text{int}} + \boldsymbol{F}_\alpha^{\text{ext}} + \boldsymbol{F}_\alpha^{\text{buoy}}.
\end{equation}
Integrating \eqref{eqn:mpm_w_drag} and \eqref{eqn:fvm_w_drag} over time will generally reduce the difference between the predicted solid phase and fluid phase velocities; however, when an explicit time integrator (e.g.\ Forward Euler) is used with a large time increment, these drag terms can `over correct' the velocity difference and produce an unstable scheme. To ensure $\boldsymbol{v}_s - \boldsymbol{v}_f$ doesn't explode numerically, we require that,
\begin{equation}
	\Delta t \leq \frac{2}{K_i^* \big[(\phi_i \rho_s)^{-1} - \langle \bar{\rho}_f \rangle_\alpha^{-1}\big]}, \quad \forall (i,\alpha) \in \big\{(j,\beta)\ |\ \mathcal{A}_{j\beta} \neq 0\big\},
\end{equation}
with $\phi_i = 1 - n_i$.

On the other hand, if \eqref{eqn:mpm_w_drag} and \eqref{eqn:fvm_w_drag} are integrated with an implicit time integrator (e.g.\ Backward Euler), substantially larger time increments could be taken without incurring an `over correction'. Such a time integration approach generally requires the construction and inversion of system spanning matrices; however, in the special case of the inter-phase drag, this inversion can be performed independently of the other forcing terms. This suggests that a semi-implicit time integrator can be developed that allows for stable time integration without requiring a complicated matrix inversion.

Here we propose such an integrator. (Note that a version of this semi-implicit drag is used in the alternative numerical algorithm described in the next section and in the simulations reported in sections \ref{sec:collapse}, \ref{sec:2d_erosion}, \ref{sec:3d_erosion}, and \ref{sec:rocket}.) This approach uses explicit time integration to determine an intermediate mixture state ($\langle \bar{\rho}_f \rangle_\alpha^*$, $\langle \bar{\rho}_f \boldsymbol{v}_f \rangle_\alpha^*$, $\langle \bar{\rho}_f E_f \rangle_\alpha^*$, and $\boldsymbol{v}_{si}^*$) between the time increments $s$ and $s+1$:
\begin{equation}\label{eqn:semi_implicit_eqn}
	\begin{aligned}
		{\small
			\begin{pmatrix}
				\langle \bar{\rho}_f \rangle_\alpha^*\\\
				\langle \bar{\rho}_f \boldsymbol{v}_f \rangle_\alpha^*\\
				\langle \bar{\rho}_f E_f \rangle_\alpha^*
			\end{pmatrix}
		} &:= {\small
		\begin{pmatrix}
			\langle \bar{\rho}_f \rangle_\alpha^s\\\
			\langle \bar{\rho}_f \boldsymbol{v}_f \rangle_\alpha^s\\
			\langle \bar{\rho}_f E_f \rangle_\alpha^s
		\end{pmatrix}
		} + \Delta t \big[ (\boldsymbol{F}_\alpha^{\text{int}})^s + 	(\boldsymbol{F}_\alpha^{\text{ext}})^s + (\boldsymbol{F}_\alpha^{\text{buoy}})^s\big],\\
		{\mathcal{M}_D}_{ii} \boldsymbol{v}_{si}^* &:= {\mathcal{M}_D}_{ii} \boldsymbol{v}_{si}^s + \Delta t \big[(\boldsymbol{f}_i^{\text{int}})^s + (\boldsymbol{f}_i^{\text{ext}})^s + (\boldsymbol{f}_i^{\text{buoy}})^s + (\boldsymbol{f}_i^{\boldsymbol{\tau}})^s\big],
	\end{aligned}
\end{equation}
with $(\cdot)^s$ denoting an evaluation of the forcing vectors at the $s$th time increment. With these intermediate states determined, the drag force vectors $({\boldsymbol{f}_i^{\text{drag}}}^*)^{s+1}$ and $({\boldsymbol{F}_\alpha^{\text{drag}}}^*)^{s+1}$ can be determined as follows,
\begin{equation}\label{eqn:implicit_mpm_drag}
	({\boldsymbol{f}_i^{\text{drag}}}^*)^{s+1} = -\tilde{K}_i^* \bigg( V_i \boldsymbol{v}_{si}^* - \sum_{\alpha=1}^{N_v} V_\alpha \mathcal{A}_{i\alpha} \langle \bar{\rho}_f \boldsymbol{v}_{f} \rangle_\alpha^* / \langle \bar{\rho}_f \rangle_\alpha^* \bigg),
\end{equation}
\begin{equation}\label{eqn:implicit_fvm_drag}
	({\boldsymbol{F}_\alpha^{\text{drag}}}^*)^{s+1} = \sum_{i=1}^{N_n} \mathcal{A}_{i\alpha} \tilde{K}^*_i \big(\boldsymbol{v}_{si}^* - \langle \bar{\rho}_f \boldsymbol{v}_{f} \rangle_\alpha^* / \langle \bar{\rho}_f \rangle_\alpha^* \big) \cdot
	\begin{pmatrix}
		0\\
		1\\
		\boldsymbol{v}^*_{si}
	\end{pmatrix},
\end{equation}
with,
\begin{equation}
	\tilde{K}_i^* \approx \frac{(K_i^*)^s}{1 + \Delta t\ (K_i^*)^s \big[\min_{\alpha\in\mathcal{V}_i}(\langle \bar{\rho}_f \rangle_\alpha)^{-1} + (\phi_i \rho_s)^{-1}\big]}, \quad \mathcal{V}_i = \{\alpha\ |\ \mathcal{A}_{i\alpha} \neq 0\}.
\end{equation}
This formulation is exact for locally uniform flows and ensures that the drag term remains stable regardless of the chosen time increment.

\setcounter{figure}{0}
\section{Alternative Numerical Algorithm}\label{sec:alternative_algorithm}
In this final section, we propose an alternative numerical algorithm for integrating the time-history of the solution coefficients in \eqref{eqn:discrete_coefficients}. (Note that this algorithm was used to generate the results described in sections \ref{sec:collapse}, \ref{sec:2d_erosion}, \ref{sec:3d_erosion}, and \ref{sec:rocket}.) We use a first-order, Forward Euler time integration method (with the update-stress-last MPM approach) for the solid phase equations of motion, we use a higher-order integration method (e.g.\ 4th-order Runge--Kutta) for the fluid phase equations, and we implement the semi-implicit drag forces described in the previous section. The benefits of this approach are a larger numerical stencil (particularly for problems where the limiting time-scale is associated with the propagation of acoustic waves in the fluid), increased accuracy in regions of the mixture without granular material, and the ability to use larger time increments when the drag forces in the mixture are relatively large.

Suppose the following parameters are known at $t^k$: (i) the mapping matrices, $[\mathcal{S}]^k$, $[\mathcal{G}]^k$, $[\mathcal{A}]$, $[\mathcal{B}_D]$, and $[\mathcal{M}_D]^k$; (ii) the solution coefficients,
$\bar{\rho}_{sp}^k,\ 
\boldsymbol{\tilde{\sigma}}_p^k,\ 
\boldsymbol{\bar{\xi}}_p^k,\ 
\langle\bar{\rho}_{f}\rangle_\alpha^k,\ 
\langle\bar{\rho}_{f}\boldsymbol{v}_f\rangle_\alpha^k,\ \text{and}\ 
\langle\bar{\rho}_{f}E_f\rangle_\alpha^k;
$ (iii) the coefficients of the material point approximation of the velocity field, $\boldsymbol{v}_{sp}^{*k}$; and (iv) the numerical representation of the material point characteristic functions, $U_p(\boldsymbol{x},t^k)$, (e.g.\
$v_p^k$ and $\boldsymbol{x}_p^k$ for basic MPM \cite{sulsky1994} and uGIMP \cite{bardenhagen2004}). We determine the value of the solution coefficient at $t^{k+1}$ according to the following steps:
\begin{enumerate}[label=(\arabic*)]
	\item Determine the mixture porosity coefficients, as in \eqref{eqn:weak_porosity}, using the diagonal matrix $[\mathcal{B}_D]$:
	\begin{equation}
		{\mathcal{B}_{D}}_{ii} (1 - n^k_i) \rho_s = \sum_{p=1}^{N_m} \mathcal{S}^k_{ip} m_p \quad \forall i \in [1,N_n].
	\end{equation}
	
	\item Determine the solid phase velocity coefficients, as in \eqref{eqn:velocity_projection}, using the diagonal matrix $[\mathcal{M}_D]$:
	\begin{equation}
		{\mathcal{M}_D}_{ii}^k \boldsymbol{v}_{si}^k = \sum_{p=1}^{N_m} \mathcal{S}^k_{ip} m_p \boldsymbol{v}_{sp}^{*k},  \qquad \forall i \in [1,N_n].
	\end{equation}
	
	\item Calculate the nodal force vectors, $(\boldsymbol{f}_i^{\text{int}})^k$ and $(\boldsymbol{f}_i^{\text{ext}})^k$, as in \eqref{eqn:momentum_evolution}.
	
	\item Estimate $({\boldsymbol{f}_i^\text{drag}}^*)^k$; $({\boldsymbol{F}_\alpha^\text{drag}}^*)^k$; and the full-step force vectors $({\boldsymbol{f}_i^\text{buoy}}^*)^k_1$, $(\boldsymbol{F}_\alpha^\text{int})^k_1$, $(\boldsymbol{F}_\alpha^\text{ext})^k_1$, and $(\boldsymbol{F}_\alpha^\text{buoy})^k_1$:
	\begin{enumerate}[label=\roman*)]
		\item Construct an approximation of the fluid phase fields, $\rho_f$, $\boldsymbol{v}_f$, $E_f$, and $\bar{\rho}_f$, within each finite volume cell, as in \eqref{eqn:barth_and_jespersen_flux_limiter}.
		
		\item Use the reconstructed fluid phase fields to calculate the nodal force vectors, $({\boldsymbol{f}_i^\text{drag}}^*)^k$ and $({\boldsymbol{f}_i^\text{buoy}}^*)^k_1$, as in \eqref{eqn:numerical_mpm_drag} and \eqref{eqn:numerical_mpm_buoyancy}, and the finite volume force vectors, $(\boldsymbol{F}_\alpha^\text{int})^k_1$, $(\boldsymbol{F}_\alpha^\text{ext})^k_1$, $({\boldsymbol{F}_\alpha^\text{drag}}^*)^k$, and $(\boldsymbol{F}_\alpha^\text{buoy})^k_1$, as in \eqref{eqn:fluid_state_vector_equation}, \eqref{eqn:discontinuous_flux_function}, and \eqref{eqn:numerical_fvm_drag}.
		
	\end{enumerate}
	
	\item Estimate the half-step force vectors $({\boldsymbol{f}_i^\text{buoy}}^*)^k_2$, $(\boldsymbol{F}_\alpha^\text{int})^k_2$, $(\boldsymbol{F}_\alpha^\text{ext})^k_2$, and $(\boldsymbol{F}_\alpha^\text{buoy})^k_2$:
	
	\begin{enumerate}[label=\roman*)]
		\item Calculate intermediate fluid phase field coefficients from the equations of motion as:
		\begin{equation}
			{\small
				\begin{pmatrix}
					\langle \bar{\rho}_f \rangle_\alpha^{k}\\
					\langle \bar{\rho}_f \boldsymbol{v}_f \rangle_\alpha^{k}\\
					\langle \bar{\rho}_f E_f \rangle_\alpha^{k}
			\end{pmatrix}}_1
			=
			{\small
				\begin{pmatrix}
					\langle \bar{\rho}_f \rangle_\alpha^{k}\\
					\langle \bar{\rho}_f \boldsymbol{v}_f \rangle_\alpha^{k}\\
					\langle \bar{\rho}_f E_f \rangle_\alpha^{k}
			\end{pmatrix}}
			+ \frac{\Delta t}{2} \big[ (\boldsymbol{F}_\alpha^{\text{int}})^k_1 + (\boldsymbol{F}_\alpha^{\text{ext}})^k_1 + (\boldsymbol{F}_\alpha^{\text{buoy}})^k_1 + ({\boldsymbol{F}_\alpha^{\text{drag}}}^*)^k \big], \quad \forall \alpha \in [1,N_v].
		\end{equation}
	
		\item Construct an intermediate approximation of the fluid phase fields, $\rho_f$, $\boldsymbol{v}_f$, $E_f$, and $\bar{\rho}_f$, within each finite volume cell, using $\{\langle \bar{\rho}_f \rangle_\alpha^k\}_1$, $\{\langle \bar{\rho}_f \boldsymbol{v}_f \rangle_\alpha^k \}_1$, and $\{\langle \bar{\rho}_f E_f \rangle_\alpha^k\}_1$.
		
		\item Use the reconstructed fluid phase fields to calculate the nodal force vector $({\boldsymbol{f}_i^\text{buoy}}^*)^k_2$, as in \eqref{eqn:numerical_mpm_buoyancy}, and the finite volume force vectors, $(\boldsymbol{F}_\alpha^\text{int})^k_2$, $(\boldsymbol{F}_\alpha^\text{ext})^k_2$, and $(\boldsymbol{F}_\alpha^\text{buoy})^k_2$, as in \eqref{eqn:fluid_state_vector_equation} and \eqref{eqn:discontinuous_flux_function}.
	\end{enumerate}

	\item Estimate the half-step force vectors $({\boldsymbol{f}_i^\text{buoy}}^*)^k_3$, $(\boldsymbol{F}_\alpha^\text{int})^k_3$, $(\boldsymbol{F}_\alpha^\text{ext})^k_3$, and $(\boldsymbol{F}_\alpha^\text{buoy})^k_3$:
	
	\begin{enumerate}[label=\roman*)]
		\item Calculate intermediate fluid phase field coefficients from the equations of motion as:
		\begin{equation}
			{\small
				\begin{pmatrix}
					\langle \bar{\rho}_f \rangle_\alpha^{k}\\
					\langle \bar{\rho}_f \boldsymbol{v}_f \rangle_\alpha^{k}\\
					\langle \bar{\rho}_f E_f \rangle_\alpha^{k}
			\end{pmatrix}}_2
			=
			{\small
				\begin{pmatrix}
					\langle \bar{\rho}_f \rangle_\alpha^{k}\\
					\langle \bar{\rho}_f \boldsymbol{v}_f \rangle_\alpha^{k}\\
					\langle \bar{\rho}_f E_f \rangle_\alpha^{k}
			\end{pmatrix}}
			+ \frac{\Delta t}{2} \big[ (\boldsymbol{F}_\alpha^{\text{int}})^k_2 + (\boldsymbol{F}_\alpha^{\text{ext}})^k_2 + (\boldsymbol{F}_\alpha^{\text{buoy}})^k_2 + ({\boldsymbol{F}_\alpha^{\text{drag}}}^*)^k \big], \quad \forall \alpha \in [1,N_v].
		\end{equation}
	
		\item Construct the second intermediate approximation of the fluid phase fields, $\rho_f$, $\boldsymbol{v}_f$, $E_f$, and $\bar{\rho}_f$, within each finite volume cell, using $\{\langle \bar{\rho}_f \rangle_\alpha^k\}_2$, $\{\langle \bar{\rho}_f \boldsymbol{v}_f \rangle_\alpha^k \}_2$, and $\{\langle \bar{\rho}_f E_f \rangle_\alpha^k\}_2$.
		
		\item Use the reconstructed fluid phase fields to calculate the nodal force vector $({\boldsymbol{f}_i^\text{buoy}}^*)^k_3$, as in \eqref{eqn:numerical_mpm_buoyancy}, and the finite volume force vectors, $(\boldsymbol{F}_\alpha^\text{int})^k_3$, $(\boldsymbol{F}_\alpha^\text{ext})^k_3$, and $(\boldsymbol{F}_\alpha^\text{buoy})^k_3$, as in \eqref{eqn:fluid_state_vector_equation} and \eqref{eqn:discontinuous_flux_function}.
	\end{enumerate}
	
	\item Estimate the full-step force vectors $({\boldsymbol{f}_i^\text{buoy}}^*)^k_4$, $(\boldsymbol{F}_\alpha^\text{int})^k_4$, $(\boldsymbol{F}_\alpha^\text{ext})^k_4$, and $(\boldsymbol{F}_\alpha^\text{buoy})^k_4$:
	
	\begin{enumerate}[label=\roman*)]
		\item Calculate the third intermediate fluid phase field coefficients from the equations of motion as:
		\begin{equation}
			{\small
				\begin{pmatrix}
					\langle \bar{\rho}_f \rangle_\alpha^{k}\\
					\langle \bar{\rho}_f \boldsymbol{v}_f \rangle_\alpha^{k}\\
					\langle \bar{\rho}_f E_f \rangle_\alpha^{k}
			\end{pmatrix}}_3
			=
			{\small
				\begin{pmatrix}
					\langle \bar{\rho}_f \rangle_\alpha^{k}\\
					\langle \bar{\rho}_f \boldsymbol{v}_f \rangle_\alpha^{k}\\
					\langle \bar{\rho}_f E_f \rangle_\alpha^{k}
			\end{pmatrix}}
			+ \Delta t \big[ (\boldsymbol{F}_\alpha^{\text{int}})^k_3 + (\boldsymbol{F}_\alpha^{\text{ext}})^k_3 + (\boldsymbol{F}_\alpha^{\text{buoy}})^k_3 + ({\boldsymbol{F}_\alpha^{\text{drag}}}^*)^k \big], \quad \forall \alpha \in [1,N_v].
		\end{equation}
	
		\item Construct the third intermediate approximation of the fluid phase fields, $\rho_f$, $\boldsymbol{v}_f$, $E_f$, and $\bar{\rho}_f$, within each finite volume cell, using $\{\langle \bar{\rho}_f \rangle_\alpha^k\}_3$, $\{\langle \bar{\rho}_f \boldsymbol{v}_f \rangle_\alpha^k \}_3$, and $\{\langle \bar{\rho}_f E_f \rangle_\alpha^k\}_3$.
		
		\item Use the reconstructed fluid phase fields to calculate the nodal force vector $({\boldsymbol{f}_i^\text{buoy}}^*)^k_4$, as in \eqref{eqn:numerical_mpm_buoyancy}, and the finite volume force vectors, $(\boldsymbol{F}_\alpha^\text{int})^k_4$, $(\boldsymbol{F}_\alpha^\text{ext})^k_4$, and $(\boldsymbol{F}_\alpha^\text{buoy})^k_4$, as in \eqref{eqn:fluid_state_vector_equation} and \eqref{eqn:discontinuous_flux_function}.
	\end{enumerate}
	
	\item Estimate the semi-implicit force vectors $({\boldsymbol{f}_i^\text{drag}}^*)^{k+1}$ and $({\boldsymbol{F}_\alpha^\text{drag}}^*)^{k+1}$:
	
	\begin{enumerate}[label=\roman*)]
		\item Calculate the intermediate fluid field coefficients for the semi-implicit calculation in \eqref{eqn:semi_implicit_eqn}:
		\begin{equation}
			\begin{aligned}
				{\small
					\begin{pmatrix}
						\langle \bar{\rho}_f \rangle_\alpha^{*}\\
						\langle \bar{\rho}_f \boldsymbol{v}_f \rangle_\alpha^{*}\\
						\langle \bar{\rho}_f E_f \rangle_\alpha^{*}
				\end{pmatrix}}
				=&\ 
				{\small
					\begin{pmatrix}
						\langle \bar{\rho}_f \rangle_\alpha^{k}\\
						\langle \bar{\rho}_f \boldsymbol{v}_f \rangle_\alpha^{k}\\
						\langle \bar{\rho}_f E_f \rangle_\alpha^{k}
				\end{pmatrix}}
				+ \frac{\Delta t}{6} \big[ (\boldsymbol{F}_\alpha^{\text{int}})^k_1 + (\boldsymbol{F}_\alpha^{\text{ext}})^k_1 + (\boldsymbol{F}_\alpha^{\text{buoy}})^k_1\big]\\
				&+ \frac{\Delta t}{3} \big[ (\boldsymbol{F}_\alpha^{\text{int}})^k_2 + (\boldsymbol{F}_\alpha^{\text{ext}})^k_2 + (\boldsymbol{F}_\alpha^{\text{buoy}})^k_2\big]\\
				&+ \frac{\Delta t}{3} \big[ (\boldsymbol{F}_\alpha^{\text{int}})^k_3 + (\boldsymbol{F}_\alpha^{\text{ext}})^k_3 + (\boldsymbol{F}_\alpha^{\text{buoy}})^k_3\big]\\
				&+ \frac{\Delta t}{6} \big[ (\boldsymbol{F}_\alpha^{\text{int}})^k_4 + (\boldsymbol{F}_\alpha^{\text{ext}})^k_4 + (\boldsymbol{F}_\alpha^{\text{buoy}})^k_4\big], \quad \forall \alpha \in [1,N_v].
			\end{aligned}
		\end{equation}
		
		\item Calculate the intermediate solid velocity coefficients for the semi-implicit calculation in \eqref{eqn:semi_implicit_eqn}:
		\begin{equation}
			\begin{aligned}
				{\mathcal{M}_D}_{ii}^k \boldsymbol{v}_{si}^* = {\mathcal{M}_D}_{ii}^k \boldsymbol{v}_{si}^k& + \Delta t \big[ (\boldsymbol{f}_i^\text{int})^k + (\boldsymbol{f}_i^\text{ext})^k + {(\boldsymbol{f}_i^{\boldsymbol{\tau}})^k}
				\big]\\
				& + \frac{\Delta t}{6} \big[ ({\boldsymbol{f}_i^\text{buoy}}^*)^k_1 + 2 ({\boldsymbol{f}_i^\text{buoy}}^*)^k_2 + 2 ({\boldsymbol{f}_i^\text{buoy}}^*)^k_3 + ({\boldsymbol{f}_i^\text{buoy}}^*)^k_4\big],\ \forall i \in [1,N_n].
			\end{aligned}
		\end{equation}
	
		\item Use these intermediate coefficients to calculate the nodal force vector $({\boldsymbol{f}_i^\text{drag}}^*)^{k+1}$, as in \eqref{eqn:implicit_mpm_drag}, and the finite volume force vector  $({\boldsymbol{F}_\alpha^\text{drag}}^*)^{k+1}$, as in \eqref{eqn:implicit_fvm_drag}.		
	\end{enumerate}
	
	\item Now calculate the updated fluid phase field coefficients from the equations of motion as:
	\begin{equation}
		\begin{aligned}
		{\small
			\begin{pmatrix}
				\langle \bar{\rho}_f \rangle_\alpha^{k+1}\\
				\langle \bar{\rho}_f \boldsymbol{v}_f \rangle_\alpha^{k+1}\\
				\langle \bar{\rho}_f E_f \rangle_\alpha^{k+!}
		\end{pmatrix}}
		=&\ 
		{\small
			\begin{pmatrix}
				\langle \bar{\rho}_f \rangle_\alpha^{k}\\
				\langle \bar{\rho}_f \boldsymbol{v}_f \rangle_\alpha^{k}\\
				\langle \bar{\rho}_f E_f \rangle_\alpha^{k}
		\end{pmatrix}}
		+ \frac{\Delta t}{6} \big[ (\boldsymbol{F}_\alpha^{\text{int}})^k_1 + (\boldsymbol{F}_\alpha^{\text{ext}})^k_1 + (\boldsymbol{F}_\alpha^{\text{buoy}})^k_1 + ({\boldsymbol{F}_\alpha^{\text{drag}}}^*)^{k+1} \big]\\
		&+ \frac{\Delta t}{3} \big[ (\boldsymbol{F}_\alpha^{\text{int}})^k_2 + (\boldsymbol{F}_\alpha^{\text{ext}})^k_2 + (\boldsymbol{F}_\alpha^{\text{buoy}})^k_2 + ({\boldsymbol{F}_\alpha^{\text{drag}}}^*)^{k+1} \big]\\
		&+ \frac{\Delta t}{3} \big[ (\boldsymbol{F}_\alpha^{\text{int}})^k_3 + (\boldsymbol{F}_\alpha^{\text{ext}})^k_3 + (\boldsymbol{F}_\alpha^{\text{buoy}})^k_3 + ({\boldsymbol{F}_\alpha^{\text{drag}}}^*)^{k+1} \big]\\
		&+ \frac{\Delta t}{6} \big[ (\boldsymbol{F}_\alpha^{\text{int}})^k_4 + (\boldsymbol{F}_\alpha^{\text{ext}})^k_4 + (\boldsymbol{F}_\alpha^{\text{buoy}})^k_4 + ({\boldsymbol{F}_\alpha^{\text{drag}}}^*)^{k+1} \big], \quad \forall \alpha \in [1,N_v].
		\end{aligned}
	\end{equation}
	
	\item Determine the solid phase acceleration coefficients associated with the finite element grid nodes from the equations of motion and appropriate boundary conditions:
	\begin{equation}
		\begin{aligned}
		{\mathcal{M}_D}_{ii}^k \boldsymbol{a}_{si}^k =&\  (\boldsymbol{f}_i^\text{int})^k + (\boldsymbol{f}_i^\text{ext})^k + ({\boldsymbol{f}_i^\text{drag}}^*)^{k+1} + \boldsymbol{f}_i^{\boldsymbol{\tau}}\\
		& +\frac{1}{6} \big[ ({\boldsymbol{f}_i^\text{buoy}}^*)^k_1 + 2 ({\boldsymbol{f}_i^\text{buoy}}^*)^k_2 + 2 ({\boldsymbol{f}_i^\text{buoy}}^*)^k_3 + ({\boldsymbol{f}_i^\text{buoy}}^*)^k_4\big], \quad \forall i \in [1,N_n].
		\end{aligned}
	\end{equation}
	
	\item Use an explicit time integrator to obtain the updated solid phase velocity coefficients associated with the finite element grid nodes; treat the grid as though it were Lagrangian:
	\begin{equation}
		\boldsymbol{v}_{si}^{k+1'} = \boldsymbol{v}_{si}^k + \Delta t\  \boldsymbol{a}_{si}^k, \quad \forall i \in [1,N_n].
	\end{equation}
	
	\item Update the solid phase material state vector, $\boldsymbol{\bar{\xi}}_p$, and effective granular stress, $\boldsymbol{\tilde{\sigma}}_p$, according to the relevant constitutive update procedure. For stability, $\boldsymbol{L}_p$, the average material point velocity gradient is often used:
	\begin{equation}
		\boldsymbol{L}_p^{k+1} = \sum_{i=1}^{N_n} \boldsymbol{v}_{si}^{k+1'} \otimes \mathcal{G}^k_{ip}, \quad \boldsymbol{\tilde{\sigma}}_p^{k+1} = \boldsymbol{T}\big( \boldsymbol{L}_p^{k+1}, \bar{\xi}_p^{k+1} \big), \quad \forall p \in [1,N_m].
	\end{equation}
	
	\item Map the solid phase nodal accelerations and velocities to the material points to update their velocity approximations, positions, and densities:
	\begin{equation}
		\begin{aligned}
			\boldsymbol{v}_{sp}^{*k+1} &= \boldsymbol{v}_{sp}^{*k} + \Delta t \sum_{i=1}^{N_n} \mathcal{S}_{ip}^k \boldsymbol{a}_{si}^{k}, \quad \forall p \in [1,N_m],\\
			\boldsymbol{x}_{p}^{k+1} &= \boldsymbol{x}_p^k + \Delta t \sum_{i=1}^{N_n} \mathcal{S}_{ip}^k \boldsymbol{v}_{si}^{k+1'}, \quad \forall p \in [1,N_m],\\
			v_{p}^{k+1} &= v_p^k \bigg( 1 + \Delta t \sum_{i=1}^{N_n} \mathcal{G}_{ip}^k \cdot \boldsymbol{v}_{si}^{k+1'}\bigg), \quad \forall p \in [1,N_m],\\[1em]
			\bar{\rho}_{sp}^{k+1} &= m_p / v_p^{k+1}, \quad \forall p \in [1,N_m].
		\end{aligned}
	\end{equation}
	
	\item Update the diagonal mass matrix, $[\mathcal{M}_D]^{k+1}$, and mapping matrices, $[\mathcal{S}]^{k+1}$ and $[\mathcal{G}]^{k+1}$, according to their definitions in \eqref{eqn:mapping_matrices}.
	
	\item Increment $k$ to $k+1$; go to (1).
	
\end{enumerate}

This algorithm is consistent with the governing equations in \eqref{eqn:mixture_equations} and has similar stability conditions to the algorithm in the main text; however, the primary advantage of this approach is improved stability in the fluid phase of the simulation and increased accuracy in regions of the mixed flow that are without granular material.

\section{Artificial Viscosity for Strong Shocks}\label{sec:artificial_viscosity}
The use of artificial (i.e.\ numerical) viscosity is common in gas dynamics simulations for problems involving strong shocks (see \cite{wilkins1980}). For the rocket exhaust impingement simulation in the main text, we use a simple shock-capturing scheme based on the works of \cite{mccorquodale2011, pandolfi2001, barter2010}. In these approaches, the numerical equations for the fluid phase of the mixture is augmented as follows:
\begin{equation}\label{eqn:fvm_artificial_viscosity}
	\frac{d}{dt}
	{\small
		\begin{pmatrix}
			\langle \bar{\rho}_f \rangle_\alpha\\\
			\langle \bar{\rho}_f \boldsymbol{v}_f \rangle_\alpha\\
			\langle \bar{\rho}_f E_f \rangle_\alpha
		\end{pmatrix}
	}
	= \boldsymbol{F}_\alpha^{\text{int}} + \boldsymbol{F}_\alpha^{\text{ext}} + \boldsymbol{F}_\alpha^{\text{drag}} + \boldsymbol{F}_\alpha^{\text{buoy}} + \boldsymbol{F}_\alpha^{v}.
\end{equation}
Here $\boldsymbol{F}_\alpha^v$ represents the additional flux arising from the artificial viscosity and takes the following form:
\begin{equation}
	\boldsymbol{F}_\alpha^{v} = \int_{\partial \Omega_\alpha} \frac{\epsilon_v}{h} (\boldsymbol{u}^+ - \boldsymbol{u}^-)\ da,
\end{equation}
with $\boldsymbol{u}^+$ and $\boldsymbol{u}^-$ the two reconstructed fluid state vectors on each side of the boundary, $\partial \Omega_\alpha$; $\epsilon_v$ the scalar artificial viscosity; and $h$ the local grid length scale. (Note that $\epsilon_v$ only appears in the fraction $\epsilon_v/h$ and is defined below.) In our implementation we use the following shock capturing form of $\epsilon_v$:
\begin{equation}
	\begin{aligned}
		\frac{\epsilon_v}{h} &= \lambda_{\text{max}} \min\bigg(\frac{(h \lambda)^2}{\beta^2 a_{\text{min}}^2},\ 1\bigg),\\
		h \lambda &= 
		\begin{cases}
			|a_f^+ - a_f^-| - (\boldsymbol{v}_f^+ - \boldsymbol{v}_f^-)\cdot \hat{\boldsymbol{n}}, & \text{if} \quad (\boldsymbol{v}_f^+ - \boldsymbol{v}_f^-)\cdot \hat{\boldsymbol{n}} < 0\\
			0, & \text{else}
		\end{cases},
	\end{aligned}
\end{equation}
with $\beta = 0.3$; $\hat{\boldsymbol{n}}$ the oriented face normal; $\boldsymbol{v}_f^+$ and $\boldsymbol{v}_f^-$ the reconstructed fluid velocities on each side of the boundary; $a_f^+$ and $a_f^-$ the local acoustic wave speeds associated with these reconstructions; $a_{\text{min}}$ the smaller of these two values; and
\begin{equation}
	\lambda_{\text{max}} = \max \big(a^+ + \|\boldsymbol{v}_f^+\|,\ a^- + \|\boldsymbol{v}_f^-\| \big).
\end{equation}

\clearpage
\bibliographystyle{elsarticle-num-names}
\bibliography{references}

\end{document}